\documentclass[a4paper,11pt]{article}
\pdfoutput=1 

\usepackage{jheppub} 
\usepackage{amsmath,amssymb,amsthm}
\usepackage{graphicx}
\usepackage{subcaption}
\usepackage[percent]{overpic}
\usepackage{color}
\usepackage{multirow}
\usepackage{slashed}
\usepackage{mathrsfs}
\usepackage{tikz}

\newcommand{\slk}{\operatorname{slk}}
\newcommand{\R}{\mathbb{R}}
\newcommand{\C}{\mathbb{C}}
\newcommand{\xUDarrow}[1]{%
 {\left\updownarrow\vbox to #1{}\right.\kern-\nulldelimiterspace}
}

\makeatletter
\newcommand\xLRarrow[2][]{%
  \ext@arrow 9999{\longleftrightarrowfill@}{#1}{#2}}
\newcommand\longleftrightarrowfill@{%
  \arrowfill@\leftarrow\relbar\rightarrow}
 \newcommand{\xdashleftrightarrow}[2][]{\ext@arrow 3359\leftrightarrowfill@@{#1}{#2}}
 \def\leftrightarrowfill@@{\arrowfill@@\leftarrow\relbar\rightarrow}
\def\arrowfill@@#1#2#3#4{%
  $\m@th\thickmuskip0mu\medmuskip\thickmuskip\thinmuskip\thickmuskip
   \relax#4#1
   \xleaders\hbox{$#4#2$}\hfill
   #3$%
}
\makeatother

\newtheorem{conjecture}{Conjecture}[section]

\definecolor{purple}{rgb}{0.7,0,0.7}

\def\be{\begin{equation}}
\def\ee{\end{equation}}
\def\IR{{\mathbb{R}}}
\def\IZ{{\mathbb{Z}}}
\def\IP{{\mathbb{P}}}
\def\IC{{\mathbb{C}}}

\def\CM{{\mathcal{M}}}
\def\CN{{\mathcal{N}}}

\def\CZ{\mathcal{Z}}

\def\tCW{\widetilde{\mathcal{W}}}
\def\Li{\mathrm{Li}}

\def\vort{\mathrm{vortex}}

\def\eff{\mathrm{eff}}

\renewcommand{\(}{\left(}
\renewcommand{\)}{\right)}

\title{Physics and geometry of knots-quivers correspondence}

\author[a,b]{Tobias Ekholm}
\author[a]{Piotr Kucharski}
\author[c,d]{Pietro Longhi}

\affiliation[a]{Department of Mathematics, Uppsala University, Box 480, 751 06 Uppsala, Sweden}
\affiliation[b]{Institut Mittag-Leffler, Aurav 17, 182 60 Djursholm, Sweden}
\affiliation[c]{Institute for Theoretical Physics, ETH Zurich, CH - 8093, Zurich, Switzerland}
\affiliation[d]{Department of Physics and Astronomy, Uppsala University, Box 516, 751 20 Uppsala, Sweden}

\emailAdd{tobias.ekholm@math.uu.se, piotr.kucharski@math.uu.se, longhip@phys.ethz.ch}

\abstract{
The recently conjectured knots-quivers correspondence \cite{Kucharski:2017poe,Kucharski:2017ogk} relates gauge theoretic invariants of a~knot $K$ in the~3-sphere to the representation theory of a~quiver $Q_{K}$ associated to the~knot. In this paper we provide geometric and physical contexts for this conjecture within the~framework of Ooguri-Vafa large $N$ duality \cite{Ooguri:1999bv}, that relates knot invariants to counts of holomorphic curves with boundary on $L_{K}$, the~conormal Lagrangian  of the~knot in the~resolved conifold, and corresponding M-theory considerations. From the~physics side, we show that the~quiver encodes a~3d $\CN=2$ theory $T[Q_{K}]$ whose low energy dynamics arises on the~worldvolume of an~M5 brane wrapping the~knot conormal and we match the~(K-theoretic) vortex partition function of this theory with the~motivic generating series of $Q_{K}$.
From the~geometry side, we argue that the~spectrum of (generalized) holomorphic curves on $L_{K}$ is generated by a~finite set of basic disks. These disks correspond to the~nodes of the~quiver $Q_{K}$ and the~linking of their boundaries to the~quiver arrows. We extend this basic dictionary further and propose a~detailed map between quiver data and topological and geometric properties of the~basic disks that again leads to matching partition functions. 
We also study generalizations of A-polynomials associated to $Q_{K}$ and (doubly) refined version of LMOV invariants \cite{Ooguri:1999bv,LM0004,LMV0010,AV1204,Fuji:2012nx}.
}

\begin{document} 

\maketitle

\section{Introduction}

Over the~last 25 years, relations between knot theory and string theory, see e.g.~\cite{Ooguri:1999bv,Witten:1992fb}, have revealed deep interconnections between physics and mathematics. 
This paper provides physical and geometric underpinnings for a~recently conjectured correspondence in this area that relates knots to quivers \cite{Kucharski:2017poe,Kucharski:2017ogk}, see also \cite{KS1608,LZ1611,Zhu1707}.

The basic incarnation of the~correspondence relates symmetrically colored HOMFLY-PT polynomials of a~knot $K\subset S^{3}$ to Poincar\'e polynomials of the quiver representation varieties of a~quiver associated to $K$, which we denote by~$Q_{K}$. Deeper aspects of the~correspondence involve relations between quiver data and knot homologies, as well as important integrality statements.
The conjectured correspondence is motivated entirely by  empirical evidence: knot data and quiver data are computed separately and shown to coincide, see~\cite{Kucharski:2017poe,Kucharski:2017ogk}. In this paper we take the~first steps toward a~conceptual understanding of the~knots-quivers correspondence, relating quiver arrows and vertices (with extra data) directly to both physical and geometric objects. 
We approach this from two different angles: on the one hand through the physics of open topological strings and gauge theories engineered by corresponding branes, and on the other hand through the geometry of holomorphic curves attached to the brane.

\begin{figure}[h!]
\begin{center}
\includegraphics[width=0.95\textwidth]{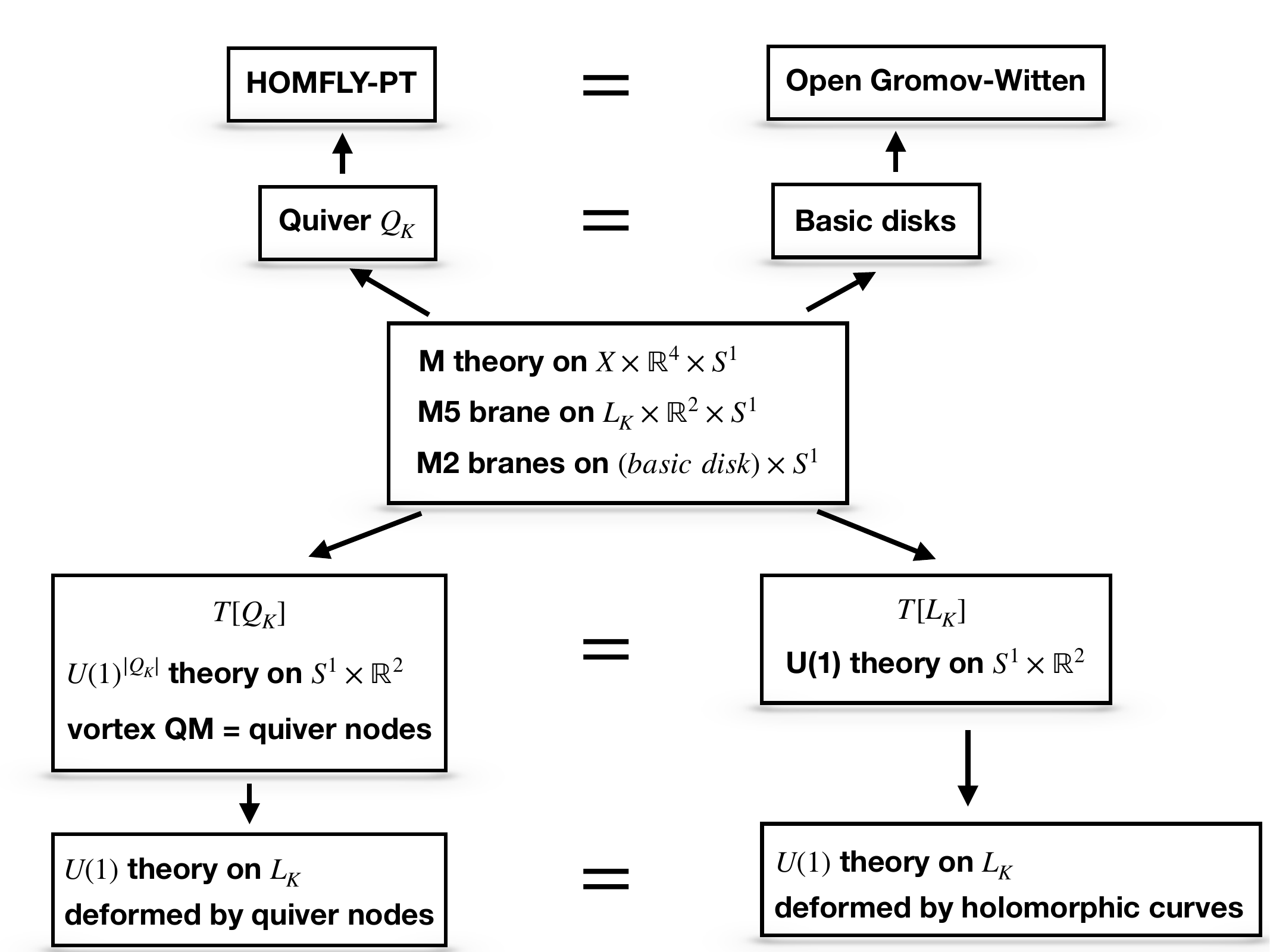}
\caption{Physics and geometry of knots and quivers -- schematic overview}
\label{Scheme}
\end{center}
\end{figure}

\subsection{Physical picture}

As shown in Figure \ref{Scheme}, the starting point of our physical and geometric considerations is the~large $N$ description of colored HOMFLY-PT polynomials as Gromov-Witten invariants of the~knot conormal $L_{K}$ in the~resolved conifold~$X$, see \cite{Ooguri:1999bv,Aganagic:2013jpa,ES}. From the~M-theory point of view, the~generating series of HOMFLY-PT polynomials counts M2-branes wrapping holomorphic curves with boundary on an~M5-brane wrapping the~knot conormal. 
The complete M-theory background is $X\times S^1\times \R^4$, where the~M5-brane wraps $L_K\times S^1\times \R^2$ and the~M2-branes wrap the~product of a~holomorphic curve and~$S^{1}$.

There are two effective descriptions of the~low energy dynamics on the~M5-brane, see Figure \ref{Scheme}. First, in terms of a~$U(1)$ Chern-Simons gauge theory on $L_K$, which from the~geometric point of view is ordinary $U(1)$ Chern-Simons theory deformed by certain embedded holomorphic disks, introducing curvature concentration along their boundaries. Second, in terms of 
a~3d~$\CN=2$ theory $T[L_K]$ on $S^1\times \IR^2$ \cite{Dimofte:2010tz,Terashima:2011qi,Dimofte:2011ju,Yag1305,LY1305,Cordova:2013cea}. 

We introduce a novel dual description of $T[L_K]$, as an~Abelian Chern-Simons-matter theory $T[Q_K]$, which features a~$U(1)$ gauge group and a~single fundamental chiral for each node of $Q_K$. 
Interactions between the~sectors corresponding to single nodes are governed by Chern-Simons couplings (encoded in the~quiver arrows) and Fayet-Ilioupoulos (FI) couplings. Recall that the~FI couplings may also be viewed as mixed Chern-Simons terms between the~gauge $U(1)$ factors and their dual topological symmetries.
Global symmetries of $T[Q_{K}]$ include rotations of the~base $\IC \IP^1$ of $X$, as well as rotations of $\IR^2\times\IR^2 \simeq \IR^4$ twisted by a~$U(1)_R$  action on $X$.  Fugacities of these symmetries are often denoted by $(a,q,t)$ in the~context of knot polynomials \cite{Fuji:2012nx,DGR0505}. In this context, the~FI couplings give the~change of variables $x_i \sim a^{a_i} q^{q_i-t_i} t^{t_i}$, where $x_{i}$ is the~variable associated to the~$i^{\rm th}$ quiver node in $T[Q_{K}]$, that identifies the~Poincar\'e generating series of quiver representation varieties of the~series of symmetric-colored superpolynomials. Here the~HOMFLY-PT polynomial is recovered by specializing to $t=-1$. In other words, FI couplings encode the~relation between global and topological symmetries of $T[Q_K]$.

The dual description $T[Q_K]$ encoded by the quiver has several advantages over descriptions of type $T[L_K]$, previously adopted in the literature. One of these is the fact that the structure of $T[Q_K]$ is fully and explicitly known, while $T[L_K]$ can be rather complicated and not fully understood even for simple knots.
Another important feature of $T[Q_K]$ is the fact that it bridges between counts of holomorphic curves (via duality with $T[L_K]$) and quiver descriptions of corresponding BPS states, hence providing a key link in the knots-quivers correspondence.
The M2-branes with boundary on the~M5-brane give rise to BPS vortices of $T[Q_K]$. It was recently observed that the~BPS vortex spectrum of certain 3d \mbox{$\CN=2$} theories admits a~quiver quantum mechanics description \cite{Hwang:2017kmk}. Here we think of the~vortex quantum mechanics of $T[Q_K]$ as the~physical origin of the~quiver $Q_K$. The existence of a~quiver description of vortices then leads to the~quiver description of knot invariants observed in \cite{Kucharski:2017poe,Kucharski:2017ogk}.
The quiver of vortices and $Q_K$ arise as effective descriptions of the~underlying M-theory system, which consists of M2-branes ending on an~M5-brane. 
The BPS vortex spectrum of the~theory $T[Q_K]$ is the~shadow (effective  description) of this theory  on the~ $\R^{2}\times S^{1}$ part of the~M5-brane, and is governed by the $\CN=2$ quantum mechanics of vortices.
The~quiver $Q_K$, on the~other hand, is the~shadow of the~theory on $X$ and describes the~ 
$\CN=4$ quantum mechanical dynamics of M2-branes that are standardized (stretched in sympletic language) near $L_K$. 

Quiver descriptions of BPS spectra are known to arise in string theory as a way to encode all BPS states as boundstates of a~finite set of basic BPS generators. Such descriptions are well understood in the~context of 4d \mbox{$\CN=2$} quantum field theories, where BPS generators are identified with nodes of the quiver, and intersections between corresponding M2-branes are encoded by arrows connecting the~nodes \cite{Denef:2002ru,Alim:2011kw,Gabella:2017hpz}. We will see that the this framework is line with interpretations of $Q_K$ given in this paper.

\subsection{Geometric picture}

Another core result of the present work is the introduction of a~geometric point of view behind the quiver description of holomorphic curves (and by extension, knot invariants). We first describe the geometric objects involved and then formulate a precise conjecture. The~quiver nodes or BPS generators correspond to basic embedded holomorphic disks with boundary on $L_{K}$ and with a framing along the boundary. The Gromov-Witten potential counts generalized holomorphic cvurves \cite{Ekholm:2018iso,iacovino1,iacovino2,iacovino3} with boundary on $L_{K}$. From the point of view of~\cite{ES}, generalized holomorphic curves are bare curves with boundary in the $U(1)$ skein module projected to homology class and linking (encoded in the $q$-power).   

For the quiver description, the holomorphic curve components of all generalized holomorphic curves lie in a~small neighborhood of the~union of $L_{K}$ with the basic disks attached. This space can be thought of as a deformation of $T^{\ast}L_{K}$, which describes a symplectic neighborhood of $L_{K}$ when there are no non-constant holomorphic disks ending on it. 
Here each sector of $T[Q_K]$, consisting of a~$U(1)$ gauge theory and its fundamental chiral, corresponds to ordinary $U(1)$ Chern-Simons theory deformed by all generalized holomorphic curves that contain at least one copy of a fixed basic holomorphic disk.

The FI couplings for the~$i^{\rm th}$ basic holomorphic disk correspond (roughly) to its homology class in $H_{2}(X)$ ($a_i$) and to the~Euler characteristics of the~curves in a~neighborhood of the~disk, $e^{q_{i}\frac12g_{s}}$ related to intersections with a~4-chain $C_{K}$ with boundary $2\cdot L_{K}$ ($q_{i}$ and $t_{i}$). The Chern-Simons couplings encoded in quiver arrows here correspond to linking and self-linking numbers of boundaries of basic holomorphic disks as embedded curves in $L_K$, where linking numbers are defined only up to a choice of longitude on the boundary torus. Then the~quiver expression for the~partition function corresponds to the~count of all generalized holomorphic curves constructed from linked configurations of the~basic disks. The 4-chain $C_{K}$ mentioned above is a~familiar object in the~context of knot contact homology \cite{Ekholm:2018iso}. From the~physical point of view, it is reminiscent of a~family of Dirac strings and we discuss an~interpretation in that spirit from the~M-theory perspective in Section~\ref{sub:4-chain-physics}. 

The above discussion can be summarized and made precise in a conjecture that we state next. 
Let $\psi_{K}(x,a,e^{\frac12g_{s}})$ denote the Gromov-Witten partition function counting all (disconnected) generalized holomorphic curves ending on $L_{K}$. This has the well-known structure of a generating series
\begin{equation}\label{eq:GW=HOMFLY}
\psi_{K}(x,a,e^{\frac12g_{s}})=\sum_{r\ge 0}P^{K}_{r}(a,q)x^{r},
\end{equation}
where $P^{K}_{r}(a,q)$ is the HOMFLY-PT polynomial of $K$ in the $r^{\rm th}$ symmetric representation \cite{Ooguri:1999bv,ES,Ekholm:2018iso}. 
We may also consider a refinement of this partition function, denoted $\Psi_{K}(x,a,q,t)$, where the coefficient of $x^{n}$ is the superpolynomial $\mathcal{P}^{K}_{r}(a,q,t)$ (Poincar\'e polynomial of the HOMFLY-PT homology \cite{DGR0505}), so that $\psi_{K}(x,a,q)=\Psi_{K}(x,a,q,t=-1)$. 

\begin{conjecture}\label{conj1}
Let $K\subset S^{3}$ be a framed knot and let $L_{K}\subset X$ be its Lagrangian conormal shifted off the zero section and considered a Lagrangian in the resolved conifold. We conjecture that there is a finite number of embedded holomorphic disks $D_{1},\dots, D_{m}$ ending on $L_{K}$ with framed boundaries such that the following holds. 

Let $Q_{K}$ be the symmetric quiver with nodes $D_{i}$ and arrows $C_{ij}$ corresponding to linking (and self-linking) numbers $\mathrm{lk}(\partial D_{i},\partial D_{j})$ measured with respect to the framing of $\partial L_{K}$. Let $x_{i}$ be the quiver variable of $D_{i}$, $a_{i}$ denote the homology class of $D_{i}$ in $H_{2}(X)\subset H_{2}(X,L_{K})$, $n_{i}$ the homology class $\partial D_{i}$ in $H_{1}(L_{K})$, and $q_{i}-C_{ii}$ the intersection number $D_{i}\cdot C_{K}$. Then we have
\begin{equation} 
\Psi_{K}(x,a,q,t)=P^{Q_K}(x_{1},\dots,x_{m},q)|_{x_{i}=x^{n_{i}}a^{a_{i}}q^{q_{i}-C_{ii}}(-t)^{C_{ii}}},
\end{equation}
where the right hand side is the quiver partition function, see \eqref{eq:derivation-3d-N2-theory}.
\end{conjecture}

We comment on possible proofs of Conjecture \ref{conj1}. For $K=0_{1}$, the unknot, the conjecture holds: after representing the unknot conormal as a toric brane there are exactly two holomorphic disks ending on it, see Section \ref{sec:Examples}. For more general knots it is likely that one can find a finite number of disks for the conjecture by degenerating the conormal to the unknot conormal as a braid representative approaches the unknot. Provided the finite collection of disks are found, the Gromov-Witten level of the conjecture could likely be proved combining \cite{ES,Ekholm:2018iso} with \cite{OP,KL}. We still lack the geometric definition and properties of the refined partition function $\Psi_{K}(x,a,q,t)$, but they are currently being investigated. 

The geometric picture in Conjecture \ref{conj1} passes several nontrivial checks, for example its behavior under changes of framing. It also raises new questions, in particular about invariance under deformations of $L_{K}$ and ways of finding all basic disks in general. For simple knots there is one basic disk for each monomial in the~HOMFLY-PT polynomial. For more complicated knots that is no longer the~case and the~quiver description is more involved. We discuss these questions with details in a~couple of explicit examples.

\subsection{Outline of the paper}

The outline of the~paper is as follows. 
Section \ref{sec:background} contains a~review of relevant material on knots, quivers, BPS states, and holomorphic disks. 
In Section \ref{sec:physics} we introduce the~quiver gauge theory $T[Q_K]$ associated to the~quiver dual to a~knot, and show that its BPS vortex spectrum is captured by the~representation theory of $Q_K$.
In Section \ref{sec:math} we provide a~geometric interpretation of quiver representations in terms of generalized holomorphic disks and discuss identification of quiver data with linking and self-linking numbers.
Section~\ref{sec:Examples} illustrates our viewpoint on the~knots-quivers correspondence on the~simplest examples of the~unknot and the~trefoil.  
We conclude in Section \ref{sec:discussion} with a~discussion and suggestions for future work.

\section*{Acknowledgements}
We would like to thank Sergei Gukov, H\'{e}lder Larragu\'{i}vel, Fabrizio Nieri, Mi{\l}osz Panfil, Du Pei, Ingmar Saberi, Marko Sto\v{s}i\'{c}, Piotr Su{\l}kowski, and Paul Wedrich for insightful discussions.
We are also grateful to the organizers of the conferences ``6th International Workshop on Combinatorics of Moduli Spaces, Cluster Algebras, and Topological Recursion'' at Steklov Mathematical Institute, and ``Quantum Fields, knots, and strings" at the University of Warsaw, where the results of this paper were presented.
Parts of the paper is based upon work supported by the National Science Foundation under Grant No. DMS-1440140 while the authors T.E. and P.K. were in residence at the Mathematical Sciences Research Institute in Berkeley, California, during the 2018 spring semester.
P.L.  thanks Aarhus QGM, the Aspen Center for Physics, ENS Paris, Caltech, The University of California at Berkeley, and Trinity College Dublin for hospitality during completion of this work.
The work of T.E. is supported by the~Knut and Alice Wallenberg Foundation and the~Swedish Reserach Council.
P.K.~acknowledges support from the~Knut and Alice Wallenberg Foundation.
The work of P.L. is supported by the~NCCR SwissMAP, funded by the~Swiss National Science Foundation. 
P.L. also acknowledges support from grants ``Geometry and Physics''and ``Exact Results in Gauge and String Theories'' from the~Knut and Alice Wallenberg Foundation during part of this work.

\section{Background}\label{sec:background}

\subsection{Knots-quivers (KQ) correspondence}\label{sub:Knots-quivers-correspondence}

If $K\subset S^{3}$ is a~knot, then its HOMFLY-PT polynomial $P^K(a,q)$ \cite{freyd1985,PT} is a~2-variable polynomial that is easily calculated from a~knot diagram (a projection of $K$ with over/under information at crossings) via the~skein relation. The polynomial is a~knot invariant, 
i.e.~invariant under isotopies and in particular independent of diagrammatic presentation. More general knot invariants are the~colored HOMFLY-PT polynomials $P^K_{R}(a,q)$, where $R$ is a~representation of the~Lie algebra ${\mathfrak{u}}(N)$. Also the~colored version admits a~diagrammatic description in terms of standard polynomial of certain satellite links of $K$. In this setting, the~original HOMFLY-PT corresponds to the~standard representation. Below, to simplify notation, we will often write simply the~HOMFLY-PT polynomial also when we refer to the~more general colored version.

From the~physical point of view, the~HOMFLY-PT polynomial is the~expectation value of the~knot viewed as a~Wilson line in $U(N)$ Chern-Simons gauge theory \cite{witten1989}, which then depends on a~choice of representation $R$ for the~Lie algebra ${\mathfrak{u}}(N)$. Here we will restrict attention to completely symmetric representations, corresponding to Young diagrams with a~single row with $r$ boxes. For each $r$-box representation we get a~polynomial $P_{r}^{K}(a,q)$ and we consider the~ \emph{HOMFLY-PT generating series} in the~variable $x$
\be\label{eq:HOMFLY-PT series}
	P^{K}(x,a,q)=\sum_{r=0}^{\infty}P_{r}^{K}(a,q)x^{r}\,.
\ee
In this setting, the~Labastida-Mari\~{n}o-Ooguri-Vafa (LMOV) invariants \cite{Ooguri:1999bv,LM0004,LMV0010} are certain numbers 
assembled into the~LMOV generating function $N^{K}(x,a,q)=\sum_{r,i,j}N_{r,i,j}^{K}x^{r}a^{i}q^{j}$ that gives the~following expression for the~HOMFLY-PT generating series
\begin{equation}\label{P^K=Exp}
P^{K}(x,a,q)=\mathrm{Exp}\left(\frac{N^{K}(x,a,q)}{1-q^{2}}\right)\,.
\end{equation}
$\textrm{Exp}$ is the~plethystic exponential -- if $f=\sum_{n}a_{n}t^{n}$, $a_{0}=0$, then
\[
\mathrm{Exp}\bigl(f\bigr)(t)=  
\exp\left(\sum_{k}\tfrac{1}{k}f(t^{k})\right) =\prod_{n}(1-t^{n})^{n \,a_{n}}.
\]
According to the \emph{LMOV conjecture} \cite{Ooguri:1999bv,LM0004,LMV0010}  $N_{r,i,j}^{K}$ are integer numbers.

The knots-quivers (KQ) correspondence introduced in \cite{Kucharski:2017poe,Kucharski:2017ogk} and mentioned in the~previous section provides a~new approach to HOMFLY-PT polynomials and LMOV invariants. We give a~brief discussion. 

A \emph{quiver} $Q$ is an~oriented graph with a~finite set of vertices $Q_0$ connected by finitely many arrows (oriented edges) $Q_1$. 
A \emph{dimension vector} for $Q$ is a~vector in the~integral lattice with basis $Q_{0}$, $\mathbf{d}\in \IZ Q_0$. We number the~vertices of $Q_{0}$ by $1,2,\dots,m=|Q_{0}|$. A~\emph{quiver representation with dimension vector $\mathbf{d}=(d_{1},\dots,d_{m})$} is the~assignment of a~vector space of dimension $d_i$ to the~node $i\in Q_0$ and of a~linear map $\gamma_{ij}\colon\IC^{d_i} \to \IC^{d_j}$ to each arrow in~$Q_{1}$ from vertex $i$ to vertex $j$. The \emph{adjacency matrix} of $Q$ is the~$m\times m$ integer matrix with entries $C_{ij}$ equal to the~number of arrows from $i$ to $j$. A quiver is symmetric if its adjacency matrix is. 

Quiver representation theory studies moduli spaces of stable quiver representations (see e.g. \cite{kirillov2016quiver} for an~introduction to this subject).
While explicit expressions for invariants describing those spaces are hard to find in general, they are quite well understood in
the case of symmetric quivers \cite{KS0811,KS1006,MR1411,FR1512,2011arXiv1103.2736E}. Important information about the~moduli space
of representations of a~symmetric quiver with trivial potential is encoded in the~\emph{motivic generating series} defined as 
\be\label{eq:Efimov}
	P^{Q}(\mathbf{x},q)=\sum_{d_{1},\ldots,d_{m}\geq0}(-q)^{\sum_{1\leq i,j\leq m}C_{ij}d_{i}d_{j}}\prod_{i=1}^{m}\frac{x_{i}^{d_{i}}}{(q^{2};q^{2})_{d_{i}}} \,,
\ee
where the~denominator is the~so-called $q$-Pochhammer symbol
\be
	(z;q^2)_r  = \prod_{s=0}^{r-1} (1-z q^{2s}) \,.
\ee
Sometimes we will call $P^{Q}(\mathbf{x},q)$ the~quiver partition function. We also point out that quiver representation theory involves the~choice of an~element, the~potential, in the~path algebra of the~quiver, and that the~trivial potential is the~zero element.

Furthermore, there are so called motivic Donaldson-Thomas (DT) invariants $\Omega_{\mathbf{d},s}^{Q}=\Omega_{(d_{1},...,d_{m}),s}^{Q}$
which can be assembled into the~DT generating function
\be
\Omega^{Q}(\mathbf{x},q)=\sum_{\mathbf{d},s}\Omega_{\mathbf{d},s}^{Q}\mathbf{x^d}q^{s}(-1)^{|\mathbf{d}|+s+1},
\ee
where $\mathbf{x^d}=\prod_{i}x_{i}^{d_{i}}$. These give the~following new expression for the~motivic generating series
\begin{equation}\label{P^Q=Exp}
P^{Q}(\mathbf{x},q)=\textrm{Exp}\left(\frac{\Omega^{Q}(\mathbf{x},q)}{1-q^{2}}\right).
\end{equation}

The DT invariants have two geometric interpretations,
either as the~intersection homology Betti numbers of the~moduli space of all
semi-simple representations of $Q$ of dimension vector $\mathbf{d}$,
or as the~Chow-Betti numbers of the~moduli space of all simple representations
of $Q$ of dimension vector $\mathbf{d}$, see \cite{MR1411,FR1512}.
In \cite{2011arXiv1103.2736E} there is a~proof that these invariants are positive integers.

The most basic version of the~conjectured knot-quiver correspondence is the~statement that for each knot $K$ there exists a~quiver $Q_{K}$ and integers $\{a_{i},q_{i}\}_{i\in {Q_{K}}_{0}}$, such that
\be\label{eq:KQ-corr-basic}
	\left.P^{Q_{K}}(\mathbf{x},q)\right|_{x_{i}=x a^{a_{i}}q^{q_{i}-C_{ii}}}=P^{K}(x,a,q) \,.
\ee
We call $x_{i}=x a^{a_{i}}q^{q_{i}-C_{ii}}$ the~\emph{KQ change of variables}. The purpose of the~shift by the~number of loops and the~meaning of $a_{i},q_{i}$ are discussed in Section \ref{sub:Knot-homologies}.

In \cite{Kucharski:2017poe,Kucharski:2017ogk} there are also refined versions of the~KQ correspondence, as well as the~one on the~level of LMOV and DT
invariants. We can obtain it by substituting \eqref{P^K=Exp} and \eqref{P^Q=Exp} into \eqref{eq:KQ-corr-basic} 
\begin{equation}\label{eq:KQDT}
\left.\Omega^{Q_{K}}(\mathbf{x},q)\right|_{x_{i}=x a^{a_{i}}q^{q_{i}-C_{ii}}}=N^{K}(x,a,q)\,.
\end{equation}
Since DT invariants are integer, this equation implies the LMOV conjecture. 
In Sections~\ref{BPSintro},~\ref{Theories T[M_K] and T[L_K]},~\ref{subsec:TofQ} we will see that the~physical meaning of (\ref{eq:KQDT}) is that DT and LMOV invariants count BPS states in dual 3d $\mathcal{N}=2$ theories. We also have a~geometrical interpretation of these invariants as counts of what we call semi-basic holomorphic disks that can loosely be described as embedded generalized holomorphic disks.

We stress that the~KQ correspondence is conjectural, and that it is currently not known how to construct the~quiver $Q_{K}$ from a~given knot $K$. Evidence for the~conjecture includes checks on infinite families of torus and twist knots. A proof for 2-bridge knots appeared recently in \cite{Stosic:2017wno}, whereas \cite{PSS1802} explores the~relation to combinatorics of counting paths. On the~other hand, \cite{PS18} contains a relation between quivers and topological strings on various Calabi-Yau manifolds.
In~this paper we study the~KQ correspondence from the~point of view of large $N$ transition and discuss how to interpret the~quiver in terms of gauge theoretic reductions of M-theory and in terms of Calabi-Yau reductions as basic holomorphic disks in the~spirit of \cite{Ooguri:1999bv,Gopakumar:1998ii,Gopakumar:1998jq}.

\subsection{BPS states}\label{BPSintro}

Both sides of the~KQ correspondence have physics counterparts schematically shown in diagram \eqref{eq:KQ-diagram}. First, knots as Wilson loops in Chern-Simons theory are related to open topological strings with branes on the~knot cononormal $L_{K}$ and the~zero-section $S^{3}$ in the~cotangent bundle $T^{\ast}S^{3}$ \cite{Witten:1992fb,ES}. Via large $N$ duality, this open string in $T^{\ast}S^{3}$  is further related to the~open string in the~resolved conifold $X$, where $S^{3}$ disappeared and only branes on $L_{K}$ remained, see \cite{Ooguri:1999bv,ES}. Second, quiver representation theory appears in the~description of how BPS states in several contexts, e.g.~in string theory and supersymmetric QFT \cite{Denef:2002ru,Douglas:1996sw,Fiol:2000wx,Alim:2011ae}, generate more general states.

\be\label{eq:KQ-diagram}
\begin{array}{ccc}
	\text{knots} & \quad \xdashleftrightarrow{\ \ \text{\footnotesize KQ corr.} \ \ } \quad & \text{quiver rep. theory} \\
	\xUDarrow{.6cm} & & \xUDarrow{.6cm}\\
	\text{topological strings} & \quad  \xLRarrow{\ \ \text{\footnotesize \phantom{KQ corr.}}\ \ } \quad & \text{BPS vortices} \\
\end{array}
\ee

In this paper we study the~origin of the~KQ correspondence from the~viewpoint of these related physics pictures. This section gives a~brief overview of these subjects, see Section~\ref{sec:physics} for new results followed by a~more detailed discussion.

A more precise characterization of the~relation between knots and topological strings is as follows.  $U(N)_\kappa$ Chern-Simons theory on $S^3$ ($\kappa$ denotes the~level) is related to topological strings on the~resolved conifold $X$ with the~following matching of parameters: 
\be
	g_{s} = \frac{2\pi i}{\kappa+N} \,,
	\qquad\qquad
	\mathtt{t} = \frac{2\pi i N}{\kappa+N} \,,
\ee
where $g_{s}$ denotes the~string coupling constant and $\mathtt{t}$ is the~K\"ahler parameter of~$X$.

The equality of vacuum partition functions for these theories led to the~conjecture that these theories are exactly equivalent in the~large $N$ limit \cite{Witten:1992fb,ES,GV9811}. 
Inserting a~Wilson loop supported on a~knot $K\subset S^3$ on the~Chern-Simons side corresponds to considering an~A-brane supported on the~Lagrangian conormal $L_K\subset T^{\ast}S^{3}$ shifted off the~zero-section and then considered as a~submanifold of~$X$. Here,   
Wilson loop expectation values correspond to open topological string amplitudes for Riemann surfaces with boundary on $L_K$. Let us multiply $X$ by $\R^{4}\times S^{1}$ and consider the~open topological string as a~reduction of M-theory. The relation to open strings can then be interpreted as a~relation between Chern-Simons knot invariants, like the~HOMFLY-PT polynomial, and the~BPS spectrum of M2-branes (corresponding to the~most basic holomorphic curves) ending on M5-branes wrapped on~$L_K$~\cite{Ooguri:1999bv}.

From the~perspective of mirror symmetry we can construct a~B-model mirror of $X$ as a~conic bundle over a~complex torus $\C^{\ast}_{\xi}\times\C^{\ast}_{\eta}$, corresponding to the~two generators $\xi$~and~$\eta$ of $H_{1}(\Lambda_{K})$, where $\Lambda_{K}$ is the~torus at infinity of $L_{K}$, that degenerates over a~curve $V_K$. (The standard notation is $(x,p)$ or $(x,y)$, see e.g.~\cite{Aganagic:2013jpa}, but we use $\xi$ and $\eta$ not to confuse with our other uses of $x$.)
This curve is known as the~mirror curve and can be thought of as the~moduli space of $L_K$ deformed by disk instanton corrections \cite{Aganagic:2000gs}.
The curve $V_K$ comes equipped with a~canonical differential $\eta d\xi$. The open topological string wavefunction 
\be
\Psi_K(\xi)=\exp\(\sum_{n} C_{n}(e^{\mathtt{t}},g_{s})e^{n\xi}\)
\ee
counts (generalized) holomorphic curves in $X$ with boundary on $L_{K}$, see \cite{Ekholm:2018iso}. Up to conventions and the~change of variables 
\be
e^{\mathtt{t}}=a^2,\qquad e^{g_s}=q^2,\qquad e^{\xi}=x\,,
\ee
the wavefunction $\Psi_K(\xi)$ is equal to the~HOMFLY-PT generating series $P^{K}(x,a,q)$, see equation \eqref{eq:GW=HOMFLY}.
 
The semiclassical limit $g_s\to 0$ of $\Psi_{K}(\xi)$ recovers the~Gromov-Witten disk potential $W_{K}(\xi)$ that is computed by the~Abel-Jacobi map on the~mirror curve \cite{Aganagic:2000gs,Aganagic:2001nx}
\be
	\Psi_K(\xi) \sim \exp \(\frac{1}{g_s} \int \eta d\xi + \dots \) = \exp \(\frac{1}{g_s} W_{K}(\xi) + \dots \)\,.
\ee
As mentioned above, basic holomorphic curves with boundary on $L_{K}$ appear as reductions of M2-branes wrapping the~curve ending on M5-branes which wrap~$L_K$. 

We next consider another reduction of M-theory: BPS counting of M2-branes can be formulated in terms of world volume dynamics on the~M5-brane wrapped on $L_K\times \IR^2 \times S^1$. 
The corresponding low energy theory $T[L_K]$ is a~3d $\CN=2$ Chern-Simons matter gauge theory on $\IR^2\times S^1$. Its field content and couplings are determined by the~geometry of $L_K\subset X$ \cite{Dimofte:2010tz,Dimofte:2011ju}.
These low energy theories have interesting spectra of BPS vortices counted by LMOV invariants. BPS vortices arise from M2-branes that wrap holomorphic curves with boundary on $L_K$ and that stretch along the~$S^{1}$-direction in $X\times\R^{4}\times S^{1}$. In fact, vortices can be localized at the~origin of the~worldvolume $\IR^2$ by turning on an~$\Omega$-background~\cite{Shadchin:2006yz}.

BPS states are the~lightest charged particles in a~supersymmetric QFT, characterized by the~requirement that their mass is linearly proportional to their charge under gauge and global symmetries \cite{Witten:1978mh}. 
In consequence, BPS states are invariant under half of the~supersymmetry (two supercharges in our case), leading to additional constraints on their dynamics. (These constraints give rise to interesting phenomena typical of BPS states, like wall-crossing.)
BPS dynamics play a~fundamental role in the~characterization of the~3d~$\CN=2$ BPS vortex spectrum. 
Generally speaking, global symmetries act on the~Hilbert space of BPS states $\mathcal{H}^{\rm BPS}$ which is therefore  naturally graded by the~corresponding charges (such as spin or magnetic flux): $\mathcal{H}^{{\rm BPS}} = \bigoplus_{\gamma} \mathcal{H}^{{\rm BPS}}_\gamma$.

In addition to the~natural grading by charges, there is a~second and more refined type of grading on $\mathcal{H}^{{\rm BPS}}$, which depends on the~details of the~theory.
While the~spectrum of BPS vortices is often infinite, under certain conditions it can be organized into boundstates of a~finite set $Q_0$ of fundamental BPS states of lowest charge, thus introducing a~$\IZ Q_0$ grading on $\mathcal{H}^{{\rm BPS}}$ \cite{Hwang:2017kmk}.
A bound state consisting of $d_i$ copies of the~$i$-th fundamental vortex is labeled by a~dimension vector $\mathbf{d}$, and its properties (e.g.~the number and spin of 'internal' configurations, or its BPS degeneracies) can be modeled by the~world line quantum mechanics of the~multi-particle system. Let $QM({\mathbf{d}})$ denote this theory. 

Supersymmetry imposes constraints on the~types of multiplets and interactions in the~quantum mechanical description, for more details see \cite{Hori:2014tda}. 
In the~case of $\CN=4$ quantum mechanics, the~interactions are governed by an~integer matrix $C_{ij}$ with $i,j\in Q_0$ and a~superpotential $W$. This data can be encoded in a~quiver, as described in the~previous section. 
In particular, the~problem of computing the~supersymmetric vacua of $QM({\mathbf{d}})$ reduces precisely to the~study of representation theory of $Q$ corresponding to the~dimension vector $\mathbf{d}$. This then gives the~rightmost arrow in diagram \eqref{eq:KQ-diagram}, connecting quivers to BPS vortices of 3d $\CN=2$ theories $T[L_K]$. In Section \ref{sec:physics} we give explicit descriptions of these theories, and of their BPS vortex spectra.

More precisely, while BPS vortices of $T[L_K]$ can be described by a~quiver, this is not yet the~one appearing in the~KQ correspondence.
In fact, theories with $\CN=2$ supersymmetry enjoy a~rich duality web, and one of the~main messages of this paper is that $T[L_K]$ has a~dual description $T[Q_K]$ whose vortex spectrum is described by the~quiver quantum mechanics of the~quiver $Q_K$ of \cite{Kucharski:2017poe,Kucharski:2017ogk}. The full extent of this relation will be explored in Section~\ref{sec:physics}.

Another important point is that the~vortex spectrum of a~3d $\CN=2$ theory is generically described by $\CN=2$ quantum mechanics. This admits a~quiver description too, albeit with more than one type of arrow connecting the~nodes \cite{Hwang:2017kmk}. 
The quiver $Q_K$ on the~other hand should be regarded as encoding data of a~$\CN=4$ quantum mechanics. 
The relation between this and the~$\CN=2$ vortex quantum mechanics may be roughly summarized by saying that the~former describes the~dynamics of holomorphic disks, and the~latter the~dynamics of vortices. 
Such a~relation between quivers with different amounts of supersymmetry appears to be novel, it is illustrated and further discussed in Section \ref{subsec:TL-TQ} with an~explicit example.

To conclude the~overview, it is worth noting that a~$\CN=4$ quiver quantum mechanics description of boundstates of M2-branes wrapping holomorphic curves with boundaries on an~M5-brane appeared in the~context of BPS spectra of 4d $\CN=2$ theories \cite{Alim:2011kw,Alim:2011ae}. In fact this is not unrelated to our setup, we will return to this point in Section \ref{sec:discussion}.

\subsection{Knot homologies and A-polynomials\label{sub:Knot-homologies}}

The physics and geometry of the KQ correspondence involves not
only HOMLFY-PT polynomials and LMOV invariants, but also other knot-theoretical
objects like HOMFLY-PT homology and A-polynomials. 

The first well understood knot homologies were introduced in \cite{Khovanov,KhR1,KhR2},
however for the~KQ correspondence the~most relevant is HOMFLY-PT
homology which was proposed in \cite{DGR0505} as a~categorification
of the~uncolored reduced HOMLFY-PT polynomial

\begin{equation}
P_{1}^{K,\textrm{reduced}}(a,q)=\sum_{i,j}(-1)^{k}a^{i}q^{j}\dim\mathscr{H}_{i,j,k}^{\textrm{reduced}}(K).\label{eq:HOMFLY-PT as Euler}
\end{equation}
Considering the~Poincar\'{e} polynomial instead of the~Euler characteristic
provides a~$t$-refinement leading to knot invariant called the~(uncolored
reduced) superpolynomial
\begin{equation}
\mathcal{P}_{1}^{K,\textrm{reduced}}(a,q,t)=\sum_{i,j,k}a^{i}q^{j}t^{k}\dim\mathscr{H}_{i,j,k}^{\textrm{reduced}}(K).\label{eq:superpolynomial as Poincare}
\end{equation}

We work with unreduced normalization and consider the~finite dimensional
unreduced homology $\mathscr{H}(K)$ \cite{GNSSS1512}, whose Poincar\'{e}
polynomial is obtained by multiplying the~(reduced) superpolynomial
by the~unknot numerator, i.e. $1+a^{2}t$. 
(Our convention for the~unknot polynomial, which
affects all normalizations of unreduced knot-theoretical objects, will be further detailed in Sections \ref{sub:Unknot}--\ref{sub:Trefoil}.)
Therefore 
\begin{equation}
\mathcal{P}_{1}^{K}(a,q,t)=\frac{\sum_{i,j,k}a^{i}q^{j}t^{k}\dim\mathscr{H}_{i,j,k}(K)}{1-q^{2}}=\frac{\sum_{i\in\mathscr{G}(K)}a^{a_{i}}q^{q_{i}}t^{t_{i}}}{1-q^{2}},\label{eq:Superpolynomial and homology generators}
\end{equation}
where we write the~sum over $\mathscr{G}(K)$ -- the set of generators of $\mathscr{H}(K)$ -- to extract powers
$a_{i}$, $q_{i}$, and $t_{i}$. The first two turn out to be integers that determine
the~KQ change of variables mentioned in Section \ref{sub:Knots-quivers-correspondence},
whereas $t_{i}$ is equal to $C_{ii}$, the~number of loops attached to the~$i$-th vertex of the~quiver \cite{Kucharski:2017poe,Kucharski:2017ogk}.
This suggest that the $t$-deformation is encoded in the~quiver and indeed can be considered as a~refined version of the~KQ correspondence
\be\label{eq:KQ-corr-ref}
	\left.P^{Q_{K}}(\mathbf{x},q)\right|_{x_{i}=x a^{a_{i}}q^{q_{i}-C_{ii}}(-t)^{C_{ii}}}=\mathcal{P}^{K}(x,a,q,t)=\sum_{r=0}^{\infty}\mathcal{P}_{r}^{K}(a,q,t)x^r \,.
\ee
Here $\mathcal{P}^{K}(x,a,q,t)$ is a~generating function of $S^r$-colored superpolynomials, 
i.e. Poincar\'{e} polynomials of the (finite dimensional unreduced version of) $S^r$-colored HOMFLY-PT homology introduced in \cite{Gukov:2011ry}.

After \cite{GGS1304}, we know that HOMFLY-PT homology
can be generalized to the quadruply graded homology.
Following once again the~idea of \cite{GNSSS1512}, we can consider
the~quadruply graded finite dimensional unreduced homology $\tilde{\mathscr{H}}(K)$,
but this time the~unknot numerator factor reads $1+a^{2}t_{r}t_{c}$. The Poincar\'{e}
polynomial is given by 
\begin{equation}\label{eq:4-graded polynomial and homology generators}
\tilde{\mathcal{P}}_{1}^{K}(a,Q,t_{r},t_{c})=\frac{\sum_{i,j,k,l}a^{i}Q^{j}t_{r}^{k}t_{c}^{l}\dim\tilde{\mathscr{H}}_{i,j,k,l}(K)}{1-t_{c}^{2}}=\frac{\sum_{i\in\tilde{\mathscr{G}}(K)}a^{a_{i}}Q^{q_{i}}t_{r}^{t_{i}}t_{c}^{t_{i}}}{1-t_{c}^{2}}\,,
\end{equation}
where $\tilde{\mathscr{G}}(K)$ is a set of generators of $\tilde{\mathscr{H}}(K)$.
We can also consider $\tilde{\mathscr{H}}^{S^r}(K)$, a~colored generalization of $\tilde{\mathscr{H}}(K)$, and we will call its Poincar\'{e}
polynomial $\tilde{\mathcal{P}}_{r}^{K}(a,Q,t_{r},t_{c})$. It is a~quadruply-graded polynomial which forms a~generating series that appears in a~doubly refined KQ correspondence \cite{Kucharski:2017poe,Kucharski:2017ogk}
\be\label{eq:KQ-corr-doubly-ref}
	\left.P^{Q_{K}}(\mathbf{x},q)\right|_{x_{i}=x a^{a_{i}}Q^{q_{i}}(-t_r)^{C_{ii}},\,q=t_c}=\tilde{\mathcal{P}}^{K}(x,a,Q,t_{r},t_{c})=\sum_{r=0}^{\infty}\tilde{\mathcal{P}}_{r}^{K}(a,Q,t_{r},t_{c})x^r \,.
\ee
Looking at the~change of variables we can see the~reason of keeping $q_{i}$ and $C_{ii}=t_{i}$ separate in \eqref{eq:KQ-corr-basic}.

Equations (\ref{eq:Superpolynomial and homology generators}--\ref{eq:KQ-corr-ref})
can be obtained from (\ref{eq:4-graded polynomial and homology generators}--\ref{eq:KQ-corr-doubly-ref})
by the~following substitution 
\begin{equation}\label{eq:4 to 3 gradings}
Q\mapsto q\,, \qquad
t_{r}\mapsto tq^{-1}\,, \qquad
t_{c}\mapsto q\,.
\end{equation}
Since $t$ comes from $t_{r}$ our refinement is called $t_{r}$.
Some authors use an~inequivalent $t_{c}$~refinement and unification
of both conventions was one of the~motivations of introducing four gradings
in \cite{GGS1304}. 

From another viewpoint, A-polynomials are also relevant for the~physics and geometry of the~KQ
correspondence. It was conjectured in \cite{AV1204}
and proved in \cite{GLL1604} that there exists a~recursion relation
for HOMFLY-PT polynomials which can be encoded in the~form
\begin{equation}
\widehat{A}^{K}P_{r}^{K}(a,q)=0,\label{eq:A-polynomial annihilating P}
\end{equation}
where the~operator $\widehat{A}^K$ is the~quantum $a$-deformed A-polynomial. See \cite{Ekholm:2018iso} for a~geometric derivation.
Further, \cite{Fuji:2012nx} provided a~$t$-refinement with the~quantum super-A-polynomial
annihilating superpolynomial $\mathcal{P}_{r}^{K}(a,q,t)$. That work predicted
that the~classical A-polynomial $A^{K}$ arises from the~semiclassical limit
($q=e^{\hbar}\rightarrow1$) of (\ref{eq:A-polynomial annihilating P})
and encodes the~supersymmetric vacua of 3d $\mathcal{N}=2$ theory. This
phenomenon will be studied in detail in Section \ref{Theories T[M_K] and T[L_K]}.
A-polynomials are also related to geometry of holomorphic disks briefly
reviewed in the~following section. $A^{K}$ is conjectured \cite{AV1204,Fuji:2012nx,Aganagic:2013jpa,FGSS1209}
to agree with augmentation polynomial introduced in \cite{Ng0407,Ng1010}.

In this paper we will mainly use the~dual classical super-A-polynomials
$\mathcal{A}^{K}$ \cite{GKS1504} which are the~$q\rightarrow1$
limits of operators annihilating the~generating function of superpolynomials
\begin{equation}
\widehat{\mathcal{A}}^{K}\mathcal{P}^{K}(x,a,q,t)=0\label{eq:dual A-polynomial annihilating P}
\end{equation}
and therefore are closer to the~KQ correspondence. Here $\mathcal{A}^{K}$
and $A^{K}$ are related by a~change of variables mentioned in Section
\ref{sub:Unknot} and explained in \cite{GKS1504}. For simplicity
we will usually skip ``dual classical'' and call $\mathcal{A}^{K}$
a super-A-polynomial.

\subsection{Holomorphic disks and generalized holomorphic curves}\label{geometric background}
In this section we give a~brief description of the~material of Section \ref{BPSintro} from a~more geometric point of view. The starting point is to view open Gromov-Witten theory of a~Maslov index zero Lagrangian submanifold $L$ in a~Calabi-Yau 3-fold $Y$ as the~holomorphic curves with boundary on $L$ deforming the~Chern-Simons theory in $L$. From a~mathematical point of view, this was recently interpreted as Gromov-Witten invariants with values in the~skein module of $L$, see \cite{ES}. Combining this viewpoint in the~case $Y=T^{\ast} S^{3}$ with a~certain deformation of almost complex structures, known as Symplectic Field Theory stretching, in fact leads to new understanding of the~geometric mechanism responsible for large $N$ duality. 

As above we consider the~Lagrangian conormal $L_{K}$ of a~knot $K\subset S^{3}$ shifted off the~zero section as a~Lagrangian in the~resloved conifold $X$. The argument above then shows as conjectured that the~HOMFLY-PT polynomials of $K$ are identified with open Gromov-Witten invariants of $L_{K}\subset X$. More precisely, the~wave function $\Psi_{K}(\xi)$ of $L_{K}$ can be written as
\be
\Psi_{K}(\xi) = \exp\left(\sum_{n\ge 1} C_{n}(e^{\mathtt{t}},g_{s})e^{n\xi}\right),
\ee
where $C_{n}(e^{\mathtt{t}},g_{s})$ is a~polynomial in $e^{\mathtt{t}}$ that counts (connected) generalized holomorphic curves with boundary in homology class $n\xi\in H_{1}(L_{K})$, where $\mathtt{t}$ corresponds to the~homology class in $H_{2}(X)$ after capping $n\xi$ off.  

In \cite{Aganagic:2013jpa,Ekholm:2018iso} rather effective indirect approaches to calculating $\Psi_{K}(\xi)$ were described. The main idea is to use punctured holomorphic curves at infinity (which are controlled by so called Morse flow trees that can be calculated combinatorially from a~braid representation of the~knot, see \cite{Ekholm:2018iso}) and their interactions with the~closed curves that contribute to the~wave function. In \cite{Aganagic:2013jpa} this led to a~calculation of the~disk potential and the~mirror curve by elimination theory in finitely many variables. The calculation also identifies the~polynomial of the~mirror curve with the~augmentation polynomial $\mathrm{Aug}_{K}$ of knot contact homology. In a~similar way the~full genus counterpart of knot contact homology gives the~quantization of $\mathrm{Aug}_{K}$, which is an~operator  $\widehat{\mathrm{Aug}}_{K}$ annihilating $\Psi_{K}$. The arguments relating curves at infinity to curves in the~bulk are of wall crossing type. Briefly, one looks at $1$-parameter family of curves that starts out at infinity, as we push the~curve into the~bulk its boundary crosses the~boundary of the~bulk curves and to understand the~moduli space we glue the~curves. The resulting cobordism then give the~equations for the~curves in the~bulk.     

In order to connect this picture to quivers, we need to introduce the~concept of \emph{generalized holomorphic curve}. 
Here we sketch the~basic idea, a~more precise characterization will be given in Section \ref{sec:math}.
Let us consider the~conormal Lagrangian in the~resolved conifold, $L_{K}\subset X$. As discussed in \cite{Aganagic:2013jpa} the~naive count of holomorphic curves is not invariant under deformations. However, if rather than counting curves, we count all potential curves keeping track of all possible gluings under deformations, then the~count is invariant. As described in \cite{Ekholm:2018iso}, such count requires extra geometric data: a~certain Morse function on $L_{K}$ and a~4-chain $C_{K}$ with boundary $\partial C_K=2\cdot L_{K}$ compatible with $f$ near its boundary. Here the~Morse function is used to construct bounding chains in $L_{K}$ for the~boundaries of the~holomorphic curves, which together with a~choice of a~longitude at infinity allows us to define the~linking number between two curve boundaries. The 4-chain is closely related and we count intersections between the~4-chain and the~interiors of the~holomorphic curves. In this context a~generalized holomorphic curve is a~graph with actual holomorphic curves at its vertices and with oriented edges corresponding to linking intersections and, when connecting to the~same vertex, to intersections with $C_{K}$. In the~language of \cite{ES} this corresponds to Gromov-Witten invariants with values in the~$U(1)$-skein module of $L_{K}$ projected to the homology class in $H_{1}(L_{K})$ and keeping track of the writhe through the $q$-power.

It is not hard to see that holomorphic disks going once around the~generator $\xi$ are generically embedded and can never be further decomposed. Assuming, in line with \cite{Gopakumar:1998ii,Gopakumar:1998jq}, that all other holomorphic curves are obtained from combinations of branched covers of these and constant curves at their boundary, the~count of curves is exactly the~quiver partition function with nodes at the~basic disks and with arrows according to linking and additional contributions from the~vertices given by $4$-chain intersections.

From this point of view, the~theory $T[Q_{K}]$ can be thought of as changing the~perspective and treating the~basic holomorphic disks as independent objects with the~Lagrangian attached. 
As we shall see below, this in particular leads to a~separation of the~effects of the~various basic disks.

\section{KQ correspondence and 3d $\bf{\mathcal{N}=2}$ physics\label{sec:physics}}

In this section we derive the~relation between a~knot $K$ and a~3d $\CN=2$ theory $T[Q_K]$, 
whose BPS vortex partition function coincides with the~quiver partition function of $Q_K$.

Our starting point will be M-theory on the~resolved conifold, with a~single M5-brane wrapping the~knot conormal $L_K$
\be\label{eq:M-theory-setup}
\begin{split}
	\text{space-time}: \quad& \IR^4 \times S^1 \times X \\
			& \cup \phantom{ \ \times S^1 \times \ \ } \cup\\
	\text{M5}: \quad & \IR^2\times S^1 \times L_K
\end{split}
\ee
The effective theory on $\IR^2\times S^1$ is expected to be a~3d $\CN=2$ theory \cite{Terashima:2011qi,Dimofte:2011ju}, which we will denote by $T[L_K]$. Roughly speaking, this theory is defined by the~requirement that its manifold of supersymmetric vacua coincides with the~moduli space of flat $U(1)$ connections on the~complement of the~boundaries of holomorphic disks on $L_K$, compare~\cite{Cordova:2013cea,Chung:2014qpa}.
This class of supersymmetric theories is characterized by a~rich duality web, and there are several dual theories with the~same moduli space.

\subsection{The theories $T[L_K]$ and $T_0[L_{K}]$}\label{Theories T[M_K] and T[L_K]}
Before studying the~theory $T[L_K]$ itself, it will be helpful to take an~intermediate step and study a~closely related theory that we denote $T_0[L_K]$. 
Just like $T[L_K]$, the~definition of $T_0[L_K]$ is based on the~requirement that its moduli space of supersymmetric vacua coincides with the~moduli space of flat connections on $L_K$ in the~complement of holomorphic disks. The basic quantity in the~theory $T[L_{K}]$  is the~holonomy of the~longitude in the~torus at infinity whereas for $T_0[L_K]$ it is that along the~meridian. 

As a~side remark, note that the~meridian is the~generator of the~first homology of the~knot complement $S^{3}\setminus K$ and the~theory $T_0[L_K]$ is therefore connected to the~study of holomorphic curves with boundary on a~Lagrangian with the~topology of the~knot complement, see \cite{Aganagic:2013jpa,2012arXiv1210.4803N}.

One way to construct a~theory such as $T_0[L_K]$ with a~suitable space of vacua was proposed in \cite{Fuji:2012nx} using the~\emph{reduced} normalization of the~superpolynomial. Here we use the~unreduced normalization which is closer to counts of holomorphic disks, and therefore more useful for explaining the~physical origin of quivers associated to knots.

The twisted superpotential of $T_0[L_K]$ is encoded by the~combined large-color and $\hbar\to 0$ limit of the~colored superpolynomial
\be\label{eq:FGS}
	\mathcal{P}^{K}_r(a,q,t)\  \mathop{\longrightarrow }^{\hbar\to 0}_{r\to\infty} \ \int \prod_{i} \frac{dz_i}{z_i} \, \exp {\frac{1}{2\hbar} \(\tCW_{T_0[L_K]}(z_i, a,t,y)+ O(\hbar) \)} \,.
\ee
Here $q=e^\hbar$ and we always work in the~regime $|q|<1$, whereas $y$ is the~value approached by  $q^{2r}$ in the~limit.
The twisted superpotential $\tCW_{T_0[L_K]}$ is a~function of several fugacities:
\begin{itemize}
\item[$y$] is associated to the~global symmetry  $U(1)_M$, whose connection arises from the~reduction of the~6d abelian 2-form along the~\emph{meridian} cycle of a~$T^2$ neihgborhood of $K$, therefore it is also identified with the~meridian holonomy of a~flat connection on $L_K$.
While the~meridian cycle is contractible in $L_K$ alone, a~connection deformed by the~presence of holomorphic curves on $L_{K}$ can have nontrivial meridian holonomy.
\item[$a$] is the~fugacity of the~global symmetry $U(1)_Q$ arising from the~internal 2-cycle in the~resolved conifold geometry. After the~geometric transition this is identified with the~base $\IC\IP^1$ of the~resolved conifold. Before the~transition it is the~2-sphere at infinity in the~$T^*S^3$ fiber.
\item[$-t$] is a~parameter associated with rotations $U(1)_F$ of the~normal bundle of $\IR^2 \subset \IR^4$. 
\item[$z_i$] are identified with fugacities for the~abelian gauge group $U(1)\times\ldots\times U(1)$ and therefore they are not unique: different dual descriptions may involve gauge groups of different ranks. The working definition of these fugacities is $z_i \sim  q^{2k_i}$ for some integer $k_i$ in the~limit $\hbar\to 0,\,r\to\infty$. We will review this below with some examples.
\end{itemize}

The twisted superpotential typically includes two main types of contributions: dilogarithms and squares of logarithms
\be\label{eq:FGS-dictionary}
\begin{split}
	\Li_2\(a^{n_Q} (-t)^{n_F} y^{n_M}  z_i^{n_i}\) \qquad & 
	\longleftrightarrow \qquad \text{(chiral field)}\,, \\
	\frac{\kappa_{ij}}{2} \log \zeta_i \cdot \log \zeta_j  \qquad & 
	\longleftrightarrow \qquad \text{(Chern-Simons coupling)}\,.
\end{split}
\ee
Each dilogarithm is interpreted as the~one-loop contribution of a~chiral superfield with charges $(n_Q,n_F,n_M,n_i)$ under the~various symmetries, while quadratic-logarithmic terms are identified with Chern-Simons couplings among the~various $U(1)$ gauge and global symmetries, with $\zeta_i$ denoting the~respective fugacities \cite{1983NuPhB.222...45D,Witten:1993yc,Hanany:1997vm,Hori:2000kt,Dimofte:2011jd}.

Integrating over the~gauge fugacities $z_i$ by a~saddle-point approximation gives the~twisted \emph{effective} superpotential of the~theory 
\be\label{eq:tCW-eff}
	\tCW^\eff_{T_0[L_K]}(a,t,y) = \tCW_{T_0[L_K]}(z_i^*,a,t,y)\,, \;\;\text{where}\;\; \left.\frac{\partial\tCW_{T_0[L_K]}(z_i, a,t,y)}{\partial {z_i}}\right|_{z_i=z_i^*} = 0 \,.
\ee
In \cite{Fuji:2012nx} it was argued that the~theory $T_0[L_K]$ defined in this way has a~moduli space of vacua that coincides with the~graph of the~super-A-polynomial
\be\label{eq:super-A-poly}
	\frac{\partial \tCW^\eff_{T_0[L_K]}(a,t,y)}{\partial\log y} = \log x^{-1} \qquad \Leftrightarrow \qquad \mathcal{A}^{K}(x,y,a,t) = 0 \,.
\ee
The slightly unconventional powers of $x^{-1}$ and $y$ arise from a~careful match with the~literature on knot invariants, as will be discussed further below. 
With these conventions, $\log y$ is interpreted as a~scalar field in the~twisted chiral multiplet corresponding to the~$U(1)_M$ field strength of the~3d $\CN=2$ theory on a~circle. 
The role of  $\log x^{-1}$ as well as the~origin of the~super-A-polynomial  are more naturally understood from the~viewpoint of the~theory $T[L_K]$, to which we now return.

To identify the~content of $T[L_K]$, we propose to consider the~generating series of $S^r$-colored superpolynomials and take a~slight variation of the~double-scaling limit (\ref{eq:FGS}):
\begin{align}\label{eq:FGS+}
	\mathcal{P}^{K}&(x,a,q,t) 
	 =  \sum_{r\geq 0} \mathcal{P}^{K}_{r}(a,q,t) x^r \\
	& \mathop{\longrightarrow}^{\hbar\to 0}_{{\footnotesize \begin{array}{c}q^{2r}\to y\\ q^{2k_i}\to z_i\end{array}}}  
	\int \frac{d y}{y}  \int \prod_{i} \frac{dz_i}{z_i}  \ \exp\frac{1}{2\hbar} \( \tCW_{T_0[L_K]}(z_i, a, t,  y) + \log y\, \log x +O(\hbar) \)\,.\nonumber
\end{align}
The 3d $\CN=2$ theory arising from this procedure differs from the~one in (\ref{eq:FGS}) by the~fact that $U(1)_M$ is now gauged, and for the~presence of a~Chern-Simons coupling between its fugacity and a~background $U(1)_L$ symmetry with fugacity~$x$ 
\be
\label{TLK-TM3}
	\tCW_{T[L_K]}(z_i, a,t,x,y) = \tCW_{T_0[L_K]}(z_i, a,t,y) + \log y\, \log x \,.
\ee
In three dimensions there is a~dual $U(1)$ ``topological'' symmetry for each $U(1)$ gauge symmetry, whose current is sourced by vortices. 
The Chern-Simons coupling between a~gauge symmetry and its dual topological symmetry is also known as a~FI coupling. This is the~interpretation of the~parameter $\log x$, corresponding to the~fact that $x$ is the~fugacity associated with $U(1)_L$.

The manifold of vacua of $T[L_K]$ is naturally identified with (\ref{eq:super-A-poly}). 
In fact, taking first the~saddle point with respect to $z_i$ leads to
\be\label{eq:FGS+ saddle}
\begin{split}
	\mathcal{P}^{K}(x,a,q,t)
	& \sim  \int \frac{d y}{y}   \ \exp\frac{1}{2\hbar} \( \tCW^\eff_{T[L_K]}(a, t, x, y)  + O(\hbar)\)\,,
\end{split}
\ee
where 
\be
\tCW^\eff_{T[L_K]}(a, t, x, y)=\tCW^\eff_{T_0[L_K]}(a, t,  y)+ \log y\, \log x.
\ee
Now integrating over $d \log y$ naturally enforces the~saddle point equation for this fugacity, which is equivalent to (\ref{eq:super-A-poly})
\be\label{eq:super-A-poly from L_K}
	\frac{\partial \tCW^\eff_{T[L_K]}(a,t,x,y)}{\partial\log y}=0 \qquad \Leftrightarrow \qquad \mathcal{A}^{K}(x,y,a,t) = 0 \,.
\ee

\subsection{Legendre transform}\label{Weff-WK}

With the~above discussion we arrived at the~definition of a~theory $T[L_K]$ described by a~twisted superpotential $\tCW_{T[L_K]}$ whose critical points coincide with those of the~Gromov-Witten disk potential $W_K$.
This has an~intuitive physical origin, which is best understood from the~viewpoint of the~general setup (\ref{eq:M-theory-setup}).
The disk potential counts holomorphic disks ending on $L_K$, which in the~physical setup are wrapped by M2-branes ending on M5.
From the~viewpoint of worldvolume of the~fivebrane, they give rise to BPS states in the~theory $T[L_K]$, namely BPS vortices counted by LMOV invariants.
These vortices couple to the~$U(1)_M$ gauge field in the~3d $\CN=2$ theory, whose fugacity we denoted by $y$. 
Since the~theory is placed on a~circle, the~worldline of these BPS particles is finite, and the~same goes for their contribution to the~effective action, which is proportional to $e^{-2\pi R \, E}$. ($E$ is the~mass, which equals the~absolute value of the~$\CN=2$ central charge for BPS states).
Therefore the~holomorphic disks deform the~theory, introducing contributions to the~effective action (this mechanism is familiar, for example, from the~context of instanton/particles in~4d/5d~\cite{Lawrence:1997jr}, and for~3d/4d~\cite{Gaiotto:2008cd}). 

As discussed in Sections \ref{BPSintro} and \ref{geometric background}, all contributions from holomorphic disks are summed in the~disk potential $W_K$, which is the~effective action. From the~viewpoint of $T[L_K]$, the~disks are sources for the~gauge field, therefore the~effective action describing their interactions is naturally computed by a~Legendre transform. The term $\log y\cdot\log x$ in (\ref{TLK-TM3}) provides the~source-current interaction, and integrating out $y$ leaves the~effective theory of the~sources, which describes interactions among holomorphic disks.

To make the~above statement concrete, let us start by recalling that the~open topological string wavefunction $\Psi_K$ is equal to the~HOMFLY-PT generating series 
and the~disk potential arises in the~$\hbar \to 0$ limit 
\be\label{eq:psiK}
\begin{split}
	P^{K}(x,a,q)  
	\mathop{\sim}^{\hbar\to 0} \exp\( \frac{1}{2\hbar} W_K(a,x) + \dots \)\,.
\end{split}
\ee
For simplicity we will work with $t=-1$ in this section, although each statement admits a~generalization to the~refined case.

On the~other hand, taking the~semiclassical limit of the~same function as defined in~(\ref{eq:FGS+}), and performing the~saddle point analysis with respect to $z_i$ leads to (\ref{eq:FGS+ saddle}). Then performing the~integral in $y$ is equivalent to taking the~saddle points $y^{*}(x)$ defined by~(\ref{eq:super-A-poly from L_K}), leading to the~promised Legendre transform
\be\label{eq:newrelation}
	W_K(a,x) =\(\tCW_{T_0[L_K]}^{\eff}(a, y)+ \log y\, \log x \)\big|_{y = y^{*}(x)}\,.
\ee
Here 
$\tCW_{T_0[L_K]}^{\eff}(a, y)$ arises as a~twisted superpotential of a~weak coupling limit of $T[L_K]$ in which we have no source-current interactions. 

The statement that $W_K$ is a~Legendre transform of $\tCW^\eff_{T_0[L_K]}$ admits a~natural generalization to a~much stronger one, which is that the~full quantum effective action of the~gauge theory coincides with the~\emph{Fourier transform} of the~topological string wavefunction $\Psi_K$. We will come back to this with a~more precise formulation in Section \ref{subsec:quiverQM}. 

The relation (\ref{eq:newrelation}) is also natural from the~geometric viewpoint.
Since $\Psi_K$ counts holomorphic curves ending on $L_K$, see Section \ref{sec:math}, it is clear that $x$ is the~holonomy of a~$U(1)$ bundle on the~M5-brane wrapping $L_K$ around the~generator of $H_1(L_K)$. 
From the~construction of conormal Lagrangians it is clear that this corresponds to the~\emph{longitudinal} cycle around $K$ \cite{Ooguri:1999bv}.
This explains the~appearance of the~Legendre transform, since both connections of $U(1)_L$ and $U(1)_M$ arise respectively as reductions of the~2-form $C_2$ on the~longitudinal and meridian cycles $(\ell,m)$ 
\be
\begin{split}
	\log x^{-1} & =  \oint_{S^1} A_L = \int_{S^1\times \ell} C_2 = \oint_\ell A\\
	\log y & =  \oint_{S^1} A_M =  \int_{S^1\times m} C_2 = \oint_m A\,.
\end{split}
\ee
Here $A$ is the~reduction of $C_2$ along $S^1$. 
Therefore $\log x^{-1}$ and $\log y$ are conjugate with respect to the~Weil-Petersson symplectic form on $\CM_{{\rm flat}}(T^2)$ (the moduli space of flat connections on the~Legendrian torus at infinity).

The Legendre transform relating $W_K$ and $\tCW^\eff_{T_0[L_K]}$ involves the~information of the~vacuum manifold given by (\ref{eq:super-A-poly}) in the~limit $t\to -1$. In turn both of them can be recovered from  the~A-polynomial or (equivalently) the~augmentation polynomial \cite{Aganagic:2013jpa}:
\begin{itemize}
\item Solving $\mathcal{A}^{K}(x,y,a) = 0 $ for $y$ gives a~function that factorizes into a~product determined by \emph{classical} LMOV invariants $b_{r,i}^{K}=\sum_{j}N_{r,i,j}^{K}$ \cite{GKS1504}
\be\label{eq:y product form}
	y^{*}(a,x) = \prod_{r\geq 0,i\in \IZ}(1-a^{i}x^r)^{r \, b^{K}_{r,i}} 
	= \(\lim_{\hbar\to 0} \frac{P^{K}(q^{2}x,a,q)}{P^{K}(x,a,q)}\)\,.
\ee
Integrating this function yields precisely the~Gromov-Witten disk potential
\be\label{eq:disk-potential-integral}
	W_K(a,x) = \int d\log x \ \log y^{*}\,.
\ee
\item Solving instead for $x^{-1}$ gives a~function $(x^{-1})^*(a, y)$ which does not necessarily factorize into a~form analogous to (\ref{eq:y product form}).
Integrating this function gives the~twisted effective superpotential of the~weak coupling limit of $T[L_K]$
\be
	\tCW^\eff_{T_0[L_K]}(a,y) = \int d\log y \ \log (x^{-1})^*\,.
\ee
\end{itemize}

We shall note that the~convention on our definition of $x,y$ is fixed by the~relation to the~(semiclassical limit of) the~HOMFLY-PT generating series $P^{K}(x,a,q)$.
In particular comparing to \cite{Fuji:2012nx} one can simply perform the~following substitutions $x \to y^{1/2},\ y \to x^{-1}$ into the~formulae of the~reference.

\subsection{The theory $T[Q_K]$}\label{subsec:TofQ}
The knot-quiver correspondence revolves around the~observation relating the~generating series  of knot polynomials and the~partition function of representation theory of a~certain quiver. As reviewed in Section \ref{sec:background}, we can write it as 
\be\label{eq:derivation-3d-N2-theory}
	\mathcal{P}^{K}(x,a,q,t)
	= \left.P^{Q_K}(\mathbf{x},q)\right|_{x_{i}=x a^{a_{i}}q^{q_{i}-C_{ii}}(-t)^{C_{ii}}} \,,
\ee
where each side can be expanded in the~following form
\[
\begin{split}
	& 
	\sum_{r\geq 0}  \mathcal{P}^{K}_{r}(a,q,t) x^r
	\\
	&
	\qquad = \left.
		\sum_{d_{1},\ldots,d_{m}\geq0}q^{\sum_{1\leq i,j\leq m}C_{ij}d_{i}d_{j}}\prod_{i=1}^{m}\frac{\((-1)^{C_{ii}}x_{i}\)^{d_{i}}}{(q^{2};q^{2})_{d_{i}}}
		\right|_{x_{i}=x a^{a_{i}}q^{q_{i}-C_{ii}}(-t)^{C_{ii}}} \,.
\end{split}
\]
On the~one hand, this is simply a~way of rewriting the~sum over symmetric representations into a~sum over quiver representations labeled by $\bf d$.
On the~other hand applying (\ref{eq:FGS+}) directly to $P^{Q_K}$ we obtain a~new 3d $\CN=2$ theory $T[Q_{K}]$
\begin{align}\label{eq:bar-P-semiclassical-limit}
	& P^{Q_K}(\mathbf{x},q)
	\mathop{\longrightarrow}^{\hbar\to 0}_{q^{2 d_i}\to y_i}  
	\int \prod_{i\in {Q_{K}}_0}  \frac{dy_i}{y_i} \, 
	\exp  \frac{1}{2\hbar} \( \tCW_{T[Q_K]}(\mathbf{x},\mathbf{y}) + O(\hbar) \) \\
	& \tCW_{T[Q_K]}(\mathbf{x},\mathbf{y})
	= \sum_{i} \Li_2(y_i) + \log \((-1)^{C_{ii}}x_{i}\) \, \log y_i  + \sum_{i,j} \frac{C_{ij}}{2} \log y_i \, \log y_j\,.   \nonumber
\end{align}
The application of the~dictionary (\ref{eq:FGS-dictionary}) gives the~following structure for $T[Q_K]$:

\begin{itemize}
\item Gauge group: $U(1)^{(1)}\times\dots\times U(1)^{(m)}$
\item Matter content: chiral fields $\Phi_i$ with charge $\delta_{ij}$ under $U(1)^{(j)}$
\item Gauge Chern-Simons couplings: $\kappa^\eff_{ij} = C_{ij}$
\item Fayet-Ilioupoulos couplings: $\log \((-1)^{C_{ii}}x_{i}\)$
\end{itemize}
We could redefine $x_i$ to absorb the~minus sign $(-1)^{C_{ii}}$, however the~resulting changes in formulas for DT invariants and A-polynomials are rather disinclining. From the~mathematical point of view, the~sign is related to the~choice of spin structure on $L_{K}$ that enters in the~orientation of the~moduli spaces of holomorphic curves. 

It is straightforward to read off the~theory $T[Q_K]$ from the~quiver: the~gauge group is a~product of  $U(1)$ factors associated to quiver nodes, the~matter content consists of a~set of chiral multiplets charged under each $U(1)$, and the~Chern-Simons couplings coincide with the~adjacency matrix of $Q_{K}$.
More precisely, $\kappa_{ij}^\eff$ denotes the~matrix of \emph{effective} Chern-Simons couplings, they are related to the~bare Chern-Simons couplings by a~diagonal shift by $1/2$ due to the~presence of charged matter \cite{Aharony:1997bx}.

The change of variables $x_{i}=x a^{a_{i}}q^{q_{i}-t_{i}}(-t)^{t_{i}}$ required by the~KQ correspondence amounts to identifying the~FI couplings of $T[Q_K]$ with specific combinations of the~physical fugacities.
Recall that Fayet-Ilioupoulos terms can be interpreted as mixed Chern-Simons couplings between each $U(1)^{(i)}$ gauge group and its dual ``topological'' global symmetry $U(1)^{(i)}_J$ \cite{Aharony:1997bx}.
The KQ change of variables signals that the~dual symmetry is partially broken to a~subgroup $U(1)_L\times U(1)_Q \times U(1)_F $, because the~fugacities $x_i$ are not all independent. It would be interesting to identify the~mechanism responsible for this breaking, in particular from a~geometric perspective. In this paper we regard it as part of the~data going into the~definition of $T[Q_{K}]$.

The Fayet-Ilioupoulos terms therefore turn into the~respective mixed Chern-Simons terms
\begin{align}
	\left.\tCW_{T[Q_K]}\right|_{x_{i}=x a^{a_{i}}q^{q_{i}-t_{i}}(-t)^{t_{i}}}  
	& = \sum_{i,j} \frac{C_{ij}}{2} \log y_i \, \log y_j  +\sum_i  \Li_2(y_i) \\
	& + \sum_{i} \log x \log y_i + a_i \log a~\log y_i + t_i \log t \log y_i  \,. \nonumber
\end{align}
With this identification $T[Q_K]$ has the~same moduli space of supersymmetric vacua as $T[L_K]$, by construction.
Among the~many dual descriptions of $T[L_K]$, the~existence of a~quiver $Q_K$ provides a~specific choice. 
Note that in taking the~semiclassical limit we left out the~parameters $q_i$, because they would contribute to subleading terms. Nevertheless, since they appear in the~definition of variables $x_i$ on the~same footing as $a_i, t_i$, they should also admit an~interpretation as couplings to a~background symmetry. 
This is the~group of rotations in the~plane $\IR^2$ wrapped by the~M5-brane, twisted by R-symmetry (see Section~\ref{subsec:quiverQM} for a~precise definition).  We will also provide a~geometric interpretation for the~origin of this symmetry in Section \ref{sec:math}.

While the~KQ change of variables is key to making contact with knot invariants, it is interesting to forget for a~moment about the~relations among various $x_i$ and contemplate the~message of the~existence of a~description like $T[Q_K]$.
The objects charged under the~topological symmetry of this theory are its BPS vortices, which in our setup are engineered by M2-branes wrapping holomorphic disks.
Adopting the~viewpoint outlined in Section \ref{Weff-WK}, $x_i$ can be regarded as fugacities for different types of sources in the~theory, with each source coupling to only one of the~gauge fields ($x_i$ couples only to $y_i$).
The topological string wavefunction given by $\mathcal{P}^{K}(x,a,q,t)$ is now replaced by the~more refined generating series $P^{Q_K}(\mathbf{x},q)$.
The semiclassical limit of this gives then a~generalization of the~Gromov-Witten disk potential, which we call the~\emph{quiver disk potential}
\be\label{eq:quiver-disk-potential}
	W_{Q_K}(\mathbf{x})   = \lim_{\hbar \to 0} \  2\hbar \cdot  \log \( P^{Q_K}(\mathbf{x},q) \). 
\ee
It is identified with the~effective action of the~theory $T[Q_{K}]$ after Legendre transform 
\be\label{eq:quiver-Legendre-transform}
	W_{Q_K}(\mathbf{x})   =  \tCW_{T[Q_K]}(\mathbf{x},\mathbf{y})\big|_{\mathbf{y} = \mathbf{y}^{*}(\mathbf{x})}\,.
\ee
The saddle point $\mathbf{y}^{*}(\mathbf{x})$ is given by 
\be\label{eq:quiver A-poly}
	\frac{\partial\tCW_{T[Q_K]}(\mathbf{x},\mathbf{y})}{\partial\log y_i}=0 \quad \Leftrightarrow \quad A_i^{Q_K}(\mathbf{x},\mathbf{y}) = 0 \quad \Leftrightarrow \qquad \frac{\partial W_{Q_{K}}(\mathbf{x})}{\partial\log x_i}=\log y_i \,.
\ee
which defines \emph{quiver A-polynomials} $\mathcal{\mathbf{A}}^{Q_K}(\mathbf{x},\mathbf{y})$ in analogy to (\ref{eq:super-A-poly from L_K}) and (\ref{eq:disk-potential-integral}). Similar objects were introduced in \cite{PSS1802,PS18,Smo2017}, however without references to physical and geometric intepretations discussed here.

Combining (\ref{eq:bar-P-semiclassical-limit}) with (\ref{eq:quiver A-poly}) we can find a~quiver A-polynomial for arbitrary~$Q_{K}$:
\be\label{eq:quiver-A-ponomial-general-formula}
A_{i}^{Q_{K}}(\textbf{x},\textbf{y})=1-y_{i}-x_{i}(-y_{i})^{C_{ii}}\prod_{j\neq i}y_{j}^{C_{ij}}\,.
\ee

The quiver disk potential describes the~interactions of a~set of basic sources (one for each quiver node) labeled by $x_i$, providing the~full count of their spectrum of boundstates. 
This viewpoint leads naturally to a~quiver description of the~BPS vortex spectrum, we will return to this below.
It is important to note that $W_{Q_{K}}(\mathbf{x})$ admits a~compact expression in terms of the~Donaldson-Thomas invariants of the~quiver. In fact since
\be\label{P^Q product form}
P^{Q_{K}}(\mathbf{x},q)=\textrm{Exp}\left(\frac{\Omega^{Q_{K}}(\mathbf{x},q)}{1-q^{2}}\right)=\prod_{\mathbf{d},s}(q^s\mathbf{x^{\mathbf{d}}};q^{2})_{\infty}^{(-1)^{|\mathbf{d}|+s}\Omega_{\mathbf{d},s}^{Q_{K}}} \,,
\ee
where 
\be
(z;q^{2})_{\infty}=\prod_{s=0}^{\infty}(1-zq^{2s})
\underset{\hbar\rightarrow0}{\sim}\exp\left(\frac{1}{2\hbar}\textrm{Li}_{2}(z)+\ldots\right)\,,
\ee
it follows that
\be\label{eq:quiver-disk-potential-from-DT}
W_{Q_{K}}(\mathbf{x})=\sum_{\mathbf{d},s}(-1)^{|\mathbf{d}|+s}\Omega_{\mathbf{d},s}^{Q_{K}}\textrm{Li}_{2}(\mathbf{x^{\mathbf{d}}})=\sum_{\mathbf{d}}(-\Omega_{\mathbf{d}}^{Q_{K}})\textrm{Li}_{2}(\mathbf{x^{\mathbf{d}}})\,.
\ee
Here $\Omega_{\mathbf{d}}^{Q_{K}}$ are \emph{numerical} Donaldson-Thomas invariants. 
Each dilogarithm in the~quiver disk potential corresponds to a~boundstate of basic holomorphic disks encoded by the~dimension vector $\mathbf{d}$, the~multiplicity of each boundstate is the~DT invariant.
Our definition 
\be
\Omega_{\mathbf{d}}^{Q_{K}}=\sum_{s}(-1)^{|\mathbf{d}|+s+1}\Omega_{\mathbf{d},s}^{Q_{K}}
\ee
differs slightly from \cite{Kucharski:2017poe,Kucharski:2017ogk} because we have
$1-q^{2}$ instead of $q^{-1}-q$ in the~denominator inside $\textrm{Exp}\left(\frac{\Omega^{Q_{K}}(\mathbf{x},q)}{1-q^{2}}\right)$.

From our perspective, exemplified by the~diagram (\ref{eq:KQ-diagram}), it is clear that these invariants actually count embedded holomorphic disks.
Indeed the~(semiclassical limit of the) KQ~change of variables translates the~quiver disk potential to the~Gromov-Witten disk potential
\begin{equation}\label{eq:W_Q=W_K}
\left.W_{Q_{K}}(\mathbf{x})\right|_{x_{i}=xa^{a_i}}=W_{K}(a,x)\,.
\end{equation}
This can be derived by rewriting the~HOMFLY-PT generating series as
\begin{equation}
P^{K}(x,a,q)=\textrm{Exp}\left(\frac{N^{K}(x,a,q)}{1-q^{2}}\right)=\prod_{r,i,j}(x^{r}a^{i}q^{j};q^{2})_{\infty}^{-N_{r,i,j}^{K}}\,,
\end{equation}
and applying (\ref{eq:psiK}) to extract an~expression for the~disk potential
\begin{equation}\label{W_K in terms of LMOV}
W_{K}(a,x)=\sum_{r,i,j}\(-N_{r,i,j}^{K}\)\textrm{Li}_{2}\(x^{r}a^{i}\)=\sum_{r,i}\(-b_{r,i}^{K}\)\textrm{Li}_{2}\(x^{r}a^{i}\)\,.
\end{equation}
This is the~usual count of holomorphic disks, from its derivation it is clear that it arises from (\ref{eq:quiver-disk-potential-from-DT})  by the~KQ change of variables (\ref{eq:derivation-3d-N2-theory}).

Equation (\ref{eq:W_Q=W_K}) implies also that we can obtain the~A-polynomial of $K$ from
the quiver A-polynomial of $Q_{K}$. Since
\begin{equation}
\begin{split}
\log y^{*}(x) &=\frac{\partial\left.W_{Q_{K}}(\textbf{x})\right|_{x_{i}=a^{a_{i}}x}}{\partial\log x}\\
&=\sum_{i}\left.\frac{\partial W_{Q_{K}}(\textbf{x})}{\partial\log x_{i}}\right|_{x_{i}=a^{a_{i}}x}=\sum_{i}\left.\log y^{*}_{i}(\textbf{x})\right|_{x_{i}=a^{a_{i}}x}\,,\label{eq:y-from-y_i-calculation}
\end{split}
\end{equation}
we have
\begin{equation}
y^*(x)=\prod_{i}\left.y^{*}_{i}(\textbf{x})\right|_{x_{i}=a^{a_{i}}x}\,,\label{eq:y-from-y_i-formula}
\end{equation}
where $y^{*}(x)$ solves $\mathcal{A}^{K}(x,y)=0$ and $y^{*}_i(\textbf{x})$
solves $A_{i}^{Q_{K}}(\textbf{x},\textbf{y})=0$.
Note that this has a~natural geometric interpretation:  $y_i^*(x)$ are meridian holonomies for the~$U(1)$ connection on $L_K$ on tubular neighborhoods of the~boundaries of basic disks (which also contain boundaries of all their boundstates), their composition adds up to the~meridian holonomy on the~torus at infinity $y^*(x)$.

From the~viewpoint of holomorphic disks, the~main message of the~quiver description of BPS vortices is that all holomorphic disks can be viewed as ``boundstates'' of a~\emph{finite} set of fundamental basic disks associated with quiver nodes. 
An analogous phenomenon is well-known to arise in the~context of BPS states of 4d $\CN=2$ theories, where the~BPS spectrum of M2-branes ending on a~fivebrane often admits a~quiver description in terms of a~finite set of ``basic'' disks -- we will return to this in Section \ref{sec:discussion}.

Finally, let us briefly comment on the~geometric interpretation of the~quiver variables~$x_i$.
The refined KQ change of variables (\ref{eq:KQ-corr-ref}) is defined by integers $a_i, q_i, t_i$ which carry a~natural geometric meaning. These variables encode topological data of  basic holomorphic disks represented by nodes of $Q_K$.
Classically, the~disks are classified by relative homology classes in $X$ with boundary on $L_K$: the~classical topological data therefore includes the~homology class of the~disk boundary $\partial\Sigma \in  H_1(L_{K})$ and the~number of wrappings around two-cycles in $H_2(X)$. Since $L_{K}$ has topology $S^1\times \IR^2$, there is only one cycle that the~disc boundary can wrap and all basic disks wrap that cycle exactly once, explaining why $x$ appears with a~unit power in (\ref{eq:KQ-corr-ref}).
Among the~two-cycles in $X$, there is of course the~resolved conifold base $\IC\IP^1$: $a_i$ counts the~number of wrappings of the~$i$-th basic disk around this $\IC\IP^1$. Since we talk about \emph{relative} homology, it is understood that the~wrappings are defined relative to a~universal (but non-canonical) choice of capping for the~disks. In other words, we choose a~reference disk with boundary on the~opposite generator of $H_{1}(L_K)$, and consider its composition with each of the~basic disks to form a~closed 2-cycle. Then $a_i$ is the~closed homology class of this closed cycle. Changing the~choice capping disk shifts all $a_i$ simultaneously by the~same amount, which can be absorbed by an~overall normalization, leaving $a_i-a_j$ as the~invariant data.

There is in fact another nontrivial two-cycle in the~geometry, which is sometimes overlooked: it is the~two-sphere linking $L_K$. Since $L_K$ supports an~M5-brane, it sources magnetic flux for the~four-form fieldstrength of eleven dimensional supergravity. In the~compactification to $X$ with the~M5 wrapped on $L_K$, this reduces to a~two-form on $X$ which has non-vanishing integral on the~two-sphere linking $L_K$ (we will return to this in Section \ref{sec:refined-CS}).
In a~situation where the~geometry is modified so as to compactify $L_K$ and where we place $M\gg 1$ branes on it, the~magnetic flux will be proportional to $M$, and indeed the~area of this $\IC\IP^1$ would arise from the~usual 't Hooft limit as $a_{L_K} = q^M$, or its refined version $a_{L_K} = t^M (t/q)^{1/2}$ \cite{Aganagic:2012hs}. 
However since we keep $M=1$ this contribution to the~holomorphic disk action is a~quantum effect in our setup, in the~sense that it is non-vanishing only for $t,q\neq 1$, and it is visible only at the~quantum level.

We therefore propose to identify $t_i = C_{ii}$ with the~wrappings of the~$i$-th basic disk on the~two-sphere linking $L_K$, whose origin is the~M2-M5 coupling via bulk fluxes. 
Finally, $q_i$ counts the~self-linking 
of the~$i$-th basic holomorphic disk. Note that it appears in the~combination $q_i-C_{ii}$ as the~power of $q$, which equals the contributions from 4-chain intersections of the disk. An explanation for this will be provided in the~next section, in terms of ``real'' (M2-M2 via M5) and ``imaginary'' (M2-M5) self-intersections.

\subsection{Vortex partition functions as quiver partition functions}\label{subsec:quiverQM}

In this section we show that the~partition function of BPS vortices of $T[Q_K]$ coincides exactly with the~motivic generating series of the~quiver $Q_K$. 
This is the~quantum uplift (to finite $\hbar$) of the~identification (\ref{eq:quiver-Legendre-transform}) between the~quiver disk potential and the~ Legendre transform of the~twisted effective superpotential of the~theory.
To put this statement into perspective let us recall again the~relation between vortices and knot theory: BPS vortices of the~3d $\CN=2$ theory $T[L_K]$ arise from M2-branes wrapping holomorphic curves ending on $L_K$; for this reason the~vortex partition function encodes open Gromov-Witten invariants and therefore knot invariants \cite{Dimofte:2010tz,ES}. 

As a~warm-up, let us start with a~simple subclass of theories of type $T[Q]$, characterized by a~quiver with diagonal adjacency matrix $C_{ij} = \kappa^\eff_{i} \delta_{ij}$. 
The  nodes of $Q$ are mutually disconnected, the~only arrows are $\kappa^\eff_i$ loops on the~$i$-th node. 
In this case $T[Q]$ is made of $m=|Q_0|$ copies of a~$U(1)_{\kappa_{i}}$ gauge theory, each with a~single chiral field with charge $+1$. 
The bare Chern-Simons level of $U(1)^{(i)}$ is related to $\kappa_i^\eff$ by a~half-integer shift induced by quantum corrections from the~charged chiral multiplets \cite{Aharony:1997bx}:
\be
	\kappa_i=\kappa^\eff_i-1/2\,.
\ee
Since $C_{ij}$ is diagonal, the~partition function factorizes:
\be
	\CZ^\vort  = \prod_{j=1}^{m} \CZ^{\vort,{(j)}} 
	= \prod_{j=1}^{m} \sum_{d_j\geq 0} z_{j}^{d_j} \CZ_{d_j}^{\vort,{(j)}} \,,
\ee
where $\CZ_{d_j}^{\vort,{(j)}}$ is the~vortex partition function of the~$j$-th sector with vorticity~$d_j$. 
This is well known to be \cite{Shadchin:2006yz,Beem:2012mb,Hwang:2012jh} 
\be\label{eq:vortex-witten-index}
\begin{split}
	\CZ_{d_j}^{\vort,{(j)}} 
	& =   \frac{e^{\kappa_{j}(d_j \mu + d_j^2 \gamma)}}{\prod_{k=1}^{d_j} \sinh \gamma(k-d_j-1)} \,.
\end{split}
\ee
Here $\gamma,\mu$ are equivariant parameters for global symmetries $U(1)_\gamma \times U(1)_\mu$ which rotate respectively the~tangent and normal bundle to the~$\IR^2\subset \IR^4$ wrapped by the~M5-brane defect, appropriately twisted by the~$U(1)_R$ symmetry (for our convention  see \cite{Hwang:2017kmk,Bullimore:2016hdc}).
It~was shown by the~authors of \cite{Hwang:2017kmk} that the~$d_j$-vortex partition function (\ref{eq:vortex-witten-index}) coincides with the~Witten index of a~$\CN=2$ quiver quantum mechanics. 
In fact the~full vortex partition function of this simple theory can be written in the~following suggestive form
\be
\begin{split}
	\CZ^\vort  
	& = \prod_{j=1}^{m} \sum_{d_j\geq 0} 
	\(2 (-e^\mu)^{\kappa^\eff_j} \, (e^{\gamma-\mu})^{1/2} \, z_{j} \)^{d_j} \, \(-e^{\gamma}\)^{\kappa^\eff_j \, d_j^2}  \prod_{s=1}^{d_j}\frac{1}{1-e^{2\gamma}} \,.
\end{split}
\ee
The similarities with (\ref{eq:Efimov}) are striking, in fact adopting the~dictionary of Section~\ref{subsec:TofQ} we can match the~two exactly.
By definition of $U(1)_\gamma$, it is natural to identify $e^\gamma = q$. 
Moreover, fugacities $z_j$ are associated to the~topological symmetries whose charges count vortices, therefore they are expected to coincide with the~FI couplings $z_i\sim x_i$ up to normalization.
This leads immediately to a~match with the~quiver partition function
\be\label{eq:Zvortex-PQ}
	\CZ^\vort = P^Q\,.
\ee
We can further compose this change of variables with the~KQ one given by~(\ref{eq:KQ-corr-basic}) (resp. one of its refined versions: (\ref{eq:KQ-corr-ref}), (\ref{eq:KQ-corr-doubly-ref})), which leads to the~identification of the~vortex partition function with the~HOMFLY-PT generating series $P^K(x,a,q)$ (resp. $\mathcal{P}^{K}(x,a,q,t)$ or $\tilde{\mathcal{P}}^{K}(x,a,Q,t_r,t_c)$) \cite{Dimofte:2010tz}. 
Note that this does not fix $e^\mu$, which we identify with $t/q$ by its physical interpretation.

The most general theory of type $T[Q]$ differs from the~class of models just considered in a~rather mild way, namely by turning on off-diagonal Chern-Simons couplings. To include their contribution we turn to an~explicit construction of the~vortex partition function in terms of \emph{holomorphic blocks} \cite{Beem:2012mb}.
The holomorphic blocks for $T[Q]$ with a~general matrix $C_{ij}$ have the~following structure:
\begin{align}\label{eq:holomorphic-blocks}
	B_\alpha \sim & 
	\int_{\Gamma_\alpha}
	\prod_{i\in Q_0} 
	\frac{dy_i}{y_i} 
	\(\prod_{i\in Q_0}
	\, (q^2 y_i  ;q^2 )_\infty 
	\)
	\cdot
	 \(
	  \prod_{i\in Q_0}
	 \theta(-y_i;q^2)^{-C_{ii}}
	\)
	\\
	& \times 
	\(
	\prod_{i<j\,\in Q_0}
	\(
	\frac{\theta(- y_i \cdot y_j;q^2)  }{\theta(-y_i;q^2)  \theta(-y_j;q^2) }
	\)^{-C_{ij}}
	\)
	\cdot
	\(
	\prod_{i\in Q_0}
	\frac{\theta(-y_i;q^2)  \theta(-\tilde x_i;q^2) }{\theta(-y_i \cdot \tilde x_i;q^2)  }
	\)\,,\nonumber
\end{align}
where the~integral is performed over all gauge fugacities, and the~$q$-Pochhammers arise from the~chiral multiplets. The $\theta$ functions are defined as 
\be
	\theta(z;q^2) = (z;q^2)_\infty  (q^2 z^{-1};q^2)_\infty  
\ee
and they arise from Chern-Simons couplings. 

The semiclassical limit of (\ref{eq:holomorphic-blocks}) is exactly (\ref{eq:bar-P-semiclassical-limit}).
On the~other hand, the~equivariant vortex partition function of $T[Q]$ is obtained by computing the~integral over the~gauge fugacities. 
For the~integral to be well-defined, a~judicious choice of contour $\Gamma_\alpha$ needs to be specified. The choice is not unique, each corresponds to a~choice of boundary condition for the~fields at infinity on $\IR^2\times S^1$. 
For the~purpose of matching the~blocks with quiver partition functions, it is convenient to work with $|q|>1$ and choose the~contour so as to pick up only contributions from the~poles of the~$q$-Pochhammers associated with chirals. In the~rest of the~paper we have been working with $|q|<1$. As noted in \cite{Beem:2012mb}, switching from $|q|<1$ to $|q|>1$ typically introduces multiplicative overall prefactors, which are unimportant for our purpose. While the~two results cannot be analytically continued into each other, they can still be compared upon recasting them as functions of $q$-Pochhammers, since the~latter have a~well-defined factorization in either regime.

For $|q|>1$, the~$q$-Pochhammer factors as $(q^2y;q^2)_\infty = \prod_{s\geq 0} (1-q^{-2s}y)^{-1}$, therefore poles are located at
\be\label{eq:fourier-variables}
	y_i = q^{2 d_i} \qquad d_i\geq 0\,.
\ee 
When taking residues, each piece of the~integrand evaluates as follows:
\be
\begin{split}
	{\rm Res}_{y = q^{2r}} (q^2 y;q^2)_\infty & = (q^{-2})^{-1}_\infty \, (q^{2})_r^{-1} \\
	\left. \theta(- y ; q^2)   \right|_{y= q^{2 r}}  & = \frac{1}{2} [q^{-2}]_\infty^{-2} q^{-r(r-1)} \\
	\left. \theta(- x y ; q^2)   \right|_{y= q^{2 r}}  & = x^{-r} q^{-r(r-1)} \theta(-x;q^2)
\end{split}
\ee
where $(x)_r = (1-x)\dots (1-x^r)$ and $[x]_r = (1+x)\dots (1+x^r)$.
The integral therefore evaluates to
\be
	B_\alpha \sim \sum_{d_1,\ldots,d_m\geq0} \(\prod_{i=1}^{m} {\tilde x_i}^{d_i}\) \, q^{\sum_i C_{ii}\, d_i (d_i-1) + 2\sum_{i<j} C_{ij} d_i d_j} \prod_{i=1}^{m} \frac{1}{(q^{2})_{d_i}}\,,
\ee
up to overall factors of $[q^{-2}]_\infty$ and $(q^{-2})_\infty$, which can be absorbed by an~overall  normalization. 
Identifying 
\be
	\tilde x_i  = x_i q^{C_{ii}}
\ee
the vortex partition function computed by the~holomorphic block matches exactly the~quiver partition function (\ref{eq:Efimov}).
Therefore the~quiver partition function, that was observed in \cite{Kucharski:2017poe,Kucharski:2017ogk} to capture knot invariants, arises as the~vortex quantum mechanics of the~3d $\CN=2$ theory $T[Q_K]$.

Another way to view the~relation between holomorphic blocks and the~quiver partition function is via a~sort of ``Fourier transform".
In the~semiclassical limit $\hbar \to 0$ this relation reduces to the~Legendre transform that relates $\tCW^\eff_{T_0[L_K]}$ to $W_K$, discussed in Section \ref{Weff-WK}. 
Here we presented the~relation between the~quiver partition function and the~gauge theory partition function to all orders in $\hbar$. The two are related by the~integral transform (\ref{eq:holomorphic-blocks}), with dual variables $\log x_i$ and $d_i$ (whose semiclassical limit is $\log y_i$) as can be evinced from the~pole structure (\ref{eq:fourier-variables}). 
Indeed, this viewpoint arises naturally by considering the~line operator identities for the~theory $T[Q_{K}]$ \cite{Beem:2012mb,Gadde:2013wq}. These give rise to operators~$\widehat{A}_i$, one for each quiver node, which annihilate the~vortex partition function of $T[Q_K]$. These identities thus provide quantum A-polynomials associated to quiver nodes, which generalize the~relation (\ref{eq:dual A-polynomial annihilating P}). In the~semiclassical limit $\hbar \to 0$ we expect them to reduce to quiver A-polynomials defined in (\ref{eq:quiver A-poly}). For the discussion of quantum A-polynomials associated to quivers in the context of various Calabi-Yau manifolds see \cite{PS18}.

The quantum line operators described by $\widehat{A}_i$ have an~interesting interpretation from the~viewpoint of the~Chern-Simons theory on $L_K$. 
Each node of the~quiver is dual to a~basic holomorphic disk bounded by $L_K$ and wrapped by a~M2-brane. 
The disk boundary lies along an~embedded curve in a~neighborhood of $K$, and sources a~holonomy for the~Chern-Simons connection. Therefore basic disks can be viewed as line-defect insertions in the~theory on $L_K$.
A natural interpretation of these defects arises in type IIA string theory on $X\times \IR^4$ with a~D4 brane on $L_K\times \IR^2$ and with D2 branes wrapped on basic holomorphic disks in $X$ times a~line in $\IR^2$ \cite{Ooguri:1999bv}. 
The boundary of a~D2 brane couples to the~magnetic 2-form on the~D4, giving rise to line defects on $L_K$ and the~corresponding line defects on $\IR^2$. 
The holonomy around the~former corresponds to the~net flux sourced by the~latter.

\subsection{Quivers $Q_K$  vs  vortex quivers}\label{subsec:TL-TQ}

The BPS vortex spectrum of $T[L_{K}]$ is expected to admit a~quantum mechanical description on physical grounds. We can view $T[L_K]$ as a~3d defect coupled to the~5d $\CN=1$ theory engineered by $X$. Then applying equivariant localization to the~path integral of this 3d-5d system in the~omega background turns the~computation of the~partition of $n$~vortices into a~(refined) Witten index computation for a~1d $\CN=2$ quantum mechanics. 
The~corresponding Hilbert space is expected to provide the~physical realization of knot homologies \cite{GNSSS1512,Gukov:2011ry,Gukov:2007ck}.
Recently it was observed that the~vortex quantum mechanics of certain 3d~$\CN=2$ theories admits a~quiver description \cite{Hwang:2017kmk}.
Here we comment on the~relation between these quivers and $Q_K$: we argue that they share important properties but they are \emph{not} equivalent. 
In fact the~relation between the~two is rather novel and worth studying in its own right.

The M-theory engineering of $T[L_{K}]$ provides a~direct link between holomorphic disks and BPS  vortices \cite{Ooguri:1999bv}, therefore a~quiver description of the~former raises the~question of a~similar description for the~latter. 
Above we argued that the~existence of a~quiver $Q_K$ implies that the~spectrum of holomorphic disks can be regarded as boundstates of a~set of ``basic disks'', a~more geometric description will be provided in the~next section.
An interesting observation made in \cite{Hwang:2017kmk} is that the~vortex spectrum also admits a~quiver description, implying that the~there is a~set of ``fundamental vortices'' that generate the~whole spectrum.
On the~one hand it is natural to identify basic holomorphic disks, which correspond to nodes of $Q_K$, with fundamental vortices. 
On the~other hand, the~two quivers generate the~same full BPS spectrum in rather different ways, suggesting that they describe different (but related) dynamics.

\begin{figure}[ht]
\begin{center}
	\includegraphics[width=.35\textwidth]{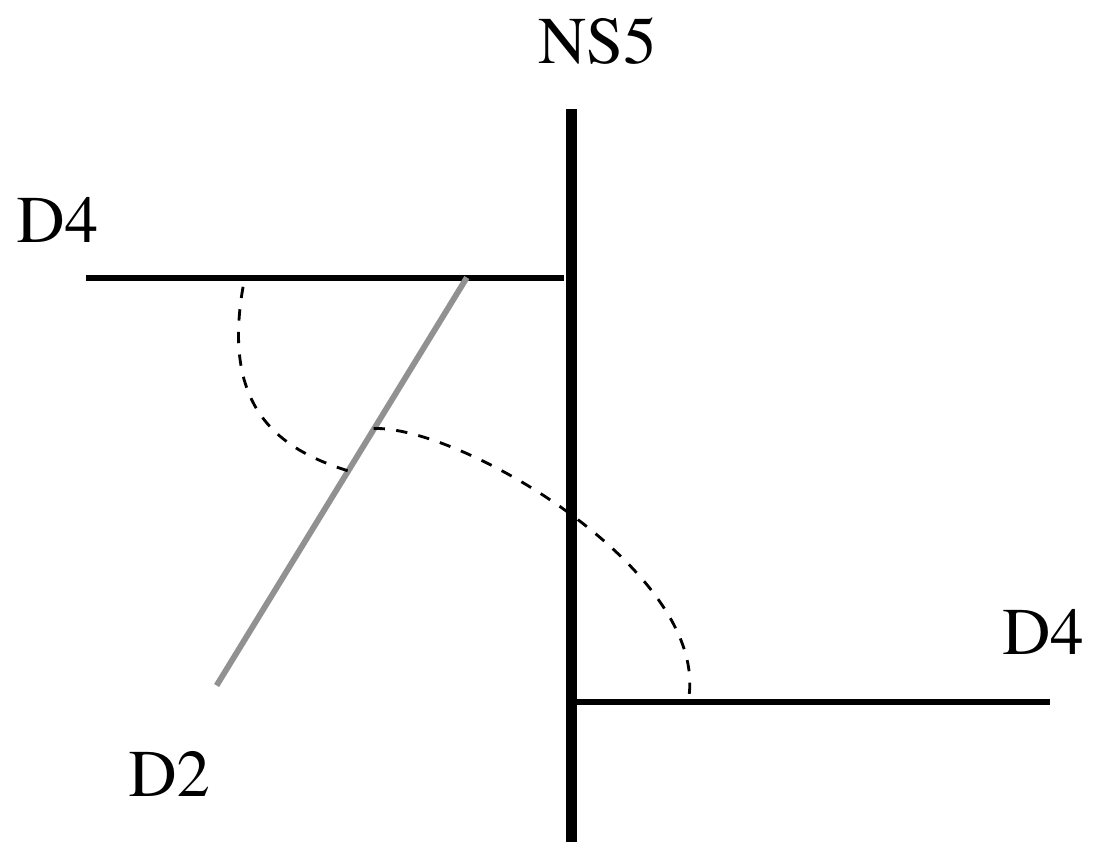}
	\caption{Brane construction of the~unknot theory.}
	\label{fig:unknot-brane}
\end{center}
\end{figure}

\begin{figure}[ht]
\begin{center}
	\includegraphics[width=.5\textwidth]{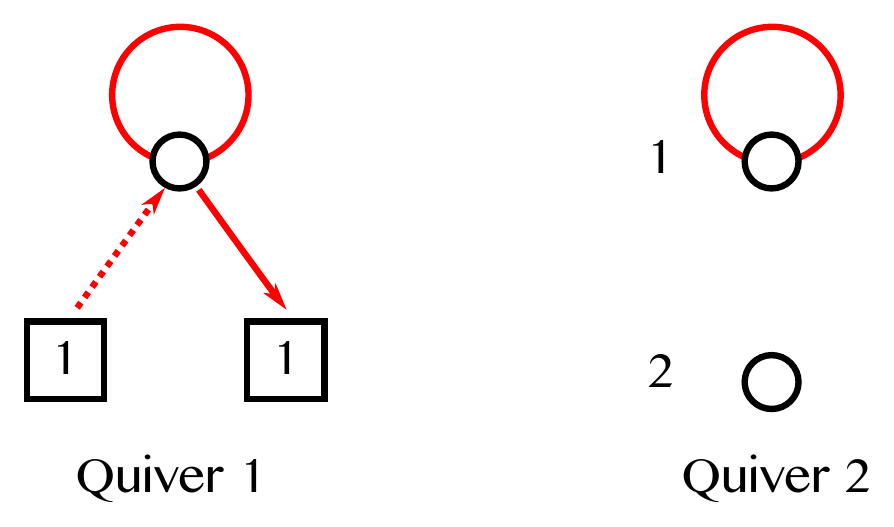}
	\caption{Quiver 1 encodes the~$\CN=2$ vortex quantum mechanics for the~unknot theory $T[L_K]$. Quiver 2 encodes the~$\CN=4$ quantum mechanics of holomorphic disks and this is  $Q_{0_1}$ analyzed in Section \ref{sub:Unknot}.}
	\label{fig:unknot-quivers}
\end{center}
\end{figure}

We do not understand the~relation between these two in general, partly because the~vortex quivers of \cite{Hwang:2017kmk} were derived from mass-deformations of 3d $\CN=4$ theories admitting a~brane construction and $T[Q_{K}]$ is generally not of this type. However, there is one example where both descriptions are available and can be compared. This is the~case of the~unknot theory, which can be described in two ways:
\begin{itemize}
\item[$T_1$:] $U(1)$ gauge theory with one fundamental chiral and one antifundamental chiral. 
\item[$T_2$:] $U(1)\times U(1)$ gauge theory with one fundamental chiral for each gauge group and effective Chern-Simons level $\kappa_\eff = 1$ for the~first gauge group
\end{itemize}
These two descriptions correspond to $T[L_K]$ and $T[Q_K]$ respectively.
Since the~unknot conormal coincides with the~standard toric brane on the~resolved conifold, the~first theory can be engineered by a~brane construction in type IIA string theory shown in Figure \ref{fig:unknot-brane}, see for example \cite{Dimofte:2010tz}. 
Moreover, the~existence of a~brane construction for this model was exploited in \cite{Hwang:2017kmk} to derive a~quiver description of the~quantum mechanics of vortices for~$T_1$, it is Quiver 1 in Figure \ref{fig:unknot-quivers}.
The circle represents a~gauge group $U(n)$, the~solid (resp. dashed) arrow represent a~1d $\CN=2$ chiral (resp. Fermi) multiplet charged under $U(n)$, cf.~{\cite[fig.~13]{Hwang:2017kmk}}.
The Witten index computes the~partition function of $n$ vortices, for a~positive choice of the~FI coupling it is given in \cite[eq. (4.7)]{Hwang:2017kmk}
\be
	I^n = a^{-n} q^{n} \frac{(a^2 ;q^2)_n}{(q^2;q^2)_n} \,.
\ee
The vortex partition function for theory $T_1$ is then
\be\label{eq:Z-vort-T1}
	\CZ^\vort[T_1] = \sum_{n\geq 0} I^n \, x^n \dot{=} \sum_{n\geq 0} x^n \frac{(a^2 ;q^2)_n}{(q^2;q^2)_n} \,,
\ee
where for notational convenience we absorbed $a^{-1} q$ into the~vortex fugacity by a~redefinition $x\rightarrow a~q^{-1}x$, which is denoted by $\dot{=}$.

On the~other hand, Quiver 2 coincides with the~quiver $Q_{0_1}$ found in \cite{Kucharski:2017poe,Kucharski:2017ogk}. It describes a~$\CN=4$ quantum mechanics with gauge group $U(d_1)\times U(d_2)$ and an~adjoint chiral multiplet charged under the~first group. The refined Witten index $I^{\mathbf{d}}$ is zero for most dimension vectors $\mathbf{d}  = (d_1,d_2)$, except for
\be
	I^{(1,0)}=-q\,,\qquad I^{(0,1)}=1\,.
\ee
These are the~motivic DT invariants of the~quiver representation theory, see Section \ref{sub:Unknot}.
The corresponding motivic generating series gives the~vortex partition function
\be\label{eq:Z-vort-T2}
	\CZ^\vort[T_2] = \textrm{Exp}\left(\frac{ I^{(1,0)} x_{1}+ I^{(0,1)} x_{2}}{1-q^{2}}\right) = \sum_{d_{1},d_{2}\geq0}(-q)^{d_{1}^{2}}\frac{x_{1}^{d_{1}}}{(q^{2};q^{2})_{d_{1}}}\frac{x_{2}^{d_{2}}}{(q^{2};q^{2})_{d_{2}}} \,.
\ee

The background Chern-Simons couplings $a_i, t_i, q_i$ of theory $T_2$ are
\be
	a_1 = 2\,,
	q_1 = 0\,,
	t_1 = 1\,,
	\qquad
	a_2 = 0\,,
	q_2 = 0\,,
	t_2 = 0\,.
\ee
These values instruct us to compare the~two expressions through the~following (KQ) change of variables:
\be\label{eq:unknot-quiver-fugacities}
	x_1 = x a^{2} q^{-1} \qquad x_2= x\,.
\ee
It is not hard to check that these relations imply
\be
	\CZ^\vort[T_1]=\CZ^\vort[T_2]\,.
\ee 

Theories $T_1$ and $T_2$ have the~same vortex spectrum, however this is described by two rather different quivers.
In the~first description there is a~single gauge node, whose fugacity is $x$: this is the~description that arises naturally from a~brane construction \cite{Dimofte:2010tz,Hwang:2017kmk} where the~$U(1)$ symmetry arises from $H_1(L_K) \simeq \IZ$.
The quiver quantum mechanics arises from the~equivariant localization of the~path integral of the~3d-5d system, as a~consequence its Witten index computes the~$n$-vortex partition function.
On the~other hand, the~second description involves two gauge groups which  appear to be a~mix of the~gauge $U(1)$ appearing in $T_1$ with the~background global symmetries. The precise combination of these $U(1)$'s is dictated precisely by $a_i, q_i, t_i$, recall (\ref{eq:unknot-quiver-fugacities}).
The Witten index of the~second quiver does not give the~vortex partition function, which is instead given by its \emph{plethystic exponential} in~(\ref{eq:Z-vort-T2}).
This is reminiscent of quiver descriptions of M2 boundstates in 4d $\CN=2$ theories~\cite{Denef:2002ru}, providing another hint that the~quiver quantum mechanics of theory $T_2$ describes the~worldvolume dynamics of two basic holomorphic disks.

Another important distinction between the~two quivers is the~amount of supersymmetry involved in each description: Quiver 1  encodes an~$\CN=2$ quantum mechanics which is the~expected description of vortices, while Quiver 2 describes a~$\CN=4$ quantum mechanics which is expected for a~description of M2-branes ending on $L_K$ \cite{Aganagic:2012hs}.
This suggests that these dual descriptions can be regarded as switching the~perspective from the~dynamics of the~M5-brane wrapped on $L_K$, to the~dynamics of holomorphic disks that end on it.
Indeed Quiver 1 arises naturally from the~brane construction \cite{Dimofte:2010tz,Hwang:2017kmk}, while the~origin of Quiver 2 is more naturally understood from the~viewpoint of interacting holomorphic disks.

From the~perspective of knot theory, it is clear that each node of the~quiver must be dual to a~generator of HOMFLY-PT homology (see Section \ref{sub:Knot-homologies}).
In turn, generators of the~homology are dual to embedded holomorphic disks ending on $L_K$. 
Since $x_i\sim x$, these are the~disks that wind exactly once around the~generator of $H_1(L_K)$. 
The rationale behind quiver descriptions of BPS spectra is that the~quiver nodes are the~``basic'' BPS states, while the~rest is generated from their boundstates.
The latter consist of more complicated holomorphic curves, winding more than once around $L_K$ -- generalized holomorphic curves introduced in Section \ref{geometric background} and further studied in Section \ref{sec:math}.
The quiver $Q_{K}$ encodes the~dynamics of interacting basic disks, which determines the~spectrum of their boundstates.
In turn, this dynamics must depend on the~geometry of the~ M5-brane wrapping $L_K$: this is non-compact and rigid, providing a~background on which the~M2-branes wrapping holomorphic disks can end and interact with each other. This viewpoint is further corroborated by considering the~Legendre transform of $T[Q_{K}]$: this gives an~effective theory associated to $W_{Q_{K}}(\mathbf{x})$ (the quiver disk potential defined in (\ref{eq:quiver-disk-potential})), which describes precisely the~interaction of sources $x_i$ corresponding to basic holomorphic disks.

Interactions among basic disks may be encoded for example by the~mutual linking numbers of disk boundaries.
When two boundaries link, the~M2 worldvolumes can interact, the~light fields localized at the~intersection of a family of Dirac strings beginning of one boundary and the other boundary give rise to $C_{ij}$ bifundamental modes (the quiver arrows) in the~quiver quantum mechanics \cite{Denef:2002ru}.
A bit more precisely, while this picture is natural for mutual disk intersections (identified with $C_{ij}$) corresponding to M2-M2 interactions, more care is needed for self-linking (identified with $C_{ii} = t_i$).
In fact this gets contribution from both M2-M2 self-interactions and M2-M5 interactions. 
For~the~example at hand, we can see this mixing appearing in the~$q$-powers of (\ref{eq:unknot-quiver-fugacities}): both $q_1=q_2=0$ are the~same since both disks end on $L_K$ without linking (a proper mathematical definition of this will be given in the~next section) and therefore both M2-branes have the~same type of interaction with M5 wrapping $L_K$; on the~other hand $t_1 \neq t_2$ because one of the~basic disks is self-linking,  and therefore a~BPS M2-brane wrapped on it experiences a~nontrivial M2-M2 self-interaction.

\section{KQ correspondence and geometry of holomorphic curves}\label{sec:math}

In this section we discuss geometric interpretations of the~knots-quivers correspondence. We first introduce necessary geometric objects, in particular generalized holomorphic curves. Then, we analyze the~meaning of nodes and arrows from different perspectives. Finally, we study refinement in the~context of LMOV invariants and geometry of Chern-Simons theory.

\subsection{Quivers and generalized holomorphic curves}\label{Q and curves}

We give a~geometric interpretation of the~quiver vertices and arrows described in Section~\ref{geometric background}. The basic idea is that all holomorphic curves arise from a~finite collection of basic holomorphic disks. Other holomorphic curves are then combinations of standard contributions from constant curves and branched covers of basic disks, and the~quiver partition function arises as the~corresponding count. For simple knots the~basic disks correspond to the~monomials in the~HOMFLY-PT polynomial. In the~general case, also other disks are needed and it is an~open problem to give an~effective characterization of when and how to find these additional disks.  

As explained in \cite{Ekholm:2018iso}, see also \cite{iacovino1,iacovino2,iacovino3} for similar earlier results, not only actual holomorphic curves, but also their composite configurations contribute to open Gromov-Witten potentials. Such configurations are called \emph{generalized holomorphic curves} and to specify them we need additional geometric data that we decribe next. Before the~description we point out that the~generalized holomorphic curves generated by basic disks are closely related to $U(1)$ Chern-Simons theory on $L_{K}$ with defects, see \cite{ES} for the~exact relation.
  
We recall the~definiton of generalized holomorphic curves with boundary on a~knot conormal $L_{K}\subset X$ in the~resolved conifold $X$, see \cite{Ekholm:2018iso}. The~additional geometric data are as follows: a~Morse function $f\colon S^{1}\times\R^{2}\to\R$ and a~$4$-chain $C$ with boundary $\partial C=2\cdot L_{K}$ and such that the~normal vector fields of $C$ along $\partial C$ equals $\pm J \cdot \nabla f$, where $J$ is the~almost complex structure. More precisely, we will use Morse functions that are small perturbations of Bott functions as follows. Consider a~function on $S^{1}\times\R^{2}$ that is independent of the~$S^{1}$ coordinate and has a~unique non-degnerate minimum in each $\R^{2}$-fiber and with radial gradient near infinity. Thinking of $S^{1}\times\R^{2}$ as $L_{K}$, such a~function has a~minimum along the~knot $K\subset L_{K}$ (the Bott-manifold) and gradient flow along the~$\IR^2$ fibers. After small perturbation by a~Morse function on $K$ with two critical points, the~resulting Morse function will have two critical points $\kappa_{0}$ of index $0$ and $\kappa_{1}$ of index $1$. The stable manifold $W^{\rm s}(\kappa_{1})$ of $\kappa_{1}$ is the~knot and the~unstable manifold $W^{\rm u}(\kappa_{1})$ is a~fiber disk. We will use Morse functions $f\colon L_{K}\to\R$ of this form.

\begin{figure}[h!]
\begin{center}
\includegraphics[width=0.5\textwidth]{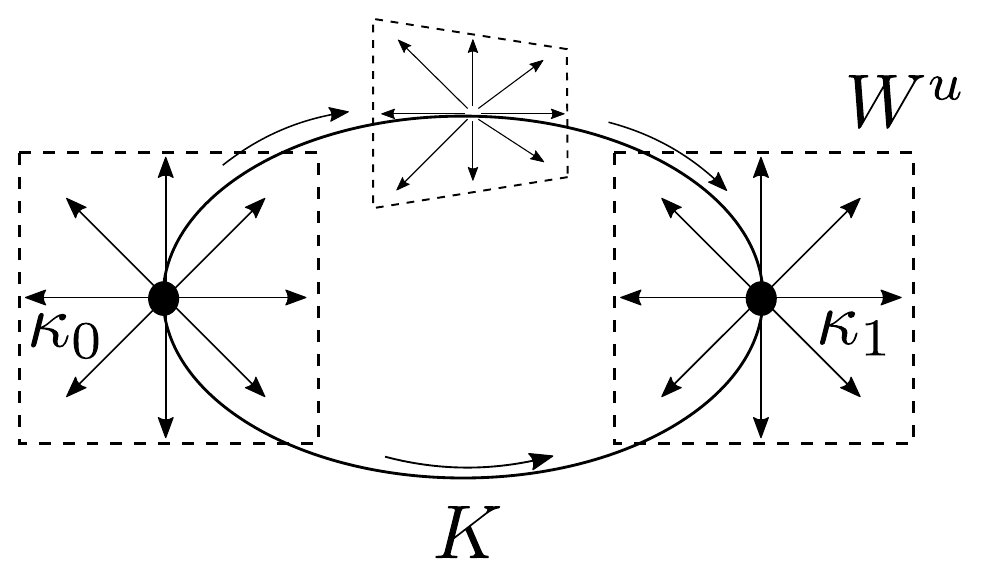}
\caption{Morse flow defining the~4-chain $C$, shown inside $L_K$.}
\label{fig:4-chain}
\end{center}
\end{figure}

We use the~Morse function $f$ to associate \emph{bounding chain} $\sigma_{u}$ to holomorphic curve with boundary on $L_{K}$, $u\colon(\Sigma,\partial \Sigma)\to (X,L_{K})$, as follows. 
Let $\sigma_{u}'$ denote the~union of all flow lines starting on $u(\partial\Sigma)$. Then the~intersection of $\sigma_{u}'$ with a~boundary torus $\Lambda_{K}$ of $L_{K}$ sufficiently close to infinity is a~curve that represents a~class $k\xi+n\eta$ where $\xi$ is the~longitude and $\eta$ the~meridian. Finally, define
\be\label{eq:bounding-chain}
\sigma_{u} = \sigma_{u}' - n\cdot W^{\rm u}(\kappa_{1}).
\ee

\begin{figure}[h!]
\begin{center}
\includegraphics[width=0.85\textwidth]{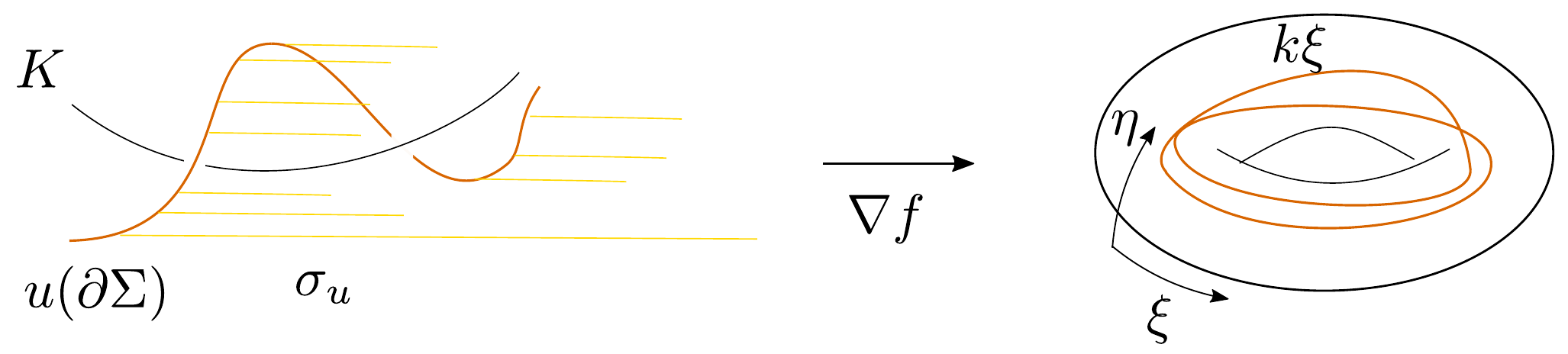}
\caption{The bounding chain $\sigma_u$ ends on the~cycle $k \xi$ on the~torus at infinity inside $L_K$.}
\label{fig:bounding-chain}
\end{center}
\end{figure}

We also consider the~following construction of an~\emph{intrinsic self linking number} of $u$. Let $\nu$ be any normal vector field of $\partial u$ and let $\partial u_{\nu}$ be a~small shift of $\partial u$. 
We extend the~vector field $J\nu$ to a~small neighborhood of $\partial u$ and use that field to shift $u$ off of $L_{K}$. We denote the~shifted curve $u_{J\nu}$ and define
\be\label{eq:slk}
\slk(u) = \partial u_{\nu}\cdot \sigma_{u} - u_{J\nu}\cdot C,
\ee
where $\cdot$ denotes algebraic intersection number. It is straightforward to check that $\slk(u)$ is independent of $\nu$: the~first intersection number changes when $\nu$ passes $\pm \nabla f$  and a~local check shows that there is then a~compensating change in the~second term.

Let us comment a~bit more informally on the~various ingredients that went into (\ref{eq:slk}). 
On an~intuitive level, the~naive self-linking of a~real curve $\partial u$ would be defined by first choosing a~pushoff of $\partial u$ and then measuring the~linking of the~original curve with its pushoff. The pushoff is provided by $\nu$, while the~linking may be defined as the~meridian winding of the~pushoff on the~$T^2$ neighborhood of the~original curve.
The role of the~Morse flow $\nabla f$ is to provide a~notion of meridian winding for $\partial u$ around $K$, which would otherwise be topologically trivial in $L_K\sim S^1\times \IR^2$. This information is encoded by the~intersection of $\sigma_u'$ with $\Lambda_K$, and eventually stored into the~topology of the~bounding chain $\sigma_u$ via (\ref{eq:bounding-chain}).
The~meridian winding of the~pushoff $\partial u_\nu$ is then its intersection with the~bounding chain $\sigma_u$. The~second piece in the~formula accounts for the~possibility that some of the~self-intersections become ``virtual'', such as through a~Reidemeister zero-type move. The role of the~four-chain is to collect these virtual contributions, and it is crucial that it is the~imaginary counterpart of the~Morse flow for this purpose.

A generalized holomorphic curve is a~directed graph with actual holomorphic curves at the~vertices. For each edge connecting two distinct curves $u$ and $v$ we pick an~intersection point in $\partial u\cap \sigma_{v}$, and for each edge connecting a~curve $u$ to itself we pick an~intersection point contributing to $\slk(u)$. Such a~generalized curve $\Gamma$ with vertices $V(\Gamma)$ and edges $E(\Gamma)$ is defined to have Euler characteristic 
\be
\chi(\Gamma_{u})=\sum_{u\in E(\Gamma)}\chi(u) -|E(\Gamma)|.
\ee

We next consider the~relation to the~quiver theory $T[Q_K]$. Let us assume that there is a~finite number of embedded holomorphic disks with boundary on $L_{K}$, $u_{1},\dots,u_{m}$, where each $u_{j}$ has boundary that goes once around the~generator of $H_{1}(L_{K})$. We assume furthermore that the~linking numbers between holomorphic disk boundaries are
\be\label{eq:quiver-linking}
C_{ij}=\partial u_{i}\cdot \sigma_{u_{j}}=\partial u_{j}\cdot \sigma_{u_{i}},\quad
C_{ii}= \partial {u_{i}}_{\nu}\cdot \sigma_{u_{i}},
\ee
where $\nu$ is the~normal vector field everywhere linearly independent with $\nabla f$.

Note that for each embedded disk, the~count of contributions from constant curves and multiple covers are exactly like for the~basic disks for the~unknot. We say that the~generalized holomorphic curves that have vertices corresponding to branched covers and constant curves of the~basic disks are the~curves \emph{generated by the~basic disks}. 

It then follows from the~count of curves for the~unknot, together with the~definition of generalized holomorphic curves, that if $Q_K$ is the~quiver with nodes $u_{1},\dots,u_{m}$ and associated quiver variables  $x_{1},\dots,x_{m}$, and adjacency matrix $C_{ij}$, then the~Gromov-Witten partition function that counts curves generated by the~basic disks and the~quiver generating function are equal (see \cite[Section~2.4]{Ekholm:2018iso}), provided we make the~substitution 
\be
x_{i}=x^{n_i}a^{a_{i}}q^{\slk(u_{i})-C_{ii}},
\ee
where $q=e^{\frac12 g_{s}}$. We will discuss this observation with more details and also including refinements of the~count in the~following sections. 

We would like to close this section with a~discussion of framing.
The framing of $L_{K}$ corresponds to the~choice of longitude curve in the~torus at infinity which affects the~definition of the~bounding chain $\sigma_{u}$ of a~holomorphic disk. More precisely, if $u$ is a~basic holomorphic disk and the~new longitude is $\xi+f\eta$, the~bounding chain for $u$ is changed by addition of $f\cdot W^{\rm u}(\kappa_{1})$.  Noting that the~holomorphic curves themselves are unaffected by the~choice of framing and that all basic disks go once around the~homology generator, it follows from (\ref{eq:quiver-linking}) that a~framing change modifies the~quiver by an~overall additive constant $C_{ij}\to C_{ij}+f,$ adding $f$ arrows between all vertices.
This matches indeed the~description of framing found in \cite{Kucharski:2017poe,Kucharski:2017ogk}.

\subsection{Contributions from quiver nodes\label{sub:Nodes}}
In this section we discuss various interpretations of quiver nodes including as
homology generators, as LMOV invariants,  and  as basic holomorphic disks.
Let us consider the~quiver
motivic generating series restricted to dimension vectors of length one:
\begin{equation}
P_{|\mathbf{d}|=1}^{Q_K}(\mathbf{x},q)=\sum_{i=1}^{m}\frac{(-q)^{C_{ii}}x_{i}}{1-q^{2}}\,.\label{eq:PQ1}
\end{equation}
Here every quiver vertex contributes, but there are no contributions from interactions between vertices since any resulting ``bound states" give terms that are at least quadratic in the~variables $x_{i}$.  

If we apply the~KQ change of variables, then we obtain the~standard HOMFLY-PT polynomial (colored by the~standard one box representation):
\begin{equation}
\left.P_{|\mathbf{d}|=1}^{Q_{K}}(\mathbf{x},q)\right|_{x_{i}=a^{a_{i}}q^{q_{i}-C_{ii}}x}  = \ P_{1}^{K}(a,q)x \ = \ P_{1}^{K}(a,q)e^{\xi}\,.
\end{equation}

We first observe a~connection to HOMFLY-PT homology. As pointed out in Section~\ref{sub:Knot-homologies}, powers $a_{i}$, $q_{i}$, $t_{i}=C_{ii}$ are equal
to degrees of natural generators of HOMFLY-PT homology $\mathscr{H}(K)$. More precisely, if  $\mathscr{G}(K)$ denotes this set of generators then 
\begin{equation}
P_{1}^{K}(a,q)=\frac{\sum_{i\in\mathscr{G}(K)}a^{a_{i}}q^{q_{i}}(-1)^{t_{i}}}{1-q^{2}}.\label{eq:PK triply homology generators}
\end{equation}
Following \cite{Kucharski:2017poe,Kucharski:2017ogk}, we consider also further refined versions of the~KQ change of variables, corresponding
to three and four gradings, and get
\begin{equation}
\begin{split}
\left.P_{|\mathbf{d}|=1}^{Q_{K}}(\mathbf{x},q)\right|_{x_{i}=a^{a_{i}}q^{q_{i}-C_{ii}}(-t)^{C_{ii}}x}
& =
\ \mathcal{P}_{1}^{K}(a,q,t)x \ 
\\
&= \ \frac{\sum_{i\in\mathscr{G}(K)}a^{a_{i}}q^{q_{i}}t^{t_{i}}}{1-q^{2}}x\,,
\end{split}
\end{equation}
\begin{equation}
\begin{split}
\left.P_{|\mathbf{d}|=1}^{Q_{K}}(\mathbf{x},q)\right|_{x_{i}=a^{a_{i}}Q^{q_{i}}(-t_{r})^{C_{ii}}x,\ q=t_{c}}
& = \ \tilde{\mathcal{P}}_{1}^{K}(a,Q,t_{r},t_{c})x \\
& = \ \frac{\sum_{i\in\tilde{\mathscr{G}}(K)}a^{a_{i}}Q^{q_{i}}t_{r}^{t_{i}}t_{c}^{t_{i}}}{1-t_{c}^{2}}x\,.
\end{split}
\end{equation}
Comparing with Section \ref{Q and curves} we see that the~splitting
of $q$ between $Q$, $t_{r}$, $t_{c}$ and other properties are perfectly consistent
with geometric interpretation in terms of holomorphic disks:
\begin{itemize}
\item Each node corresponds to a~holomorphic disk.  	
\item	
The power $r=1$ in $x^{r}=e^{r\xi}$ corresponds to the~homology class of the~boundary of the~holomorphic curve in $H_{1}(L_{K})$.
\item 
After picking a~reference disk $\delta$ capping the~generator $\xi$ of $H_{1}(L_{K})$ the~power $a_{i}$ in $a^{a_{i}}$ is the~homology class of the~holomorphic curve capped by $-r\delta$ in $H_{2}(X)$.  	
\item The power $q_{i}$ in $Q^{q_{i}}$ is the~self-linking $\textrm{slk}(u_{i})$.
\item The exponent $t_{i}=C_{ii}$ in $t_{r}^{t_{i}}=t_{r}^{C_{ii}}$ equals the~intersection number $\partial {u_{i}}_{\nu}\cdot \sigma_{u_{i}}$. 
\item The powers of $t_{c}$ are encoded in the~off-diagonal entries in the~adjacency matrix $C_{ij}$ and correspond to to the~number
of intersection points between $\partial u$ and $\sigma_{v}$: $C_{ij}=\partial u_{i}\cdot \sigma_{u_{j}}$.
\end{itemize}
In conclusion, we observe that there is a~natural correspondence between basic holomorphic disks and topological data associated to them and homology generators and their degrees. 

Since uncolored HOMFLY-PT homology is directly related to LMOV invariants
\cite{GSV0412}, we can add another ingredient to our picture. Let us
start by expressing (\ref{eq:PQ1}) in terms of the~DT generating
function

\begin{equation}
P_{|\mathbf{d}|=1}^{Q_K}(\mathbf{x},q)=\frac{\Omega_{|\mathbf{d}|=1}^{Q_K}(\mathbf{x},q)}{1-q^{2}}\,.\label{eq:PQ1 DT}
\end{equation}
After applying the~KQ change of variables, we obtain
\begin{equation}
P_{1}^{K}(a,q)x=\frac{N_{1}^{K}(a,q)x}{1-q^{2}}\,.\label{eq:PK1 LMOV}
\end{equation}
LMOV invariants count BPS states in 3d $\mathcal{N}=2$ theory $T[L_{K}]$
described in Section~\ref{Theories T[M_K] and T[L_K]}, whereas DT invariants
give numbers of BPS states in 3d $\mathcal{N}=2$ theory $T[Q_K]$ from
Section~\ref{subsec:TofQ}. Therefore we can
reinterpret our holomorphic disks and their topological data in terms
of BPS states and their quantum numbers. This is in line with the
brane construction from \cite{Ooguri:1999bv}. $T[L_{K}]$ arises as effective
theory on the~surface of M5-brane and its BPS particles originate
from M2-branes ending on M5, which brings us back to holomorphic disks
ending on the~Lagrangian submanifold. Here it is natural to interpret the~adjacency matrix $C_{ij}$ as corresponding to M2-M2 interactions and the~parameter $q_{i}-C_{ii}$ as corresponding to M2-M5 interactions.

Looking back at equations (\ref{eq:PQ1 DT}--\ref{eq:PK1 LMOV}), we
can ask what happens if we apply refined KQ changes of variables.
We will come back to this question in Section~\ref{sub:refinement-of-LMOV}.

In this section we were dealing with different interpretations of
quiver nodes. They correspond to holomorphic disks, homology generators
and BPS states. The KQ change of variables can be encoded in topological
data of disks, degrees of generators, or quantum numbers of BPS states.
Let us stress that we considered dimension vectors of lenght one,
which corresponds to fundamental representation or disks winding
around $L_{K}$ once. In this case there is no interaction between
nodes, so all observations from this section can be summarized as
describing internal properties of considered objects. (Here internal should be understood as involving M2 self interactions and M2-M5 interactions.)

\subsection{Arrows: how basic disks generate all curves\label{sub:Arrows}}
After considering internal properties of objects corresponding to
quiver nodes, we are ready to look at their interactions encoded in
quiver arrows.

The motivic generating series is built
by combinations of factors from the~nodes determined by dimension
vector $\mathbf{d}$ and weighted according to the~adjacency matrix $C_{ij}$
\begin{equation}
P^{Q}(\mathbf{x},q)=\sum_{d_{1},\ldots,d_{m}\geq0}(-q)^{\sum_{1\leq i,j\leq m}C_{ij}d_{i}d_{j}}\prod_{i=1}^{m}\frac{x_{i}^{d_{i}}}{(q^{2};q^{2})_{d_{i}}}.
\end{equation}
The KQ change of variables hides this structure, but does
not break it
\begin{align}
\left.P^{Q_{K}}(\mathbf{x},q)\right|_{x_{i}=a^{a_{i}}q^{q_{i}-C_{ii}}x} & =P^{K}(x,a,q)\\
 & =\sum_{d_{1},\ldots,d_{m}\geq0}(-q)^{\sum_{1\leq i,j\leq m}C_{ij}d_{i}d_{j}}\prod_{i=1}^{m}\frac{(a^{a_{i}}q^{q_{i}-C_{ii}}x)^{d_{i}}}{(q^{2};q^{2})_{d_{i}}}.\nonumber 
\end{align}
We then see that quiver arrows encode the~way of building
the whole spectrum of objects (counted by the~motivic generating series
or the~HOMFLY-PT generating function) from the~basic ones which correspond
to nodes.

Interpreting the~nodes as holomorphic disks we find that the~adjacency matrix exactly encodes how to build all generalized holomorphic curves from the~basic disks. The dimension vector encodes the~number of copies of the~disk, the~matrix $C_{ij}$ counts in how many ways the~various curves can be combined into generalized curves, and the~transformations of self linking and topological degrees behaves in accordance with this interpretation. 

In fact, the~quiver partition function counts all disconnected generalized holomorphic curves generated by the~basic disks $u$. To see this, we first note that all embedded disks have a~standard neighborhood and then we can infer the~count of multiple covers and contributions from constant curves attached to these from that of disks on the~unknot, where we have the~recursion that determines the~factors. The contributions from linking and self-linking follows from the~description of generalized holomorphic curves above.

The holomorphic disks corresponding to the~nodes of the~quiver are embedded. Generalized holomorphic curves are constructed in order for the~count of all curves to remain invariant under deformations. One can also consider the~counts of generalized embedded disks. Basically, new embedded disks appear when two embedded disks intersect along the~boundary. Therefore, in order to count these objects we should count trees of nodes and arrows rather than arbitrary graphs. We call such generalized holomorphic curves \emph{semi-basic disks}. Arguing as above, we then find that the~generating function of semi-basic disks equals the~DT generating function, i.e.~the count of BPS states of the~quiver.

Looking at (\ref{eq:Efimov}) and (\ref{P^Q product form}) we can see that generalized holomorphic curves counted by quiver partition function are generated by basic disks in the~``sum way" or characterized by semi-basic disks leading to a~partition function written in the~``product way". From the~point of view of states in $T[Q_K]$ we would say that all states are generated by basic states (those corresponding to $|\textbf{d}|=1$) in the~``sum way" or associated to BPS states that give the~partition function in the~``product way".

A similar distinction on the~knot side was described in \cite{KS1608} using combinatorics on words. The ``sum way" means direct generation of all words in a~given formal language, whereas the~the ``product way" is represented by obtaining all words by concatenation of Lyndon words. We shall stress that the~analogy is not perfect because in \cite{KS1608} the~analysed generating fuction was the~ratio $P^K(q^2 x,a,q)/P^K(x,a,q)$, not the~HOMFLY-PT generating series itself. However, the~main structure is preserved, only building blocks are different. In our case they are basic disks corresponding to terms in $P^K_1(a,q)$, whereas in \cite{KS1608} they are one-letter words representing terms in the~quantum A-polynomial. 

Let us  consider consequences of the~result on generated holomorphic curves for the~relation between the~size of 
colored and uncolored homology. Since the~KQ correspondence relates quiver nodes with generators of $\tilde{\mathscr{H}}(K)$ and quiver partition function with Poincar\'{e} polynomial of $\tilde{\mathscr{H}}^{S^r}(K)$, we can translate the~generation of all generalized holomorphic curves by the~basic disks to the~exponential growth property \cite{Gukov:2011ry,GGS1304,Wed1602}
\be
\dim\tilde{\mathscr{H}}^{S^r}(K)=\(\dim\tilde{\mathscr{H}}(K)\)^r\,.
\ee
This statement shows limitations of the~KQ correspondence with quiver nodes corresponding to generators of $\tilde{\mathscr{H}}(K)$. In Section \ref{sub:9_42} we study the~case of the~knot $9_{42}$ which does not have this property and discuss possible generalizations of \cite{Kucharski:2017poe,Kucharski:2017ogk} that would cure this problem.

\subsection{Quiver-based refinement of LMOV invariants\label{sub:refinement-of-LMOV}}
Let us recall the~relation between LMOV invariants of a~knot $K$ and
DT invariants of the~corresponding quiver $Q_K$:
\begin{equation}
N^{K}(x,a,q)=\left.\Omega^{Q_K}(\mathbf{x},q)\right|_{x_{i}=a^{a_{i}}q^{q_{i}-C_{ii}}x}\,.
\end{equation}
As suggested in \cite{Kucharski:2017poe,Kucharski:2017ogk}, we can define refined LMOV invariants by applying the~refined KQ change of
variables to $\Omega^{Q_K}(\mathbf{x},q)$:
\begin{equation}
N^{K}(x,a,q,t)=\sum_{r,i,j,k}N_{r,i,j,k}^{K}x^{r}a^{i}q^{j}t^{k}=\left.\Omega^{Q_K}(\mathbf{x},q)\right|_{x_{i}=a^{a_{i}}q^{q_{i}-C_{ii}}(-t)^{C_{ii}}x}\,.
\end{equation}
Applying the~plethystic exponential to this formula gives the~generating function of
unreduced superpolynomials:
\begin{align}
\textrm{Exp}\left(\frac{N^{K}(x,a,q,t)}{1-q^{2}}\right) & =\left.\textrm{Exp}\left(\frac{\Omega^{Q_K}(\mathbf{x},q)}{1-q^{2}}\right)\right|_{x_{i}=a^{a_{i}}q^{q_{i}-C_{ii}}(-t)^{C_{ii}}x}\nonumber \\
 & =\left.P^{Q_K}(\mathbf{x},q)\right|_{x_{i}=a^{a_{i}}q^{q_{i}-C_{ii}}(-t)^{C_{ii}}x}\\
 & =\mathcal{P}^{K}(x,a,q,t).\nonumber 
\end{align}

Since for symmetric quivers DT invariants are integer \cite{2011arXiv1103.2736E},
the integrality of refined LMOV invariants defined in this way holds
by definition. In the~limit $q\rightarrow1$ they reduce
to classical refined LMOV invariants introduced in \cite{GKS1504} (see Section
\ref{sec:Examples} for examples and Appendix \ref{sub:Comparison-of-t-refinement} for comparison of conventions). A slightly
different refinement was proposed in \cite{KN1703} where $t_{c}$ is used as
$t$ instead of $t_{r}$. 

These two possibilities of consistent refinement of HOMFLY-PT homology was
in fact one of the~sources of inspiration for introducing a~quadruply graded homology
in \cite{GGS1304}. Our framework also enables a~natural definition
of doubly refined LMOV invariants:
\begin{equation}
\begin{split}
N^{K}(x,a,Q,t_{r},t_{c})
&=\sum_{r,i,j,k,l}N_{r,i,j,k,l}^{K}x^{r}a^{i}Q^{j}t_{r}^{k}t_{c}^{l}
\\
&=\left.\Omega^{Q_K}(\mathbf{x},q)\right|_{x_{i}=a^{a_{i}}Q^{q_{i}}(-t_{r})^{C_{ii}}x,\ q=t_{c}}
\end{split}
\end{equation}
In this case the~plethystic exponential gives the~generating function of unreduced
quadruply-graded polynomials:
\begin{align}
\textrm{Exp}\left(\frac{N^{K}(x,a,Q,t_{r},t_{c})}{1-q^{2}}\right) & =\left.\textrm{Exp}\left(\frac{\Omega^{Q_K}(\mathbf{x},q)}{1-q^{2}}\right)\right|_{x_{i}=a^{a_{i}}Q^{q_{i}}(-t_{r})^{C_{ii}}x,\ q=t_{c}}\nonumber \\
 & =\left.P^{Q_{K}}(\mathbf{x},q)\right|_{x_{i}=a^{a_{i}}Q^{q_{i}}(-t_{r})^{C_{ii}}x,\ q=t_{c}}\\
 & =\sum_{d_{1},\ldots,d_{m}\geq0}(-t_{c})^{\sum_{1\leq i,j\leq m}C_{ij}d_{i}d_{j}}\prod_{i=1}^{m}\frac{(a^{a_{i}}Q^{q_{i}}t_{r}^{C_{ii}}x)^{d_{i}}}{(t_{c}^{2};t_{c}^{2})_{d_{i}}}\nonumber \\
 & =\mathcal{\tilde{P}}^{K}(x,a,Q,t_{r},t_{c}).\nonumber 
\end{align}
Note that the~form of the~motivic generating function induces the~$(t_{c}^{2};t_{c}^{2})_{r}$
denominator of~$\mathcal{\tilde{P}}_{r}^{K}$, which is the~source
of the~nonstandard definition in Section~\ref{sub:Knot-homologies}.

Explicit calculations of refined and doubly refined LMOV invariants for the~unknot and the~trefoil are presented in Section \ref{sub:Unknot} and \ref{sub:Trefoil} respectively.

\subsection{The geometry of refined Chern-Simons}\label{sec:refined-CS}
In this section we discuss the~geometry of refined Chern-Simons theory. Our approach is closely related to that of \cite{Aganagic:2012hs}. We will also indicate how this relates to self linking and 4-chain intersections above. The details of that depend on \cite{ES} and will not be fully discussed here.

Let $K=K_{1}\cup\dots\cup K_{n}$ be an~$n$-component link and let $L_{K}=L_{K_{1}}\cup\dots\cup L_{K_{n}}$ be its Lagrangian conormal considered as a~Lagrangian submanifold of the~resolved conifold as above. We will consider holomorphic curves with boundary on $L_{K}$. As above they are all rigid. We apply Symplectic Field Theory stretching near each component of $L_{K}$. We fix a~metric on the~components of $L_{K}$ (topology $S^{1}\times\R^2$) with exactly one closed geodesic $S^{1}\times 0$. Under such stretching, the~curves subdivide into components in the~complement $X\backslash L_{K}$ asymptotic to multiples of the~Reeb orbit which is the~lift of multiples of the~geodesic, and components on the~inside also asymptotic to the~Reeb orbits and with boundary on the~$0$-section. A straightforward index argument shows that the~count of curves in the~complement with fixed asymptotics is invariant under deformations. Fixing a~capping disk in $X\backslash L_{K}$ (which is simply connected) for each Reeb orbit we cap each holomorphic curve to a~2-cycle in $X\backslash L_{K}$. 

A straightforward calculation shows that 
\be
H_{2}(X\backslash L_{K})=\mathbb{Z}[\C \IP^{1}]\oplus \mathbb{Z}[S_{1}]\oplus\dots  \oplus \mathbb{Z}[S_{n}],
\ee
where $S_{j}$ is the~fiber 2-sphere in the~boundary of a~tubular neihgborhood of $L_{K_{j}}$. This gives a~refined Gromov-Witten potential:
\be
\Psi_{K}(a,a_{(1)},\dots,a_{(n)}),
\ee
where $a_{(j)}$ keeps track of the~homology class (it is \emph{not} $a_i$ from the KQ change of variables). 

For calculations from infinity, it is useful to glue back the~curves and consider curves with boundaries. Following \cite{ES}, we should view the~Gromov-Witten invariant as taking values in the~skein module of $L_{K}$, where $(a_{(j)},q)$ are the~variables in the~skein module of~$L_{K_{j}}$ and $q=e^{\frac12 g_{s}}$. This is a~useful perspective since it allows for using Legendrian SFT at infinity to compute the~refined Schr\"{o}dinger equation by elimination as for the~unrefined colored HOMFLY-PT in \cite{Ekholm:2018iso}. From this perspective, the refined partition function corresponds to $U(1)$ Chern-Simons theory on $L_{K}$, which means that $a_{(j)}=q$ and refinement concern a splitting of this $q$-variable into $q$ and $t$, where one $t$ corresponds to the self-linking of the boundary of the holomorphic disk and the other $q$ accounts for genus and for $4$-chain intersections. The exact nature of this splitting is not yet understood.

We end this section with a~discussion of the~distinctions between the~refinements $t_{c}$ and~$t_{r}$ above. Here the~variable $t_{c}$ is related to linking (and self-linking) between boundaries of holomorphic curves inside $L_{K}$ and is hence an~effect of M2-M2 interactions. The variable~$t_{r}$ on the~other hand counts intersections with the~$4$-chain and is thus an~effect of M2-M5 interactions. 
Looking at the~curves generated by a~given set of disks, it is clear that one transforms quadratically and the~other linearly in the~number of disks in a~boundstate (the~dimension vector~${\mathbf{d}}$).

\section{Examples\label{sec:Examples}}

In this section we present the~ideas and interpretations described
in Sections \ref{sec:physics} and \ref{sec:math} with explicit examples
of knots and corresponding quivers. Since every case demands a~long discussion,
we will focus on the~two simplest ones: the~unknot and the~trefoil.

\subsection{Unknot\label{sub:Unknot}}
According to the~KQ correspondence \cite{Kucharski:2017poe,Kucharski:2017ogk}, the~unknot quiver contains
two nodes and one loop:
\begin{equation}
\label{eq:Qunknot}
Q_{0_{1}}=\raisebox{-5ex}{\includegraphics[width=0.1\textwidth]{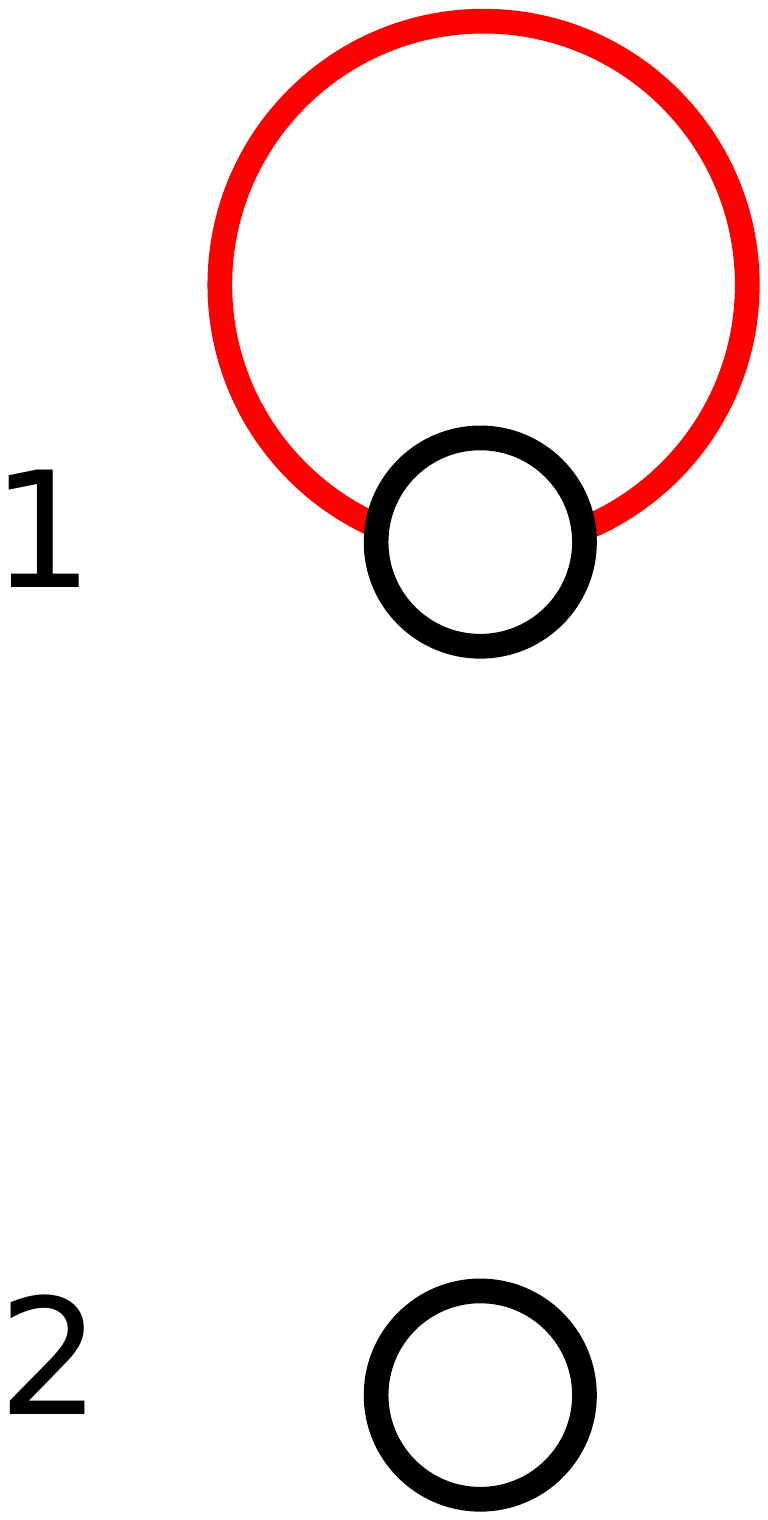}}
\qquad\Longleftrightarrow\qquad C=\left[\begin{array}{cc}
1 & 0\\
0 & 0
\end{array}\right]\,.
\end{equation}
The motivic generating series is therefore given by
\begin{equation}\label{eq:PQunknot}
P^{Q_{0_{1}}}(\mathbf{x},q)=\sum_{d_{1},d_{2}\geq0}(-q)^{d_{1}^{2}}\frac{x_{1}^{d_{1}}}{(q^{2};q^{2})_{d_{1}}}\frac{x_{2}^{d_{2}}}{(q^{2};q^{2})_{d_{2}}}\,.
\end{equation}
Note that in order to compare with formulas in \cite{Kucharski:2017poe,Kucharski:2017ogk} one has to substitute
$x_{i}\rightarrow qx_{i}$. On the~other hand, \cite{PSS1802} uses the~same
convention as we do.

Since
\begin{align}
P^{Q_{0_{1}}}(\mathbf{x},q)= & \textrm{Exp}\left(\frac{\Omega^{Q_{0_{1}}}(\mathbf{x},q)}{1-q^{2}}\right)\,,\\
\textrm{Exp}\left(\frac{-qx_{1}+x_{2}}{1-q^{2}}\right)= & \textrm{Exp}\left(\frac{\sum_{\mathbf{d},s}\Omega_{\mathbf{d},s}^{Q_{0_{1}}}\mathbf{x}^{\mathbf{d}}q^{s}(-1)^{|\mathbf{d}|+s+1}}{1-q^{2}}\right)\,,\nonumber 
\end{align}
we have only two nonzero DT invariants
\begin{equation}
\Omega_{(1,0),1}^{Q_{0_{1}}}=1\,,\qquad\Omega_{(0,1),0}^{Q_{0_{1}}}=1\,,
\end{equation}
and numerical DT invariants
\begin{equation}\label{eq:numerical-DT-unknot}
\Omega_{(1,0)}^{Q_{0_{1}}}=-1\,,\qquad\Omega_{(0,1)}^{Q_{0_{1}}}=1\,.
\end{equation}
All of them correspond to $\mathbf{|d|}=1$, so for $T[Q_{0_{1}}]$
all BPS states are basic states.

The KQ change of variables
\begin{equation}
x_{1}=a^{2}q^{-1}x,\qquad x_{2}=x\label{eq:KQ change of vars unknot}
\end{equation}
translates the~motivic generating series to the~(unreduced) HOMFLY-PT generating
series of the~unknot:
\begin{align}
\left.P^{Q_{0_{1}}}(\mathbf{x},q)\right|_{x_{1}=a^{2}q^{-1}x,\; x_{2}=x}= & P^{0_{1}}(x,a,q)\label{eq:HOMFLY gen series unknot}\\
= & \sum_{r=0}^{\infty}\frac{(a^{2};q^{2})_{r}}{(q^{2};q^{2})_{r}}x^r\,,\nonumber 
\end{align}
which confirms (\ref{eq:Qunknot}). Note that in the~knot theory literature
(including \cite{Kucharski:2017poe,Kucharski:2017ogk}) usually there is a~prefactor $a^{-r}q^{r}$ before
$\frac{(a^{2};q^{2})_{r}}{(q^{2};q^{2})_{r}}$ in $P_{r}^{0_{1}}(a,q)$
and often (e.g. in \cite{Fuji:2012nx}) $a^{1/2}$ and $q^{1/2}$ are used instead
of $a$ and $q$. Our convention is designed for the~most natural
description of knots-quivers correspondence.

We next look at this calculation from the~point of view of holomorphic disks. We first note that from the~toric picture of the~unknot conormal it is clear that there are two basic holomorphic disks \cite{Ooguri:1999bv}. Furthermore, we see from the~HOMFLY-PT polynomial that they contribute the~same to generalized holomorphic disk. The only point that remains to explain is the~difference in $t$-powers, or in other words the~difference between $C_{00}$ and $C_{11}$. 
To explain this, we look at moduli spaces of curves with one positive puncture, as in knot contact homology. At infinity there are four such disks in homology classes $1$, $e^{\eta}$, $e^{\xi}$, and $a^{2}e^{\xi}e^{\eta}$ (these are the~terms in the~operator equation (\ref{eq:A-polynomial annihilating P})). 
By the~main result of \cite{ES}, the~boundary of holomorphic disks in the~skein module does not change under deformation provided the~skein variable $(q,t)$ satisfies $q=e^{\frac12 g_{s}}$, and that 4-chain intersections contribute $t^{\pm 1}$. The deformation invariance then implies that the~boundaries of a~1-dimensional moduli space equals zero in the~skein module. This can then be used to understand relevant boundaries of holomorphic disks.

We use this argument here. First, consider the~disks with boundaries that represent the~trivial class in $H_{1}(L_{K})$. There are two such curves $1$ and $e^{\eta}$ that come with opposite signs and hence cancel in the~skein module of $L_{K}$. Consider next the~disks that goes once around the~generator and represent the~trivial homology class in $H_{2}(X)$. There are then the~disks $e^{\xi}$ at infinity and the~disks $e^{\eta}\cdot u_{0}$, see \cite{Ekholm:2018iso}. The corresponding disks for $a^{2}$ are $a^{2}e^{\eta}e^{\xi}$ at infinity and $e^{\eta}\cdot u_{1}$.
Write the~boundary of the~basic disk with minimal $a$ power $u_{0}$ in the~skein module as $p(q,t)\cdot e^{\xi}$, where we think of $e^{\xi}$ as the~standard longitude with one twist. Then, using the~skein relation (note that \cite{ES} uses the~skein relation with $q-q^{-1}$ instead of $1-q^{2}$) we have
\[ 
(1-e^{\eta})e^{\xi} = (1-q^{2})e^{\xi}
\]
and we find that $u_{0}=\frac{1}{1-q^{2}}$. The only difference in the~calculation for the~maximal disk is that $e^{\xi}$ is replaced by the~disk $e^{\xi}e^{\eta}=te^{\xi}$ and we get $u_{1}=\frac{t}{1-q^{2}}$. It follows that $C_{11}=C_{00}+1$.

\subsubsection*{Homologies and BPS states}
Generators of uncolored HOMFLY-PT homology $\mathscr{H}(0_{1})$, i.e. elements of $\mathscr{G}(0_{1})$, have degrees given by the~following vectors:
\begin{align}
\mathbf{a}= & (a_{1},a_{2})=(2,0)\,,\nonumber \\
\mathbf{q}= & (q_{1},q_{2})=(0,0)\,,\label{eq:HOMFLY-PT homology degrees}\\
\mathbf{t}= & (t_{1},t_{2})=(1,0)\,.\nonumber 
\end{align}
On the~other hand, the~application of KQ change of variables to $\mathbf{|d|}=1$ restriction of (\ref{eq:PQunknot}) leads to
\begin{equation}
\left.P_{|\mathbf{d}|=1}^{Q_{0_{1}}}(\mathbf{x},q)\right|_{x_{1}=a^{2}q^{-1}x,\ x_{2}=x}=\frac{1-a^{2}}{1-q^{2}}x=\frac{\sum_{i\in\mathscr{G}(0_{1})}a^{a_{i}}q^{q_{i}}(-1)^{t_{i}}}{1-q^{2}}x\,,
\end{equation}
which shows a~perfect consistency. Note that all comments about HOMFLY-PT conventions apply to degrees
of $\mathscr{H}(0_{1})$ as well.

We can apply KQ change of variables also to the~DT generating
function in order to obtain the~LMOV generating
function:
\begin{align}
\left.\Omega^{Q_{0_{1}}}(\mathbf{x},q)\right|_{x_{1}=a^{2}q^{-1}x,\; x_{2}=x}= & N^{0_{1}}(x,a,q)\\
x-a^{2}x= & \sum_{r,i,j}N_{r,i,j}^{0_{1}}x^{r}a^{i}q^{j}\,,\nonumber 
\end{align}
which leads to 
\begin{equation}
N_{1,0,0}^{0_{1}}=1\,,\qquad N_{1,2,0}^{0_{1}}=-1\,.
\end{equation}
Classical LMOV invariants are therefore given by
\begin{equation}\label{eq:Classical-LMOV-unknot}
b_{1,0}^{0_{1}}=1\,,\qquad b_{1,2}^{0_{1}}=-1\,.
\end{equation}
The shift in $a$ variable with respect to usual knot theory conventions
(e.g. \cite{KS1608}) comes from the~lack of $a^{-r}q^{r}$ prefactor in $P_{r}^{0_{1}}(a,q)$.
In $q$ variable this effect is compensated by $1-q^{2}$ denominator
in the~definition of the~LMOV generating function (\ref{P^K=Exp}), instead of the usual $q-q^{-1}$ present e.g. in \cite{KS1608},
so at the~end of the~day only the~sign flip remains.

We can consider also the refined KQ change of variables:
\begin{equation}
x_{1}=a^{2}q^{-1}(-t)x,\qquad x_{2}=x\label{eq:refined KQ change of vars unknot}
\end{equation}
which translates the~motivic generating series to the~generating series of
superpolynomials:
\begin{align}
\left.P^{Q_{0_{1}}}(\mathbf{x},q)\right|_{x_{1}=a^{2}q^{-1}(-t)x,\ x_{2}=x}= & \mathcal{P}^{0_{1}}(x,a,q,t)\label{eq:super gen series unknot}\\
= & \sum_{r=0}^{\infty}\frac{(-a^{2}t;q^{2})_{r}}{(q^{2};q^{2})_{r}}x^{r}\,.\nonumber 
\end{align}
Restriction to $\mathbf{|d|}=1$ shows consistency
of refined KQ change of variables with degrees of $\mathscr{H}(0_{1})$
given in (\ref{eq:HOMFLY-PT homology degrees}):
\begin{equation}
\left.P_{|\mathbf{d}|=1}^{Q_{0_{1}}}(\mathbf{x},q)\right|_{x_{1}=a^{2}q^{-1}(-t)x,\ x_{2}=x}=\frac{1+a^{2}t}{1-q^{2}}x=\frac{\sum_{i\in\mathscr{G}(0_{1})}a^{a_{i}}q^{q_{i}}t^{t_{i}}}{1-q^{2}}x\,.
\end{equation}
Applying refined KQ change of variables to the~DT generating function we obtain 
\begin{align}
\left.\Omega^{Q_{0_{1}}}(\mathbf{x},q)\right|_{x_{1}=a^{2}q^{-1}(-t)x,\ x_{2}=x}= & N^{0_{1}}(x,a,q,t)\\
x+a^{2}tx= & \sum_{r,i,j}N_{r,i,j,k}^{0_{1}}x^{r}a^{i}q^{j}t^{k}\,,\nonumber 
\end{align}
so refined LMOV invariants read
\begin{equation}
N_{1,0,0,0}^{0_{1}}=1\,,\qquad N_{1,2,0,1}^{0_{1}}=1\,.
\end{equation}
This is consistent with \cite{GKS1504} after taking into accout the~fact
that they use the~convention of~\cite{Fuji:2012nx} for the~$t_{r}$ refinement, with an~extra $(-a^{-2}q^{2}t^{-3})^{r/2}$
prefactor, and use $a^{2}t^{3}$ instead of $a^{2}t$. These prefactors were introduced
to provide more symmetric expressions -- we drop them in order to
obtain the~KQ correspondence in a~natural way. Our choice of $a^{2}t$ in
$\mathcal{P}_{r}^{0_{1}}(a,q,t)$ is consistent with conventions of  \cite{GGS1304} and
\cite{GNSSS1512} on the~$t_{r}$ refinement. 
The~$t_{c}$~refinement adopted in \cite{KN1703} introduces $(q^{2};q^{2}t^{2})_{r}$
in the~denominator, which is inconsistent with the~form of the~motivic generating
series.

We can generalize our results to both $t_{r}$ and $t_{c}$ by considering
doubly refined KQ change of variables:
\begin{equation}
x_{1}=a^{2}(-t_{r})x\,,\qquad x_{2}=x\,,\qquad q=t_{c}\,.\label{eq:doubly refined KQ change of vars unknot}
\end{equation}
It translates the~motivic generating series to the~generating series of
quadruply-graded polynomials:
\begin{align}
\left.P^{Q_{0_{1}}}(\mathbf{x},q)\right|_{x_{1}=a^{2}(-t_{r})x,\ x_{2}=x,\ q=t_{c}}= & \tilde{\mathcal{P}}^{0_{1}}(x,a,Q,t_{r},t_{c})\label{eq:quadruple gen series unknot}\\
= & \sum_{r=0}^{\infty}\frac{(-a^{2}t_{r}t_{c};t_{c}^{2})_{r}}{(t_{c}^{2};t_{c}^{2})_{r}}x^{r}\,.\nonumber 
\end{align}
Note that in \cite{GGS1304} and \cite{GNSSS1512} there is a~$a^{-r}Q^{r}$ prefactor
and $(Q^{2};t_{c}^{2})_{r}$ in the~denominator. This discrepancy is analogous
to the~one with \cite{KN1703} and is forced by demanding consistency
with the~motivic generating series.

The doubly refined KQ change of variables is consistent with degrees of
generators of the~quadruply-graded homology $\tilde{\mathscr{H}}(0_{1})$. 
We can see it by comparing (\ref{eq:HOMFLY-PT homology degrees}) to the~$\mathbf{|d|}=1$ restriction of (\ref{eq:quadruple gen series unknot}):
\begin{equation}
\left.P_{|\mathbf{d}|=1}^{Q_{0_{1}}}(\mathbf{x},q)\right|_{x_{1}=a^{2}(-t_{r})x,\ x_{2}=x,\ q=t_{c}}=\frac{1+a^{2}t_{r}t_{c}}{1-t_{c}^{2}}x=\frac{\sum_{i\in\tilde{\mathscr{G}}(0_{1})}a^{a_{i}}Q^{q_{i}}t_{r}^{t_{i}}t_{c}^{t_{i}}}{1-q^{2}}x\,.
\end{equation}
Note that for uncolored homology degrees of $t_{r}$ and $t_{c}$
are the~same, so they are both given by the~vector $\mathbf{t}$ from
(\ref{eq:HOMFLY-PT homology degrees}). All comments about quadruply-graded
polynomial conventions apply to degrees of $\tilde{\mathscr{H}}(0_{1})$
as well.

According to our reasoning from Section \ref{sub:refinement-of-LMOV},
we can define doubly refined LMOV invariants by applying (\ref{eq:doubly refined KQ change of vars unknot})
to the~DT generating function:
\begin{align}
\left.\Omega^{Q_{0_{1}}}(\mathbf{x},q)\right|_{x_{1}=a^{2}(-t_{r})x,\ x_{2}=x,\ q=t_{c}}= & N^{0_{1}}(x,a,Q,t_{r},t_{c})\\
x+a^{2}t_{r}t_{c}x= & \sum_{r,i,j,k,l}N_{r,i,j,k,l}^{0_{1}}x^{r}a^{i}Q^{j}t_{r}^{k}t_{c}^{l}\,,\nonumber 
\end{align}
which leads to
\begin{equation}
N_{1,0,0,0,0}^{0_{1}}=1\,,\qquad N_{1,2,0,1,1}^{0_{1}}=1\,.
\end{equation}

\subsubsection*{3d $\mathcal{N}=2$ theories}

As discussed in Section \ref{sec:physics}, we can use the~generating series of superpolynomials and the~motivic generating
series to obtain twisted superpotentials of $T[L_{0_{1}}]$ and  $T[Q_{0_{1}}]$ theories, whose BPS states
are counted by LMOV and DT invariants respectively. 
Before considering them, let us follow Section~\ref{Theories T[M_K] and T[L_K]} and focus on the~theory $T_0[L_{0_{1}}]$.
It is obtained from the~limit (\ref{eq:FGS}), which for the~unknot
reads
\begin{align}
 & \mathcal{P}_{r}^{0_{1}}(a,q,t)\underset{q^{2r}\rightarrow y}{\overset{\hbar\rightarrow0}{\longrightarrow}}\exp\left[\frac{1}{2\hbar}\left(\widetilde{\mathcal{W}}_{T_0[L_{0_{1}}]}(a,t,y)+O(\hbar)\right)\right],\label{eq:W knot complement}\\
 & \widetilde{\mathcal{W}}_{T_0[L_{0_{1}}]}(a,t,y)=\textrm{Li}_{2}\left(y\right)-\textrm{Li}_{2}\left(-a^{2}ty\right)+\textrm{Li}_{2}\left(-a^{2}t\right)\,.\nonumber 
\end{align}
We can compare the~above expression with the~twisted superpotential for the~knot complement theory in \cite{Fuji:2012nx}. After changing variables
$y\rightarrow x,\; a^{2}\rightarrow a,\; q^{2}\rightarrow q$, taking into account the~fact that they have a~$(-a^{-1}qt^{-3})^{r/2}$
prefactor and $-t^{3}$ instead of $-t$ in the~$q$-Pochhammer, and noting that we have 
dropped the~irrelevant $\textrm{Li}_{2}\left(1\right)$ term, we can see that they are consistent.

The moduli
space of vacua of $T_0[L_{0_{1}}]$ is described by the~zero-locus of the
super-A-polynomial. 
According to (\ref{eq:super-A-poly}), it is given by
\begin{equation}
\log x^{-1}=\frac{\partial\widetilde{\mathcal{W}}_{T_0[L_{0_{1}}]}(a,t,y)}{\partial\log y}=-\log(1-y)+\log(1+a^{2}ty)
\end{equation}
so
\begin{equation}
\mathcal{A}^{0_{1}}(x,y,a,t)=1-x-y-a^{2}txy=0.\label{eq:superA unknot}
\end{equation}
Note that according to terminology of \cite{KS1608,GKS1504}, $\mathcal{A}^{0_1}(x,y,a,t)$
is a~dual super-A-polynomial, so in order to compare with a~super-A-polynomial
from \cite{Fuji:2012nx} we have to change variables $y\rightarrow x,\; x\rightarrow y^{-1}$.
Of course we have to include other conventional differences which
were already discussed, namely add the~prefactor $(-a^{-1}t^{-3})^{1/2}$
and change the~term $a^{2}t$ to $at^{3}$. Comparison of (\ref{eq:superA unknot})
with dual A-polynomials from \cite{KS1608,GKS1504} demands only usual
restriction $t=-1$ and rescaling $x\rightarrow a^{-1}x$.

We can obtain theory $T[L_{0_{1}}]$ by considering the~generating
function of superpolynomials.
In the~case of the~unknot the~limit (\ref{eq:FGS+}) reads
\begin{align}
& \mathcal{P}^{0_{1}}(x,a,q,t)  \underset{q^{2r}\rightarrow y}{\overset{\hbar\rightarrow0}{\longrightarrow}}\int dy\exp\left[\frac{1}{2\hbar}\left(\widetilde{\mathcal{W}}_{T[L_{0_{1}}]}(a,t,x,y)+O(\hbar)\right)\right]\label{eq:W L_K}\\
& \widetilde{\mathcal{W}}_{T[L_{0_{1}}]}(a,t,x,y) =\textrm{Li}_{2}\left(y\right)-\textrm{Li}_{2}\left(-a^{2}ty\right)+\textrm{Li}_{2}\left(-a^{2}t\right)+\log x\log y\nonumber 
\end{align}
This twisted superpotential corresponds to $U(1)$ gauge theory with
one fundamental chiral and one antifundamental chiral as it was noted
in Section \ref{subsec:TL-TQ}. Comparing (\ref{eq:W L_K}) with (\ref{eq:W knot complement}),
we can see that $U(1)_{M}$ symmetry corresponding to fugacity $y$
is gauged and there is an~extra Fayet-Ilioupoulos term $\log x\log y$,
which is line with (\ref{TLK-TM3}). It also confirms that the~weak coupling limit of $T[L_{0_{1}}]$ is $T_0[L_{0_{1}}]$.

Since for the~unknot there are no gauge fuacities other than $y$ (denoted
by $z_{i}$ in Section~\ref{Theories T[M_K] and T[L_K]}), we have
\begin{align}
\widetilde{\mathcal{W}}_{T[L_{0_{1}}]}^{\textrm{eff}}(a,t,x,y) & =\widetilde{\mathcal{W}}_{T[L_{0_{1}}]}(a,t,x,y)\,,\\
\widetilde{\mathcal{W}}_{T_0[L_{0_{1}}]}^{\textrm{eff}}(a,t,y)&=\widetilde{\mathcal{W}}_{T_0[L_{0_{1}}]}(a,t,y)\,.\nonumber 
\end{align}

According to (\ref{eq:newrelation}), we can obtain the~Gromov-Witten
disk potential $W_{0_{1}}$ as a~Legendre transform of $\widetilde{\mathcal{W}}_{T_0[L_{0_{1}}]}$.
It is equivalent to integrating out $y$ in $\widetilde{\mathcal{W}}_{T[L_{0_{1}}]}$,
which can be done by saddle-point approximation. Solving 
\begin{equation}
0=\frac{\partial\widetilde{\mathcal{W}}_{T[L_{0_{1}}]}(a,t,x,y)}{\partial\log y}=-\log(1-y)+\log(1+ya^{2}t)+\log x
\end{equation}
we obtain 
\begin{equation}
y^{*}(x)=\frac{1-x}{1+a^{2}tx}\,,
\end{equation}
which is equivalent to the~super-A-polynomial (\ref{eq:superA unknot}). Therefore
\begin{align}
W_{0_{1}}(a,x)= & \left.\left(\widetilde{\mathcal{W}}_{T_0[L_{0_{1}}]}(a,t,y)+\log y\log x\right)\right|_{y^{*}(x),\ t=-1}\nonumber \\
= & \left.\widetilde{\mathcal{W}}_{T[L_{0_{1}}]}(a,t,x,y)\right|_{y^{*}(x),\ t=-1}\label{eq:W-unknot}\\
= & \textrm{Li}_{2}\left(\frac{1-x}{1-a^{2}x}\right)-\textrm{Li}_{2}\left(\frac{a^{2}-a^{2}x}{1-a^{2}x}\right)+\textrm{Li}_{2}\left(a^{2}\right)+\log x\log\left(\frac{1-x}{1-a^{2}x}\right)\nonumber\\
= & \textrm{Li}_{2}\left(a^{2}x\right)-\textrm{Li}_{2}\left(x\right)\,,\nonumber 
\end{align}
where we used the~pentagon relation for dilogarithms. As a~cross-check
let us calculate the~Gromov-Witten disk potential from classical LMOV
invariants (\ref{eq:Classical-LMOV-unknot}) using  (\ref{W_K in terms of LMOV}). We obtain
\begin{equation}
W_{0_{1}}(a,x)=\left(-b_{1,0}^{0_{1}}\right)\textrm{Li}_{2}\left(x^{1}a^{0}\right)+\left(-b_{1,2}^{0_{1}}\right)\textrm{Li}_{2}\left(x^{1}a^{2}\right)=\textrm{Li}_{2}\left(a^{2}x\right)-\textrm{Li}_{2}\left(x\right)\,,
\end{equation}
which reproduces (\ref{eq:W-unknot}). We can also see that equation
\begin{equation}
\log y=\frac{\partial W_{0_{1}}(a,x)}{\log x}=-\log(1-a^{2}x)+\log(1-x)
\end{equation}
gives the~A-polynomial being the specialization $t=-1$ of (\ref{eq:superA unknot}),
as suggested by~(\ref{eq:disk-potential-integral}).

Let us now follow Section \ref{subsec:TofQ} and focus on the~quiver side.
The limit (\ref{eq:bar-P-semiclassical-limit})
for the~unknot quiver is given by
\begin{align}
 & P^{Q_{0_{1}}}(\textbf{x},q)\underset{q^{2d_{i}}\rightarrow y_{i}}{\overset{\hbar\rightarrow0}{\longrightarrow}}\int dy_{1}dy_{2}\exp\left[\frac{1}{2\hbar}\left(\widetilde{\mathcal{W}}_{T[Q_{0_{1}}]}(\textbf{x},\textbf{y})+O(\hbar)\right)\right]\\
 & \widetilde{\mathcal{W}}_{T[Q_{0_{1}}]}(\textbf{x},\textbf{y})=\textrm{Li}_{2}\left(y_{1}\right)+\textrm{Li}_{2}\left(y_{2}\right)\\
 & \phantom{\widetilde{\mathcal{W}}_{T[Q_{0_{1}}]}(\textbf{x},\textbf{y})} +\log(- x_{1})\log y_{1}+\log x_{2}\log y_{2}+\frac{1}{2}\log y_{1}\log y_{1}\,.\nonumber 
\end{align}
We can see that $T[Q_{0_{1}}]$ is $U(1)^{(1)}\times U(1)^{(2)}$
gauge theory with one chiral field for each group and effective Chern-Simons
level one for $U(1)^{(1)}$ as it was noted in Section \ref{subsec:TL-TQ}. 

According to (\ref{eq:quiver A-poly}), the~critical point of quiver
twisted superpotential given by
\begin{align}
0=\frac{\partial\widetilde{\mathcal{W}}_{T[Q_{0_{1}}]}}{\partial\log y_{1}} & =\log(-x_{1})+\log y_{1}-\log(1-y_{1})\,,\\
0=\frac{\partial\widetilde{\mathcal{W}}_{T[Q_{0_{1}}]}}{\partial\log y_{2}} & =\log x_{2}+\log y_{2}-\log(1-y_{2})\nonumber 
\end{align}
defines the~quiver A-polynomial 
\be
A_{1}^{Q_{0_{1}}}(\textbf{x},\textbf{y})  =1-y_{1}+x_{1}y_{1}\,,\qquad\label{eq:unknot-quiver-A-poly}
A_{2}^{Q_{0_{1}}}(\textbf{x},\textbf{y})  =1-y_{2}-x_{2}\,,
\ee
which is consistent with (\ref{eq:quiver-A-ponomial-general-formula}). $A_{1}^{Q_{0_{1}}}$ and $A_{2}^{Q_{0_{1}}}$ are decoupled, which
originates from the~lack of arrows between vertices $1$ and $2$
in $Q_{0_{1}}$. 

It is also interesting to note that $\textbf{A}^{Q_{0_{1}}}(\textbf{x},\textbf{y})$
correspond to the~unknot \emph{extremal} A-polynomials from \cite{GKS1504}. The change of variables $x_{1}=x_{2}=x$, $y_{1}=y_{2}=y^{2}$
translates (\ref{eq:unknot-quiver-A-poly}) to
\be
\mathcal{A}^{+}(x,y)  =1-y^{2}+xy^{2}\,,\qquad
\mathcal{A}^{-}(x,y)  =1-y^{2}-x\,,
\ee
where $\mathcal{A}^{+}$ and $\mathcal{A}^{-}$ are maximal and minimal
A-polynomials respectively. These expressions coincides exactly with 
\cite[eq. (4.7)]{GKS1504}. We can understand $x_{1}=x_{2}=x$ as the~classical
limit ($q\rightarrow1$) of the~KQ change of variables (\ref{eq:KQ change of vars unknot})
preceded by the~ rescaling $x_{1}\rightarrow a^{-2}x_{1}$ that is characteristic
for extremal A-polynomials. The change of power in $y_{1}=y_{2}=y^{2}$ is purely a~matter of conventions.

In order to explain this connection, let us calculate $\textbf{A}^{Q_{0_{1}}}(\textbf{x},\textbf{y})$
from the~quiver disk potential
\begin{equation}
W_{Q_{0_{1}}}(\textbf{x})=\textrm{Li}_{2}\left(x_{1}\right)-\textrm{Li}_{2}\left(x_{2}\right)\label{eq:W-unknot-quiver}
\end{equation}
obtained using (\ref{eq:quiver-disk-potential-from-DT}) and numerical
DT invariants given in (\ref{eq:numerical-DT-unknot}). Indeed,
\be
\log y_{1}=\frac{\partial W_{Q_{0_{1}}}(\textbf{x})}{\partial\log x_{1}}  =-\log(1-x_{1})\,, \quad
\log y_{2}=\frac{\partial W_{Q_{0_{1}}}(\textbf{x})}{\partial\log x_{2}}  =\log(1-x_{2})
\ee
agrees with (\ref{eq:unknot-quiver-A-poly}), which provides a~nice
consistency check. Now we can see that each component of $\textbf{A}^{Q_{0_{1}}}(\textbf{x},\textbf{y})$
encodes one BPS state (represented by numerical DT invariant being
the coefficient next to $\textrm{Li}_{2}\left(x_{i}\right)$ in $W_{Q_{0_{1}}}$), which corresponds to one extremal BPS state 
(counted by extremal LMOV invariant) encoded by each extremal A-polynomial, see~\cite{GKS1504}. In that work the~relation between A-polynomials and classical LMOV invariants was discussed without using disk potentials, however product formulas like (\ref{eq:y product form}) which were considered instead are completely equivalent.

We can obtain also the~full (not extremal) A-polynomial from the~quiver A-polynomial. The solution of $\textbf{A}^{Q_{0_{1}}}(\textbf{x},\textbf{y})=0$ is given by
\be
y^*_{1}(x_1)=\frac{1}{1-x_{1}}\,, \qquad
y^*_{2}(x_2)=1-x_{2}\,.
\ee
According to (\ref{eq:y-from-y_i-formula}) we have
\be
y^{*}(x)=\left.y^*_{1}(x_1)y^*_{2}(x_2)\right|_{x_{1}=a^{2}x,\ x_{2}=x}=\frac{1-x}{1-a^{2}x}\,,
\ee
which reproduces the specialization to $t=-1$ of the~super-A-polynomial (\ref{eq:superA unknot}). As discussed in Section \ref{subsec:TofQ}, this is a~consequence of the~correspondence between knot and quiver disk potentials (\ref{eq:W_Q=W_K}), which for the~unknot reads
\begin{align}
W_{0_{1}}(a,x)  &=\left.W_{Q_{0_{1}}}(\textbf{x})\right|_{x_{1}=a^{2}x,\ x_{2}=x} \\
\textrm{Li}_{2}\left(a^{2}x\right)-\textrm{Li}_{2}\left(x\right)  &=\left.\textrm{Li}_{2}\left(x_{1}\right)-\textrm{Li}_{2}\left(x_{2}\right)\right|_{x_{1}=a^{2}x,\ x_{2}=x}\nonumber\,.
\end{align}
Here one of the~disks has zero linking an~no intersection with the~4-chain. The other has self linking one and one intersection with the~4-chain.

\subsection{Trefoil\label{sub:Trefoil}}
According to knots-quivers correspondence \cite{Kucharski:2017poe,Kucharski:2017ogk}, the~adjacency
matrix of the~(unreduced) trefoil quiver is given by
\begin{equation}
C=\left[\begin{array}{cccccc}
0 & 0 & 1 & 2 & 1 & 2\\
0 & 1 & 1 & 2 & 1 & 2\\
1 & 1 & 2 & 2 & 2 & 3\\
2 & 2 & 2 & 3 & 2 & 3\\
1 & 1 & 2 & 2 & 3 & 3\\
2 & 2 & 3 & 3 & 3 & 4
\end{array}\right]\,.\label{eq:Qtrefoil}
\end{equation}
The motivic generating series is therefore given by a~sum over six-dimensional non-negative  dimension vectors
\begin{equation}
P^{Q_{3_{1}}}(\mathbf{x},q)=\sum_{d_{1},\ldots,d_{6}\geq0}(-q)^{\sum_{1\leq i,j\leq6}C_{ij}d_{i}d_{j}}\prod_{i=1}^{6}\frac{x_{i}^{d_{i}}}{(q^{2};q^{2})_{d_{i}}}\,.\label{eq:PQtrefoil}
\end{equation}
In contrary to the~unknot quiver, the~spectrum of DT invariants for
$Q_{3_{1}}$ is infinite. Using
\begin{equation}
P^{Q_{3_{1}}}(\mathbf{x},q)=\textrm{Exp}\left(\frac{\Omega^{Q_{3_{1}}}(\mathbf{x},q)}{1-q^{2}}\right)=\textrm{Exp}\left(\frac{\sum_{n=1}^{\infty}\Omega_{|\mathbf{d}|=n}^{Q_{3_{1}}}(\mathbf{x},q)}{1-q^{2}}\right)
\end{equation}
we obtain
\begin{align}
\Omega_{|\mathbf{d}|=1}^{Q_{3_{1}}}(\mathbf{x},q)= & x_{1}-qx_{2}+q^{2}x_{3}-q^{3}x_{4}-q^{3}x_{5}+q^{4}x_{6}\,,\label{eq:DT-trefoil-quiver}\\
\Omega_{|\mathbf{d}|=2}^{Q_{3_{1}}}(\mathbf{x},q)= & -q^{2}x_{1}x_{3}+q^{3}x_{2}x_{3}-q^{4}x_{3}^{2}+q^{3}x_{1}x_{4}+q^{5}x_{1}x_{4}-q^{4}x_{2}x_{4}\nonumber \\
 &-q^{6}x_{2}x_{4}+q^{5}x_{3}x_{4} +q^{7}x_{3}x_{4}-q^{8}x_{4}^{2}+q^{3}x_{1}x_{5}-q^{4}x_{2}x_{5}\nonumber \\
 &+q^{5}x_{3}x_{5}+q^{7}x_{3}x_{5}-q^{6}x_{4}x_{5}-q^{8}x_{4}x_{5}-q^{8}x_{5}^{2}-q^{4}x_{1}x_{6}\nonumber \\
 &-q^{6}x_{1}x_{6}+q^{5}x_{2}x_{6}+q^{7}x_{2}x_{6}-q^{6}x_{3}x_{6}-q^{8}x_{3}x_{6}\nonumber \\
 &-q^{10}x_{3}x_{6}+q^{7}x_{4}x_{6}+q^{9}x_{4}x_{6}+q^{11}x_{4}x_{6}+q^{7}x_{5}x_{6}\nonumber \\
 &+q^{9}x_{5}x_{6}+q^{11}x_{5}x_{6}-q^{8}x_{6}^{2}-q^{12}x_{6}^{2}\,.\nonumber 
\end{align}

The KQ change of variables
\begin{align}
x_{1}=a^{2}q^{-2}x\,,\qquad & x_{2}=a^{4}q^{-3}x\,,\qquad x_{3}=a^{2}x\,,\label{eq:KQ change of vars trefoil}\\
x_{4}=a^{4}q^{-1}x\,,\qquad & x_{5}=a^{4}q^{-3}x\,,\qquad x_{6}=a^{6}q^{-4}x\nonumber 
\end{align}
translates the~motivic generating series to the~HOMFLY-PT generating series
of the~trefoil:
\begin{align}
\left.P^{Q_{3_{1}}}(\mathbf{x},q)\right|_{(\ref{eq:KQ change of vars trefoil})}= & P^{3_{1}}(x,a,q)\label{eq:HOMFLY-PT gen series trefoil}\\
= & \sum_{r=0}^{\infty}\frac{(a^{2};q^{2})_{r}}{(q^{2};q^{2})_{r}}\left(\frac{a^{2r}}{q^{2r}}\sum_{k=0}^{r}\left[\begin{array}{c}
r\\
k
\end{array}\right]_{q^{2}}q^{2k(r+1)}(a^{2}q^{-2};q^{2})_{k}\right)x^{r}\,,\nonumber 
\end{align}
which confirms (\ref{eq:Qtrefoil})\,. Let us explain that
\begin{equation}
\left[\begin{array}{c}
r\\
k
\end{array}\right]_{q^{2}}=\frac{(q^{2};q^{2})_{r}}{(q^{2};q^{2})_{k}(q^{2};q^{2})_{r-k}}
\end{equation}
is a~$q$-binomial and stress that we use the~unreduced normalization -- the~first factor corresponds
to the~unknot, the~second (inside the~bracket) is the~reduced HOMFLY-PT polynomial for the~trefoil found
in \cite{Fuji:2012nx,FGSS1209}. Note that in \cite{Fuji:2012nx} $a^{1/2}$ and $q^{1/2}$ are used instead of $a$~and~$q$. Conventions for the~unknot were discussed in Section \ref{sub:Unknot}.

\subsubsection*{Homologies and BPS states}

Let us compare the~KQ change of variables with uncolored HOMFLY-PT homology. We use the~finite-dimensional unreduced homology, basing on the~idea from
\cite{GNSSS1512} and reduced HOMFLY-PT homology for the~trefoil
given in \cite{DGR0505}.
Degrees of generators of $\mathscr{H}(3_{1})$ are
given by vectors
\begin{align}
\mathbf{a} & = (a_{1},\ldots,a_{6})=(2,4,2,4,4,6)\,,\nonumber \\
\mathbf{q} & =(q_{1},\ldots,q_{6})=(-2,-2,2,2,0,0)\,,\label{eq:HOMFLY-PT homology degrees trefoil}\\
\mathbf{t} & =(t_{1},\ldots,t_{6})=(0,1,2,3,3,4)\,.\nonumber 
\end{align}
On the~other hand, reducing (\ref{eq:HOMFLY-PT gen series trefoil}) to $\mathbf{|d|}=1$
we obtain
\begin{align}
\left.P_{|\mathbf{d}|=1}^{Q_{3_{1}}}(\mathbf{x},q)\right|_{(\ref{eq:KQ change of vars trefoil})} & =P_{1}^{3_{1}}(a,q)x\nonumber \\
 & =\frac{1-a^{2}}{1-q^{2}}\left(a^{2}q^{-2}+a^{2}q^{2}-a^{4}\right)x\\
 & =\frac{\sum_{i\in\mathscr{G}(3_{1})}a^{a_{i}}q^{q_{i}}(-1)^{t_{i}}}{1-q^{2}}x\,,\nonumber 
\end{align}
which shows that (\ref{eq:KQ change of vars trefoil}) and (\ref{eq:HOMFLY-PT homology degrees trefoil}) are consistent.

The~KQ change of variables can be applied also to the~generating function
of motivic DT invariants leading to the~LMOV generating function
\begin{equation}
\left.\sum_{n=1}^{\infty}\Omega_{|\mathbf{d}|=n}^{Q_{3_{1}}}(\mathbf{x},q)\right|_{(\ref{eq:KQ change of vars trefoil})}=\left.\Omega^{Q_{3_{1}}}(\mathbf{x},q)\right|_{(\ref{eq:KQ change of vars trefoil})}=N^{3_{1}}(x,a,q)=\sum_{r=1}^{\infty}N_{r}^{3_{1}}(a,q)x^{r}\,.
\end{equation}
For $r=1,2$ it gives 
\begin{align}
N_{1}^{3_{1}}(a,q)= & a^{2}q^{-2}-a^{4}q^{-2}+a^{2}q^{2}-a^{4}q^{2}-a^{4}+a^{6}\,,\label{eq:LMOV-trefoil}\\
N_{2}^{3_{1}}(a,q)= & -a^{4}+2a^{6}-2a^{8}+2a^{10}-a^{12}+a^{6}q^{-2}-2a^{8}q^{-2}\nonumber \\
 &+a^{10}q^{-2}+2a^{6}q^{2}-4a^{8}q^{2} +2a^{10}q^{2}-a^{4}q^{4}+2a^{6}q^{4}\nonumber \\
 &-2a^{8}q^{4}+2a^{10}q^{4}-a^{12}q^{4}+a^{6}q^{6}-2a^{8}q^{6}+a^{10}q^{6}\,.\nonumber 
\end{align}

We can repeat our analysis with the~refined KQ change of variables
\begin{align}
x_{1}=a^{2}q^{-2}x\,,\qquad & x_{2}=-a^{4}q^{-3}tx,\qquad x_{3}=a^{2}t^{2}x,\label{eq:refined KQ change of vars trefoil}\\
x_{4}=-a^{4}q^{-1}t^{3}x,\qquad & x_{5}=-a^{4}q^{-3}t^{3}x,\quad\;\; x_{6}=a^{6}q^{-4}t^{4}x\,,\nonumber 
\end{align}
which transforms the~motivic generating series to the~generating
series of superpolynomials found in \cite{Fuji:2012nx,FGSS1209}
\begin{align}
\left.P^{Q_{3_{1}}}(\mathbf{x},q)\right|_{(\ref{eq:refined KQ change of vars trefoil})}= & \mathcal{P}^{3_{1}}(x,a,q,t)\label{eq:super gen series trefoil}\\
= & \sum_{r=0}^{\infty}\frac{(a^{2};q^{2})_{r}}{(q^{2};q^{2})_{r}}\left(\frac{a^{2r}}{q^{2r}}\sum_{k=0}^{r}\left[\begin{array}{c}
r\\
k
\end{array}\right]_{q^{2}}q^{2k(r+1)}t^{2k}(-a^{2}q^{-2}t;q^{2})_{k}\right)x^{r}\,.\nonumber 
\end{align}
After restricting this equation to $\mathbf{|d|}=1$, we can see the
consistency of refined KQ change of variables with degrees of $\mathscr{H}(3_{1})$
given in (\ref{eq:HOMFLY-PT homology degrees trefoil}):
\begin{align}
\left.P_{|\mathbf{d}|=1}^{Q_{3_{1}}}(\mathbf{x},q)\right|_{(\ref{eq:refined KQ change of vars trefoil})} & =\mathcal{P}_{1}^{3_{1}}(a,q,t)x\nonumber \\
 & =\frac{1+a^{2}t}{1-q^{2}}\left(a^{2}q^{-2}+a^{2}q^{2}t^{2}+a^{4}t^{3}\right)x\\
 & =\frac{\sum_{i\in\mathscr{G}(3_{1})}a^{a_{i}}q^{q_{i}}t^{t_{i}}}{1-q^{2}}x\,.\nonumber 
\end{align}

Let us apply refined KQ change of variables to the~DT generating function
\begin{equation}
\left.\Omega^{Q_{3_{1}}}(\mathbf{x},q)\right|_{(\ref{eq:refined KQ change of vars trefoil})}=N^{3_{1}}(x,a,q,t)\,.
\end{equation}
Restricting our attention to the~simplest cases of $r=1,2$, we can
see that refined LMOV invariants are given by 
\begin{align}
N_{1}^{3_{1}}(a,q,t)= & a^{2}q^{-2}+a^{4}q^{-2}t+a^{2}q^{2}t^{2}+a^{4}t^{3}+a^{4}q^{2}t^{3}+a^{6}t^{4}\,,\\
N_{2}^{3_{1}}(a,q,t)= & -a^{4}t^{2}-2a^{6}t^{3}-a^{6}q^{-2}t^{3}-a^{6}q^{2}t^{3}-2a^{8}t^{4}-2a^{8}q^{-2}t^{4}\nonumber \\
 &-a^{8}q^{2}t^{4}-a^{4}q^{4}t^{4} -a^{10}t^{5}-a^{10}q^{-2}t^{5}-a^{6}q^{2}t^{5}-2a^{6}q^{4}t^{5}\nonumber \\
 &-a^{6}q^{6}t^{5}-3a^{8}q^{2}t^{6}-2a^{8}q^{4}t^{6} -2a^{8}q^{6}t^{6}-a^{10}t^{7}-2a^{10}q^{2}t^{7}\nonumber \\
 &-2a^{10}q^{4}t^{7}-a^{10}q^{6}t^{7}-a^{12}t^{8}-a^{12}q^{4}t^{8}\,.\nonumber 
\end{align}
After taking into accout differences in conventions, we can check that
this result is consistent with \cite{GKS1504}. Details of calculation can
be found in  Appendix~\ref{sub:Comparison-of-t-refinement}. This
cross-check confirms that quiver-based refinement of LMOV invariants
gives the~same results (up to conventions) as the~usual one, however
the former is much simpler and more natural. Moreover, according to
expectations $N_{r,i,j,k}$ for fixed $r$ carry the~same sign. Usually they are expected to be positive, but this can absorbed by
the change of convention.

We can generalize our results to four gradings by considering the~doubly
refined KQ change of variables
\begin{align}
x_{1}=a^{2}Q^{-2}x\,,\qquad x_{2} & =-a^{4}Q^{-2}t_{r}x\,,\qquad x_{3}=a^{2}Q^{2}t{}_{r}^{2}x\,,\label{eq:doubly refined KQ change of vars trefoil}\\
x_{4}=-a^{4}Q^{2}t{}_{r}^{3}x\,,\qquad x_{5} & =-a^{4}t_{r}^{3}x\,,\qquad\quad\;\;\,\, x_{6}=a^{6}t_{r}^{4}x\,,\nonumber \\
q & =t_{c}\,.\nonumber 
\end{align}
It translates the~motivic generating series to the~generating series of
quadruply-graded polynomials
\begin{equation}
\left.P^{Q_{3_{1}}}(\mathbf{x},q)\right|_{(\ref{eq:doubly refined KQ change of vars trefoil})}=\tilde{\mathcal{P}}^{3_{1}}(x,a,Q,t_{r},t_{c})\,.\label{eq:quadruple gen series trefoil}
\end{equation}
Restriction to $\mathbf{|d|}=1$
gives
\begin{align}
\left.P_{|\mathbf{d}|=1}^{Q_{3_{1}}}(\mathbf{x},q)\right|_{(\ref{eq:doubly refined KQ change of vars trefoil})} & =\tilde{\mathcal{P}}_{1}^{3_{1}}(a,Q,t_{r},t_{c})x\nonumber \\
 & =\frac{1+a^{2}t_{r}t_{c}}{1-t_{c}^{2}}\left(a^{2}Q^{-2}+a^{2}Q^{2}t_{r}^{2}t_{c}^{2}+a^{4}t_{r}^{3}t_{c}^{3}\right)x\\
 & =\frac{\sum_{i\in\tilde{\mathscr{G}}(3_{1})}a^{a_{i}}Q^{q_{i}}t_{r}^{t_{i}}t_{c}^{t_{i}}}{1-q^{2}}x\,,\nonumber 
\end{align}
which is consistent with $\tilde{\mathscr{H}}(3_{1})$ given in \cite{GGS1304}. Differences come solely from the~other conventions for the~unknot -- they were explained in Section \ref{sub:Unknot}.

According to our reasoning from Section \ref{sub:refinement-of-LMOV},
we can define doubly refined LMOV invariants by applying change of
variables (\ref{eq:doubly refined KQ change of vars trefoil}) to
the~DT generating function:
\begin{equation}
\left.\Omega^{Q_{3_{1}}}(\mathbf{x},q)\right|_{(\ref{eq:doubly refined KQ change of vars trefoil})}=N^{3_{1}}(x,a,Q,t_{r},t_{c})\,.
\end{equation}
For the~simplest cases of $r=1,2$ it gives
\begin{align}
N_{1}^{3_{1}}(a,Q,t_{r},t_{c})= & a^{2}Q^{-2}+a^{4}Q^{-2}t_{r}t_{c}+a^{2}Q^{2}t_{r}^{2}t_{c}^{2}+a^{4}t_{r}^{3}t_{c}^{3}\nonumber \\
 &+a^{4}Q^{2}t_{r}^{3}t_{c}^{3}+a^{6}t_{r}^{4}t_{c}^{4},\\
N_{2}^{3_{1}}(a,Q,t_{r},t_{c})= & -a^{4}t_{r}^{2}t_{c}^{2}-2a^{6}t_{r}^{3}t_{c}^{3}-a^{6}Q^{-2}t_{r}^{3}t_{c}^{3}-a^{6}t_{r}^{3}t_{c}^{5}-a^{8}t_{r}^{4}t_{c}^{4}\nonumber \\
 &-a^{8}Q^{-2}t_{r}^{4}t_{c}^{6} -2a^{8}Q^{-2}t_{r}^{4}t_{c}^{4}-a^{8}t_{r}^{4}t_{c}^{6}-a^{4}Q^{4}t_{r}^{4}t_{c}^{4}\nonumber \\
 &-a^{10}Q^{-2}t_{r}^{5}t_{c}^{7}-a^{10}Q^{-2}t_{r}^{5}t_{c}^{5} -a^{6}Q^{2}t_{r}^{5}t_{c}^{5}-a^{6}Q^{4}t_{r}^{5}t_{c}^{5}\nonumber \\
 &-a^{6}Q^{2}t_{r}^{5}t_{c}^{7}-a^{6}Q^{4}t_{r}^{5}t_{c}^{7}-2a^{8}Q^{2}t_{r}^{6}t_{c}^{6}-a^{8}t_{r}^{6}t_{c}^{8}\nonumber \\
 &-2a^{8}Q^{2}t_{r}^{6}t_{c}^{8}-a^{8}Q^{4}t_{r}^{6}t_{c}^{8}-a^{8}Q^{2}t_{r}^{6}t_{c}^{10}-a^{10}t_{r}^{7}t_{c}^{7}\nonumber \\
 &-a^{10}Q^{2}t_{r}^{7}t_{c}^{7} -a^{10}t_{r}^{7}t_{c}^{9}-a^{10}Q^{2}t_{r}^{7}t_{c}^{9}-a^{10}t_{r}^{7}t_{c}^{11}\nonumber \\
 &-a^{10}Q^{2}t_{r}^{7}t_{c}^{11}-a^{12}t_{r}^{8}t_{c}^{8}-a^{12}t_{r}^{8}t_{c}^{12}\,.\nonumber 
\end{align}

\subsubsection*{3d $\mathcal{N}=2$ theories}

Let us focus on theories whose BPS states are counted by DT and LMOV
invariants, namely $T[Q_{3_{1}}]$ and $T[L_{3_{1}}]$. However, as in the~unknot case, let us follow Section
\ref{sec:physics} and discuss $T_{0}[L_{3_{1}}]$ theory first. It is obtained from the~limit
(\ref{eq:FGS}), which for the~trefoil reads
\begin{align}
\mathcal{P}_{r}^{3_{1}}(a,q,t) & \underset{q^{2r}\rightarrow y,\, q^{2k}\rightarrow z}{\overset{\hbar\rightarrow0}{\longrightarrow}}\int dz\exp\left[\frac{1}{2\hbar}\left(\widetilde{\mathcal{W}}_{T_{0}[L_{3_{1}}]}(a,t,y,z)+O(\hbar)\right)\right]\nonumber\\
\widetilde{\mathcal{W}}_{T_{0}[L_{3_{1}}]}(a,t,y,z)& =  -\textrm{Li}_{2}\left(-a^{2}ty\right)+\textrm{Li}_{2}\left(yz^{-1}\right) +\textrm{Li}_{2}\left(z\right)+2\textrm{Li}_{2}\left(-a^{2}t\right) \\
 &\phantom{=}\; -\textrm{Li}_{2}\left(-a^{2}tz\right)+\log a^{2}\log y+\log y\log z+\log(-t)^{2}\log z\,,\nonumber 
\end{align}
and is consistent with \cite{Fuji:2012nx}. In order to see it, we have to change variables
$y\rightarrow x,\; a^{2}\rightarrow a,\; q^{2}\rightarrow q$
and take into account the~fact that they use reduced normalization, have $(-a^{-1}qt^{-3})^{r/2}$ prefactor, and use $-t^{3}$ instead
of $-t$ in the~$q$-Pochhammer. We also dropped irrelevant $\textrm{Li}_{2}\left(1\right)$ factors.

In contrast to the~case of the~unknot, 
we now have the~gauge fugacity $z$ which we integrate out using the~saddle-point
approximation:
\begin{equation}
\frac{\partial\widetilde{\mathcal{W}}_{T_{0}[L_{3_{1}}]}(a,t,y,z)}{\partial z}=0\,.\label{eq:saddle point approx trefoil}
\end{equation}
In order to avoid tedious computations, let us not calculate the~critical
point $z^{*}$ and $\widetilde{\mathcal{W}}^{\textrm{eff}}_{T_{0}[L_{3_{1}}]}(a,t,y)$
explicitly, but combine (\ref{eq:saddle point approx trefoil}) with
\begin{equation}
\frac{\partial\widetilde{\mathcal{W}}_{T_{0}[L_{3_{1}}]}(a,t,y,z)}{\partial\log y}=\log x^{-1}
\end{equation}
and eliminate $z$ to obtain the~super-A-polynomial. The exponent of
this system of equations reads
\begin{align}
\frac{t^{2}y(y-z)(1+a^{2}tz)}{z(z-1)}= & 1,\\
-\frac{a^{2}(1+a^{2}ty)z^{2}}{y-z}= & x^{-1}\,,\nonumber 
\end{align}
so
\begin{align}
\mathcal{A}^{3_{1}}(x,y,a,t)= & -1+a^{2}x+y-a^{2}t^{3}y+a^{4}txy-a^{2}t^{2}xy+a^{2}t^{3}y^{2}\\
 & +2a^{2}t^{2}xy^{2}+a^{4}t^{3}xy^{2}+2a^{4}t^{3}xy^{3}+2a^{6}t^{4}xy^{3}+a^{4}t^{5}xy^{3}\nonumber \\
 & +a^{4}t^{4}x^{2}y^{3}+2a^{6}t^{6}xy^{4}+2a^{6}t^{5}x^{2}y^{4}+a^{8}t^{7}xy^{5}+a^{8}t^{6}x^{2}y^{5}\,,\nonumber 
\end{align}
which is consistent with \cite{GKS1504}. Restriction to $t=-1$, substitution $x\rightarrow a^{-1}x,\;y\rightarrow y^2$, and factorization gives
eq. (4.44) from that paper.

Let us move to the~theory $T[L_{3_{1}}]$. It can be obtain from the
limit (\ref{eq:FGS+}), which for the~trefoil reads
\begin{align}
\mathcal{P}^{3_{1}}(x,a,q,t) & \underset{q^{2r}\rightarrow y,\, q^{2k}\rightarrow z}{\overset{\hbar\rightarrow0}{\longrightarrow}}\int dydz\exp\left[\frac{1}{2\hbar}\left(\widetilde{\mathcal{W}}_{T[L_{3_{1}}]}(a,t,x,y,z)+O(\hbar)\right)\right]\nonumber\\
\widetilde{\mathcal{W}}_{T[L_{3_{1}}]}(a,t,x,y,z)&=  -\textrm{Li}_{2}\left(-a^{2}ty\right)+\textrm{Li}_{2}\left(yz^{-1}\right) +\textrm{Li}_{2}\left(z\right) \label{eq:W L_3_1}\\
 &\quad +2\textrm{Li}_{2}\left(-a^{2}t\right)-\textrm{Li}_{2}\left(-a^{2}tz\right)+\log a^{2}\log y\nonumber \\
& \quad+\log y\log z+\log(-t)^{2}\log z+\log x\log y\,. \nonumber
\end{align}
This twisted superpotential corresponds to $U(1)_{M}\times U(1)_{z}$ gauge theory
with background symmetry $U(1)_L$ and global symmetries $U(1)_Q \times U(1)_F$. We have six fundamental chirals and their charges can be read from powers of the~respective fugacities as described in Section \ref{Theories T[M_K] and T[L_K]}.

Let us now move to the~quiver side and analyze $T[Q_{3_{1}}]$ theory. The limit (\ref{eq:bar-P-semiclassical-limit}) for the~trefoil quiver
is given by
\begin{align}
 & P^{Q_{3_{1}}}(\mathbf{x},q)\underset{q^{2d_{i}}\rightarrow y_{i}}{\overset{\hbar\rightarrow0}{\longrightarrow}}\int dy_{1}\ldots dy_{6}\exp\left[\frac{1}{2\hbar}\left(\widetilde{\mathcal{W}}_{T[Q_{3_{1}}]}(\mathbf{x},\mathbf{y})+O(\hbar)\right)\right]\label{eq:T[Q] trefoil}\\
 & \widetilde{\mathcal{W}}_{T[Q_{3_{1}}]}(\mathbf{x},\mathbf{y})=\sum_{i=1}^{6}\left[\textrm{Li}_{2}\left(y_{i}\right)+\log\((-1)^{C_{ii}} x_{i}\)\log y_{i}\right]+\sum_{i,j=1}^{6}\frac{C_{ij}}{2}\log y_{i}\log y_{j}\,.\nonumber 
\end{align}
We can see that $T[Q_{3_{1}}]$ is $U(1)^{(1)}\times\ldots\times U(1)^{(6)}$
gauge theory with one chiral field for each group and effective Chern-Simons
level given by the~trefoil quiver adjacency matrix, which can be found in (\ref{eq:Qtrefoil}). 

Applying this $C_{ij}$ to (\ref{eq:quiver-A-ponomial-general-formula}), we find that
\begin{align*}
A_{1}^{Q_{3_{1}}}(\mathbf{x},\mathbf{y}) & =1-y_{1}-x_{1}y_{3}y_{4}^{2}y_{5}y_{6}^{2}  &
A_{2}^{Q_{3_{1}}}(\mathbf{x},\mathbf{y}) & =1-y_{2}+x_{2}y_{2}y_{3}y_{4}^{2}y_{5}y_{6}^{2} \nonumber\\
A_{3}^{Q_{3_{1}}}(\mathbf{x},\mathbf{y}) & =1-y_{3}-x_{3}y_{1}y_{2}y_{3}^{2}y_{4}^{2}y_{5}^{2}y_{6}^{3} &
A_{4}^{Q_{3_{1}}}(\mathbf{x},\mathbf{y}) & =1-y_{4}+x_{4}y_{1}^{2}y_{2}^{2}y_{3}^{2}y_{4}^{3}y_{5}^{2}y_{6}^{3} \\
A_{5}^{Q_{3_{1}}}(\mathbf{x},\mathbf{y}) & =1-y_{5}+x_{5}y_{1}y_{2}y_{3}^{2}y_{4}^{2}y_{5}^{3}y_{6}^{3} &
A_{6}^{Q_{3_{1}}}(\mathbf{x},\mathbf{y}) & =1-y_{6}-x_{6}y_{1}^{2}y_{2}^{2}y_{3}^{3}y_{4}^{3}y_{5}^{3}y_{6}^{4}\,. \nonumber
\end{align*}
In contrary to the~unknot case, we have arrows connecting different vertices and equations for different $A_{i}^{Q_{3_{1}}}(\mathbf{x},\mathbf{y})$ are coupled.

Also the~quiver disk potential is much more complicated than in the
unknot quiver case. For the~trefoil quiver it is infinite (as the~BPS spectrum) and below
we write the~first terms corresponding to numerical DT invariants
for $|\mathbf{d}|=1$, which can be read from (\ref{eq:DT-trefoil-quiver}):
\begin{equation}\label{eq:W_Q-trefoil}
W_{Q_{3_{1}}}(\mathbf{x})=-\textrm{Li}_{2}\left(x_{1}\right)+\textrm{Li}_{2}\left(x_{2}\right)-\textrm{Li}_{2}\left(x_{3}\right)+\textrm{Li}_{2}\left(x_{4}\right)+\textrm{Li}_{2}\left(x_{5}\right)-\textrm{Li}_{2}\left(x_{6}\right)+\ldots
\end{equation}

Basing on (\ref{eq:W_Q=W_K}), we can find the~first terms (corresponding to $r=1$) of the~Gromov-Witten disk potential 
\[
W_{3_{1}}(a,x)=\left.W_{Q_{3_{1}}}(\mathbf{x})\right|_{(\ref{eq:KQ change of vars trefoil}),\,q\rightarrow 1}=-2\textrm{Li}_{2}\left(a^{2}x\right)+3\textrm{Li}_{2}\left(a^{4}x\right)-\textrm{Li}_{2}\left(a^{6}x\right)+\ldots
\]
We can do a~cross-check and obtain the~same result by extracting classical LMOV invariants from the~$q\rightarrow 1$ limit in (\ref{eq:LMOV-trefoil}) and putting them into equation (\ref{W_K in terms of LMOV}).

\section{Discussion and open questions}\label{sec:discussion}

We conclude with a~few remarks on our main results: on limits of the~KQ correspondence,  potential directions related to BPS counting in 4d $\CN=2$ theories, and to the~role of the~four-chain in physics.

\subsection{The knot $9_{42}$ and the~KQ correspondence\label{sub:9_42}}

The knots-quivers correspondence in the~original form proposed
in \cite{Kucharski:2017poe,Kucharski:2017ogk} is not valid for the
knot $9_{42}$. In order to see that, we analyze uncolored and $S^{2}$
HOMFLY-PT homology. 

Reduced uncolored HOMFLY-PT homology of $9_{42}$ was given in \cite{DGR0505}
and has 9~generators, so the~corresponding unreduced quiver would have
18 vertices (doubling comes from the~unknot factor). Assuming that we know $C_{ij},\: a_{i},\: q_{i}$, we would
write 
\begin{align}
\left.P_{|\mathbf{d}|=1}^{Q_{9_{42}}}(\mathbf{x},q)\right|_{x_{i}=a^{a_{i}}q^{q_{i}-C_{ii}}(-t)^{C_{ii}}x} & =\frac{\sum_{i\in\mathscr{G}(9_{42})}a^{a_{i}}q^{q_{i}}t^{t_{i}}}{1-q^{2}}x\nonumber \\
\left.\sum_{i=1}^{18}\frac{(-q)^{C_{ii}}x_{i}}{1-q^{2}}\right|_{x_{i}=a^{a_{i}}q^{q_{i}-C_{ii}}(-t)^{C_{ii}}x} & =\mathcal{P}_{1}^{9_{42}}(a,q,t)x,
\end{align}
where 
\[
\begin{split}
\mathcal{P}_{1}^{9_{42}}(a,q,t)=\frac{1+a^{2}t}{1-q^{2}} &\left(  a^{-2}q^{-2}t^{-2}+a^{-2}q^{2}+q^{-4}t^{-1}+1 \right.\\
&\ \ \left.+2t+q^{4}t^{3}+a^{2}q^{-2}t^{2}+a^{2}q^{2}t^{4}\right).
\end{split}
\]

For $S^{2}$ we would have
\begin{equation}
\left.P_{|\mathbf{d}|=2}^{Q_{9_{42}}}(\mathbf{x},q)\right|_{x_{i}=a^{a_{i}}q^{q_{i}-C_{ii}}(-t)^{C_{ii}}x}=\frac{\sum_{i\in\mathscr{G}^{S^{2}}(9_{42})}a^{a_{i}}q^{q_{i}}t^{t_{i}}}{\left(1-q^{2}\right)\left(1-q^{4}\right)}x=\mathcal{P}_{2}^{9_{42}}(a,q,t)x.\label{eq:KQ for S^2  9_42}
\end{equation}
Let us compare the~number of terms on both sides. On the~quiver side
we have 
\begin{align}
P_{|\mathbf{d}|=2}^{Q_{9_{42}}}(\mathbf{x},q) & =\sum_{d_{i}=2}\frac{(-q)^{C_{ii}d_{i}^{2}}x_{i}^{d_{i}}}{(q^{2};q^{2})_{d_{i}}}+\sum_{d_{i},d_{j}=1}\frac{(-q)^{(C_{ij}+C_{ji})d_{i}d_{j}}x_{i}^{d_{i}}x_{j}^{d_{j}}}{(q^{2};q^{2})_{d_{i}}(q^{2};q^{2})_{d_{j}}}\\
 & =\frac{1}{\left(1-q^{2}\right)\left(1-q^{4}\right)}\left[\sum_{i=1}^{18}q^{4C_{ii}}x_{i}^{2}+\sum_{1\leq i<j\leq18}q^{2C_{ij}}x_{i}x_{j}\left(1+q^{2}\right)\right],\nonumber 
\end{align}
so there are 18 terms coming from $d_{i}=2$ and $18\times17$ corresponding to $d_{i},d_{j}=1$. In total it gives $18^{2}=324$
terms, in line with the~exponential growth property. On the~other hand, reduced $S^{2}$ HOMFLY-PT homology of $9_{42}$
is known to have $401$ terms \cite{GGS1304}, so including the~unknot
factor the~number of generators of $\mathscr{H}^{S^{2}}(9_{42})$
is equal to $4\times401=1604$. Since we are on the~refined level,
 there is no chance for cancellations and (\ref{eq:KQ for S^2  9_42})
cannot be true. 

We expect that for $9_{42}$ and other knots that do not satisfy the~exponential
growth property the~KQ correspondence is still valid, but with more nodes and modified
change of variables
\begin{equation}
x_{i}=x^{n_{i}}a^{a_{i}}q^{q_{i}-C_{ii}}\,.
\end{equation}
The exponent $n_{i}$ is the~number of windings around the~generator
of $H_{1}(L_{K})$, that can now take values other than $1$.
For example, we expect to have $1280$ additional vertices of $Q_{9_{42}}$ corresponding
to disks that wind around $L_{K}$ twice. Of course, the~inclusion of nodes with $n_{i}=3,4,5,\ldots$ may also be neccesary. However, we expect
that this procedure (and therefore the~quiver) is finite.
We give two arguments in this direction. First, the~exponential growth property
holds asymptotically for all knots \cite{GGS1304}. In other words,
it breaks only for the~finite $r$ and this is exactly the~region
where we need extra vertices. Second,
there is a~finite recursion relation for $P_{r}^{K}(a,q)$ for every
knot, see \cite{GLL1604}. This means that we can encode the~information
about all symmetric representations in the~finite quantum A-polynomial
annihilating $P_{r}^{K}(a,q)$, which indicate that the~data of the~quiver should be 
finite as well.

\subsection{Relation to 4d $\CN=2$ quivers}\label{sec:class-S}

Quivers are familiar tools for describing BPS spectra in 4d $\CN=2$ theories \cite{Alim:2011kw,Alim:2011ae}. In fact this setup is not too far from the~theories considered in this paper, and some of the~connections provide a~useful (albeit heuristic) validation of some of our results.

Let us briefly recall the~physical data encoded by quivers from the~viewpoint of quantum field theory.
For this purpose we consider a~low-energy $U(1)$ gauge theory, featuring a~spectrum of BPS particles carrying both electric and magnetic charges.
The BPS quiver~$Q$ that describes the~interactions of an~electric BPS particle of charge $(n,0)$ and a~magnetic BPS particle of charge $(0,m)$ is known as the~$k$-Kronecker quiver. 
It is a~quiver made of two nodes, with $k = m n$ oriented arrows connecting them.
$k$ arises as the~quantum number of the~angular momentum of the~electromagnetic field sourced by the~two particles, which is integer with proper normalization. 
The BPS boundstates with $d_1$ particles of the~first type and $d_2$ particles of the~second type correspond to zero-modes of a~1d $\CN=4$ quantum mechanics encoded by $Q$, featuring $U(d_1)\times U(d_2)$ gauge group and $k$ bifundamental chirals.
Despite the~simplicity of $Q$, the spectrum of boundstates has a~rich structure \cite{2009arXiv0909.5153G,Galakhov:2013oja}. 

A useful way to think about the~low energy theory is to consider an~M5-brane wrapping a~Riemann surface $\Sigma$ known as the~Seiberg-Witten curve \cite{Seiberg:1994aj,Witten:1997sc}. The gauge theory $T[\Sigma]$ then lives on the~four transverse directions on the~fivebrane worldvolume. 
Charges of BPS particles are classified by oriented cycles in $\gamma\in H_1(\Sigma)$ (we can choose $\Sigma$ to be a~two-torus in our example), and $k$ is their intersection number. The sign of $k$ determines the~orientation of arrows in $Q$.
The relation between BPS states and cycles can be understood geometrically, since BPS states arise from M2-branes wrapping disks ending on $\Sigma$, and stretching in time.
Nodes of the~quiver correspond to the~``basic'' BPS states, and arise from M2-branes wrapping simple disks, whereas boundstates arise from M2-branes (possibly) more complicated curves.
By charge conservation, the~boundary of a~boundstate curve must be (as an~element of $H_1(\Sigma)$) the~sum of of the~constituent charges. In fact the~curves corresponding to BPS boundstates can be studied quite explicitly in this case using spectral networks \cite{Gaiotto:2009hg,Gaiotto:2012rg}, and they can be seen to arise by connecting basic disks at their intersections. This is reminiscent of the~notion of generalized holomorphic curves we introduced. 
A precise relation between basic disks and quivers in this setting was proposed recently in~\cite{Gabella:2017hpz}.

At this point the~analogies between BPS states of 4d $\CN=2$ theories and our BPS vortices are already rather suggestive, but we can draw ties even closer. 
Suppose that 4d spacetime is $\IR^3\times \IR_\geq 0$, where at the~3d boundary we replace $\Sigma$ with a~three-manifold $M$ with $\partial M = \Sigma$.
The boundary degrees of freedom are described by a~3d $\CN=2$ theory $T[M]$, whose content is (loosely speaking) a~reduction of the~four-dimensional theory. Such theories have been studied for example in  \cite{Terashima:2011qi,Dimofte:2011ju}.
We can push the~two BPS particles to the~boundary, where they look like BPS vortices for $T[M]$. 
When we push these particles to the~boundary, the~cycles $\gamma$ cease to live on $\Sigma$ and begin to move into $M$.
The skew-symmetric intersection pairing of two cycles (disk boundaries) on $\Sigma$ then gets replaced by the~linking number of two cycles in $M$ (this lift is generally non-canonical, it depends on how we transport particles to the~boundary theory).
Recall that intersection of cycles on $\Sigma$ determines arrows of $Q$ between two basic discs.
When pushing cycles into $M$, the~interactions between the~M2-branes are expected to change, leading to a~new quiver description with arrows now determined by the~linking number, which is naturally symmetric.
It would be interesting to check this prediction by studying the~worldvolume dynamics of the~4d BPS states as they move into the~3d boundary, along the~lines of \cite{Denef:2002ru}. 
This may provide another perspective on how \emph{symmetric} quivers arise in the~context of 3d $\CN=2$ theories associated to knots.

\subsection{Physics of the~four-chain\label{sub:4-chain-physics}}
The definition of self-linking number, and therefore of physical charges of holomorphic disks, involves a~certain four-chain $C$ whose boundary is twice $L_K$. 
To the~best of our knowledge, this object has not been encountered previously in the~physics literature on knot theory and open topological strings. 
Here we would like to provide some intuition for its potential role in physics, with the~hope to motivate further work on this question.

The low energy limit of M-theory is eleven-dimensional supergravity, whose degrees of freedom include the~metric and a~three-form $A_3$. M2-branes are electrically charged under $A_3$, while M5-branes are magnetic objects. (For mathematicians, M2-branes have 3-dimensional worldsheets and the~electric action is $\int_{M2}A_{3}$. The Hodge dual $\star dA_{3}$ of $dA_3$ is an~$11-4=7$-form. Its potential is a~$6$-form $\tilde A_{6}$, $d\tilde A_{6}=\star dA_{3}$, and the~magnetic action on the~M5-brane with 6-dimensional worldsheet is $\int_{M5}\tilde A_{6}$.)
Consider a~compactification on $X\times S^1\times\IR^4$ where $X$ is a~toric Calabi-Yau threefold, and where M5 wraps a~Lagrangian $L_K\times S^1\times \IR^2$. 
By turning on an~omega-background in the~flat directions they get effectively compactified, and one can reduce the~eleven-dimensional description either to the~five-dimensions of $S^1\times \IR^4$, or to the~six-dimensions of the~Calabi-Yau $X$.

The second viewpoint features a~three-dimensional magnetic object on $L_K\subset X$, sourcing a~field in the~three transverse dimensions, which couples to two-dimensional electric probes wrapping holomorphic curves. For holomorphic curves in the~complement of $L_{K}$ this is a~standard coupling, but for curves with boundary of $L_{K}$ one needs to make choices. This derives from M2-branes that end on M5-branes, where certain lifts of the~4-chain appears in the~action. A proper description of the~six-dimensional M2-M5 dynamics requires a~careful analysis of the~reduction from eleven dimensions, and this is beyond the~scope of this paper. We point out though that from a~linking perspective, the~flux of the~M5-brane can be encoded in a~family of Dirac strings emanating from the~worldsheet of the~M5-brane which gives a~7-chain. The primitive $A_{3}$ of the~flux form then measures linking with this 7-chain, which then can be encoded in yet another family of Dirac strings emanating from the~7-chain and it is likely that the~4-chain is the~reduction of the~latter family of Dirac strings. Again we must consider extra effects when the~M2 has boundary on the~M5, and we expect that the~extra chains enter as parts of the~action.
Instead of going into M-theory, we consider a~simple toy model which shares some qualitative features with the~one of interest.

Consider a~magnetic monopole at rest in four dimensions, its worldline extending in time and sourcing a~static and radially symmetric magnetic field in space. 
The action of a~probe electron moving in its field includes the~term $S_e = \int A\wedge j_e$ where $j_e$ is a~three-form with delta-function support on the~electric worldline $\gamma$. (Often in the~physics literature the~convention is to use $\star j$ instead.) Taking $\gamma$ to be closed, we may therefore write it as $S_e = e \oint_\gamma A$, where $e$ is the~electric charge.
The Maxwell equation in presence of a~magnetic monopole are
\be
	d F = j_m\,, \qquad d\star F = j_e\,.
\ee
Introducing a~Dirac string supported on a~line starting at the~monopole, we may write $F = dA + \theta_D$, where $\theta_D$ is a~two-form with support on the~Dirac string.
The action of the~electron may then further be expressed as $S_e = \int_{S}(F-\theta_D)$ on a~surface $S$ bounded by $\gamma$. The Dirac string contributes for each time it intersects $S$, and ensures that the~action is well-defined under deformations of $S$, including across the~monopole: the~action depends only on $\gamma$.
However, if we deform $\gamma$, the~action changes. 
The action is well-defined as long as  $\gamma$ does not intersect the~Dirac string,  in fact when $\gamma $ \emph{crosses}  the~string, the~action jumps by $4\pi e m$ where $m$ is the~magnetic charge of the~monopole. The Dirac string is a~sort of branch cut for the~action $S_e$, viewed as a~function of $\gamma$.
Multi-valuedness of the~action is invisible to the~quantum physics, as long as charges are quantized.

However, the~factor of $4\pi$ arises from the~integral over a~2-sphere around the~monopole, which is swept by $\gamma$ to obtain the~monodromy. In presence of the~omega-background the~solid angle is modified to $4\pi-\epsilon$ and the~quantum action $e^{i S_e}$ picks up a~nontrivial phase $q e^{i S_e}$ under monodromy, where $\log q \sim \epsilon$.

Let us next consider a~scalar field $\varphi$ living on the~monopole worldline $L$. 
We can let the~electron end on the~monopole, if we couple $\varphi$  to the~charges attached to the~endpoints of the~electric worldline. 
The modified Maxwell equations read
\be
	d F = j_m
	\qquad
	d\star F = j_e + j_m \wedge \varphi \,.
\ee
The electric action is now 
\be
	S_e = \int_\gamma A + \int_{\partial\gamma} \varphi = \int_\gamma A +\varphi(p_2) - \varphi(p_1)
\ee
where we denoted $\partial \gamma = p_2-p_1$ the~endpoints of the~electric worldline on the~magnetic one.
The first term in the~action is very sensitive to the~behavior of $\gamma$ near $L$, it is especially important to keep track of how $\gamma$ ends on $L$. We will denote $v_{in}, v_{out}$ the~vectors in the~normal bundle to the~magnetic worldline.
Now suppose we want to study an~ensemble of electrons with endpoints on $L$.
If we require all electrons to have the~same endpoints $p_1, p_2$ and also the~same behavior near $L$, dictated by $v_{in}, v_{out}$, it is possible to compare their actions in a~canonical way.
Sufficiently close to $L$ any two such paths $\gamma,\gamma'$ must eventually coincide, and therefore we can define the~closed path  $\gamma-\gamma' = \eta$. The difference of the~two actions can then be defined as the~integral along $\eta$
\be
	S_\gamma- S_\gamma' = \(\int_\gamma A + \int_{p_2-p_1} \varphi \) - \(\int_{\gamma'} A + \int_{p_2-p_1} \varphi\)
	=  \oint_\eta A \,.
\ee
This is an~integral on a~closed cycle in the~bulk, like the~ones considered previously. 
As we have argued, the~integral is multi-valued, with a~quantized shift appearing whenever $\eta$ is deformed across the~Dirac string and brought back to its original shape. 
This means that two open electric paths differing only by an~infinitesimal curl near $L$ (while both ending along $v_{in}, v_{out}$) will have actions differing by $e^{i S_{\gamma'}}\sim  q^{1/2}\, e^{i S_{\gamma}}$.

While the~situation with holomorphic disks ending on Lagrangians is different in several regards, it is still the~case that in order to define the~action of M2-branes one needs to  specify how they end on the~M5, especially in the~case of BPS states. 
Only if all M2-branes have the~same behavior near the~M5, their actions can be properly compared to one another.
This information, in particular the~normal direction of the~M2, is stored in the~geometry of the~four-chain.

\newpage
\appendix
\section{Cross-check of $\mathbf{t}$ refinement\label{sub:Comparison-of-t-refinement}}
In order to compare refined LMOV invariants for trefoil obtained in
Section \ref{sub:Trefoil} with \cite{GKS1504} we have to adjust conventions
and take care about some subtleties. 

Authors of \cite{GKS1504} use the convention of $t_{r}$ refinement with
$-a^{2}t^{3}$ instead of $-a^{2}t$ in the~$q$-Pochhammer. Moreover,
in section 4.7 concerning refined classical LMOV invariants they consider
``super-A-polynomials as $T$-deformation and $a$-deformation of
bottom A-polynomials (that arise for $a=0$ and $T=1$), where $t=-T^{2}$'',
which is equivalent to rescaling of $N_{r}^{3_{1}}(a,q,t)$ by the
overall $a$ factor in such a~way that it starts from $1$. Therefore,
the refined LMOV generating function for
the fundamental representation should be transformed as follows
\[
\begin{split}
N_{1}^{3_{1}}(a,q,t)&=\left(1+a^{2}t\right)\left(a^{2}q^{-2}+a^{2}q^{2}t^{2}+a^{4}t^{3}\right)\\
&\rightsquigarrow\left(1+a^{2}t^{3}\right)\left(1+q^{4}t^{2}+a^{2}q^{2}t^{3}\right)
\end{split}
\]
In order to see $1$ clearly, we rescaled also the~$q$ power, which
however does not matter due to the~semiclassical limit. It will also hide
the $q$-shift between definitions of denominators of $N_{r}(a,q,t)$
 (we have $1-q^{2}$, they have $q-q^{-1}$), however the~sign
difference will stay. In consequence 
\[
N_{1}^{3_{1}}(a,q=1,t)\rightsquigarrow-\left(1+a^{2}t^{3}\right)\left(1+t^{2}+a^{2}t^{3}\right).
\]
Finally, authors of \cite{GKS1504} put $t=-T^{2}$ and
then rescale $a\rightarrow a^{1/2},\ T\rightarrow T^{1/2}$ to reduce the~volume of Table 10 in the~reference, which contains classical refined LMOV invariants.
Unfortunately the~last rescaling is not explicitly stated, for which
the common author of the~two papers apologizes. They consider only terms
up to power $5$, so summing up
\[
N_{1}^{3_{1}}(a,q=1,t)\rightsquigarrow-\left(1-aT^{3}\right)\left(1+T^{2}-aT^{3}\right)=-1-T^{2}+2aT^{3}+aT^{5}+\ldots
\]
which is exactly equal to $\sum_{i,j}\tilde{b}_{1,i,j}a^{i}T^{j}$
from  \cite[Table 10]{GKS1504} ($\tilde{b}_{r,i,j}$ denotes classical
refined LMOV invariants in representation $r$ corresponding to power
$i$ of $a$ and power $j$ of $T$). 

We can repeat above steps for $r=2,3,4$ to obtain
\begin{align*}
N_{2}^{3_{1}}(a,q=1,t)\rightsquigarrow & T^{2}-aT^{3}+T^{4}-5aT^{5}+\ldots\\
N_{3}^{3_{1}}(a,q=1,t)\rightsquigarrow & -2T^{4}+7aT^{5}+\ldots\\
N_{4}^{3_{1}}(a,q=1,t)\rightsquigarrow & T^{4}-3aT^{5}+\ldots
\end{align*}
which again matches results from  \cite[Table 10]{GKS1504} perfectly.

\nocite{*}


\begin{thebibliography}{10}

\bibitem{Kucharski:2017poe}
P.~Kucharski, M.~Reineke, M.~Stosic, and P.~Sulkowski, {\it {BPS states, knots
  and quivers}},  {\em Phys. Rev.} {\bf D96} (2017), no.~12 121902,
  [\href{http://arxiv.org/abs/1707.02991}{{\tt arXiv:1707.02991}}].

\bibitem{Kucharski:2017ogk}
P.~Kucharski, M.~Reineke, M.~Stosic, and P.~Sulkowski, {\it {Knots-quivers
  correspondence}},   {\em Adv. Theor. Math. Phys.} {\bf 23} (2019), no.~7
 1849--1902, [\href{http://arxiv.org/abs/1707.04017}{{\tt
  arXiv:1707.04017}}].

\bibitem{Ooguri:1999bv}
H.~Ooguri and C.~Vafa, {\it {Knot invariants and topological strings}},  {\em
  Nucl. Phys.} {\bf B577} (2000) 419--438,
  [\href{http://arxiv.org/abs/hep-th/9912123}{{\tt hep-th/9912123}}].

\bibitem{LM0004}
J.~M.~F. Labastida and M.~Marino, {\it Polynomial invariants for torus knots
  and topological strings},  {\em Comm. Math. Phys.} {\bf 217} (2001), no.~2
  423--449, [\href{http://arxiv.org/abs/hep-th/0004196}{{\tt hep-th/0004196}}].

\bibitem{LMV0010}
J.~M.~F. Labastida, M.~Marino, and C.~Vafa, {\it Knots, links and branes at
  large {$N$}},  {\em JHEP} {\bf 11} (2000) 007,
  [\href{http://arxiv.org/abs/hep-th/0010102}{{\tt hep-th/0010102}}].

\bibitem{AV1204}
M.~Aganagic and C.~Vafa, {\it Large {N} duality, mirror symmetry, and a
  {Q}-deformed {A}-polynomial for knots},
  \href{http://arxiv.org/abs/1204.4709}{{\tt arXiv:1204.4709}}.

\bibitem{Fuji:2012nx}
H.~Fuji, S.~Gukov, and P.~Sulkowski, {\it {Super-A-polynomial for knots and BPS
  states}},  {\em Nucl. Phys.} {\bf B867} (2013) 506--546,
  [\href{http://arxiv.org/abs/1205.1515}{{\tt arXiv:1205.1515}}].

\bibitem{Witten:1992fb}
E.~Witten, {\it {Chern-Simons gauge theory as a string theory}},  {\em Prog.
  Math.} {\bf 133} (1995) 637--678,
  [\href{http://arxiv.org/abs/hep-th/9207094}{{\tt hep-th/9207094}}].

\bibitem{KS1608}
P.~Kucharski and P.~Sulkowski, {\it {BPS} counting for knots and combinatorics
  on words},  {\em JHEP} {\bf 11} (2016) 120,
  [\href{http://arxiv.org/abs/1608.06600}{{\tt arXiv:1608.06600}}].

\bibitem{LZ1611}
W.~Luo and S.~Zhu, {\it Integrality structures in topological strings {I}:
  framed unknot},  \href{http://arxiv.org/abs/1611.06506}{{\tt
  arXiv:1611.06506}}.

\bibitem{Zhu1707}
S.~Zhu, {\it Topological strings, quiver varieties and {Rogers-Ramanujan}
  identities}, {\em Ramanujan J.}  {\bf 48} (2019) 2, 399--421 [\href{http://arxiv.org/abs/1707.00831}{{\tt arXiv:1707.00831}}].

\bibitem{Aganagic:2013jpa}
M.~Aganagic, T.~Ekholm, L.~Ng, and C.~Vafa, {\it Topological strings,
  {D}-model, and knot contact homology},  {\em Adv. Theor. Math. Phys.} {\bf
  18} (2014), no.~4 827--956, [\href{http://arxiv.org/abs/1304.5778}{{\tt
  arXiv:1304.5778}}].

\bibitem{ES}
T.~Ekholm and V.~Shende, {\it {Skeins on branes}},  \href{http://arxiv.org/abs/1901.08027}{{\tt
  arXiv:1901.08027}}.

\bibitem{Dimofte:2010tz}
T.~Dimofte, S.~Gukov, and L.~Hollands, {\it Vortex counting and {Lagrangian}
  3-manifolds},  {\em Lett. Math. Phys.} {\bf 98} (2011) 225--287,
  [\href{http://arxiv.org/abs/1006.0977}{{\tt arXiv:1006.0977}}].

\bibitem{Terashima:2011qi}
Y.~Terashima and M.~Yamazaki, {\it {SL(2,R) Chern-Simons, Liouville}, and gauge
  theory on duality walls},  {\em JHEP} {\bf 08} (2011) 135,
  [\href{http://arxiv.org/abs/1103.5748}{{\tt arXiv:1103.5748}}].

\bibitem{Dimofte:2011ju}
T.~Dimofte, D.~Gaiotto, and S.~Gukov, {\it Gauge theories labelled by
  three-manifolds},  {\em Commun. Math. Phys.} {\bf 325} (2014) 367--419,
  [\href{http://arxiv.org/abs/1108.4389}{{\tt arXiv:1108.4389}}].

\bibitem{Yag1305}
J.~Yagi, {\it 3d {TQFT} from 6d {SCFT}},  {\em JHEP} {\bf 08} (2013) 017,
  [\href{http://arxiv.org/abs/1305.0291}{{\tt arXiv:1305.0291}}].

\bibitem{LY1305}
S.~Lee and M.~Yamazaki, {\it 3d {Chern-Simons} theory from {M5-branes}},  {\em
  JHEP} {\bf 12} (2013) 035, [\href{http://arxiv.org/abs/1305.2429}{{\tt
  arXiv:1305.2429}}].

\bibitem{Cordova:2013cea}
C.~Cordova and D.~L. Jafferis, {\it {Complex Chern-Simons} from {M5}-branes on
  the squashed three-sphere},  {\em JHEP} {\bf 11} (2017) 119,
  [\href{http://arxiv.org/abs/1305.2891}{{\tt arXiv:1305.2891}}].

\bibitem{DGR0505}
N.~M. Dunfield, S.~Gukov, and J.~Rasmussen, {\it The superpolynomial for knot
  homologies},  {\em Experiment. Math.} {\bf 15} (2006), no.~2 129--159,
  [\href{http://arxiv.org/abs/math/0505662}{{\tt math/0505662}}].

\bibitem{Hwang:2017kmk}
C.~Hwang, P.~Yi, and Y.~Yoshida, {\it Fundamental vortices, wall-crossing, and
  particle-vortex duality},  {\em JHEP} {\bf 05} (2017) 099,
  [\href{http://arxiv.org/abs/1703.00213}{{\tt arXiv:1703.00213}}].

\bibitem{Denef:2002ru}
F.~Denef, {\it {Quantum quivers and Hall / hole halos}},  {\em JHEP} {\bf 10}
  (2002) 023, [\href{http://arxiv.org/abs/hep-th/0206072}{{\tt
  hep-th/0206072}}].

\bibitem{Alim:2011kw}
M.~Alim, S.~Cecotti, C.~Cordova, S.~Espahbodi, A.~Rastogi, and C.~Vafa, {\it
  {$\mathcal{N} = 2$ quantum field theories and their BPS quivers}},  {\em Adv.
  Theor. Math. Phys.} {\bf 18} (2014), no.~1 27--127,
  [\href{http://arxiv.org/abs/1112.3984}{{\tt arXiv:1112.3984}}].

\bibitem{Gabella:2017hpz}
M.~Gabella, P.~Longhi, C.~Y. Park, and M.~Yamazaki, {\it {BPS} graphs: From
  spectral networks to {BPS} quivers},  {\em JHEP} {\bf 07} (2017) 032,
  [\href{http://arxiv.org/abs/1704.04204}{{\tt arXiv:1704.04204}}].


\bibitem{OP}
A.~Okounkov, R.~Pandharipande,
{\it Hodge integrals and invariants of the unknot}
{\em Geom. Topol.},
{\bf 8} (2004) 675--699.

\bibitem{KL}
S.~Katz, C.-C.~M.~Liu,
{\it Enumerative geometry of stable maps with Lagrangian boundary conditions and multiple covers of the disc}, {\em Adv. Theor. Math. Phys.}  {\bf 5} (2001) no. 1, 1--49.

\bibitem{Ekholm:2018iso}
T.~Ekholm and L.~Ng, {\it {Higher genus knot contact homology and recursion for
  colored HOMFLY-PT polynomials}},  \href{http://arxiv.org/abs/1803.04011}{{\tt
  arXiv:1803.04011}}.

\bibitem{iacovino1}
V.~Iacovino, {\it Open Gromov-Witten theory on Calabi-Yau three-folds I},
  \href{http://arxiv.org/abs/0907.5225}{{\tt arXiv:0907.5225}}.

\bibitem{iacovino2}
V.~Iacovino, {\it Open Gromov-Witten theory on Calabi-Yau three-folds II},
  \href{http://arxiv.org/abs/0908.0393}{{\tt arXiv:0908.0393}}.

\bibitem{iacovino3}
V.~Iacovino, {\it Frame ambiguity in open Gromov-Witten invariants},
  \href{http://arxiv.org/abs/1003.4684}{{\tt arXiv:1003.4684}}.

\bibitem{freyd1985}
P.~Freyd, D.~Yetter, J.~Hoste, W.~B.~R. Lickorish, K.~Millett, and A.~Ocneanu,
  {\it A new polynomial invariant of knots and links},  {\em Bull. Amer. Math.
  Soc. (N.S.)} {\bf 12} (1985), no.~2 239--246.

\bibitem{PT}
J.~Przytycki and P.~Traczyk, {\it Invariants of links of {Conway} type},  {\em
  Kobe J. Math.} {\bf 4} (1987) 115--139.

\bibitem{witten1989}
E.~Witten, {\it Quantum field theory and the {Jones} polynomial},  {\em Comm.
  Math. Phys.} {\bf 121} (1989), no.~3 351--399.

\bibitem{kirillov2016quiver}
A.~Kirillov, {\em Quiver representations and quiver varieties}.
\newblock Graduate Studies in Mathematics. American Mathematical Society, 2016.

\bibitem{KS0811}
M.~Kontsevich and Y.~Soibelman, {\it Stability structures, motivic
  {Donaldson-Thomas} invariants and cluster transformations},
  \href{http://arxiv.org/abs/0811.2435}{{\tt arXiv:0811.2435}}.

\bibitem{KS1006}
M.~Kontsevich and Y.~Soibelman, {\it Cohomological {Hall} algebra, exponential
  {Hodge} structures and motivic {Donaldson-Thomas} invariants},  {\em
  Commun.Num.Theor.Phys.} {\bf 5} (2011) 231--352,
  [\href{http://arxiv.org/abs/1006.2706}{{\tt arXiv:1006.2706}}].

\bibitem{MR1411}
S.~Meinhardt and M.~Reineke, {\it {Donaldson-Thomas} invariants versus
  intersection cohomology of quiver moduli},
  \href{http://arxiv.org/abs/1411.4062}{{\tt arXiv:1411.4062}}.

\bibitem{FR1512}
H.~Franzen and M.~Reineke, {\it Semi-stable {Chow-Hall} algebras of quivers and
  quantized {Donaldson-Thomas} invariants},
  \href{http://arxiv.org/abs/1512.03748}{{\tt arXiv:1512.03748}}.

\bibitem{2011arXiv1103.2736E}
A.~I. {Efimov}, {\it {Cohomological Hall algebra of a symmetric quiver}}, {\em Compos. Math.} {\bf 148} (2012) 4, 1133-1146
  [\href{http://arxiv.org/abs/1103.2736}{{\tt arXiv:1103.2736}}].

\bibitem{Stosic:2017wno}
M.~Stosic and P.~Wedrich, {\it {Rational links and DT invariants of quivers}}, {\em Int. Math. Res. Notices}, {\bf rny289} (2019)
  [\href{http://arxiv.org/abs/1711.03333}{{\tt arXiv:1711.03333}}.]

\bibitem{PSS1802}
M.~Panfil, P.~Sulkowski, and M.~Stosic, {\it {Donaldson-Thomas} invariants,
  torus knots, and lattice paths},  {\em Phys. Rev.} {\bf D98} (2018), no.~2
  026022, [\href{http://arxiv.org/abs/1802.04573}{{\tt arXiv:1802.04573}}].

\bibitem{PS18}
P.~Sulkowski and M.~Panfil, {\it Topological strings, strips and quivers}, {\em
  JHEP} {\bf 01} (2019) 124,  [\href{http://arxiv.org/abs/1811.03556}{{\tt arXiv:1811.03556}}].

\bibitem{Gopakumar:1998ii}
R.~Gopakumar and C.~Vafa, {\it M-theory and topological strings -- {I}},
  \href{http://arxiv.org/abs/hep-th/9809187}{{\tt hep-th/9809187}}.

\bibitem{Gopakumar:1998jq}
R.~Gopakumar and C.~Vafa, {\it M-theory and topological strings -- {II}},
  \href{http://arxiv.org/abs/hep-th/9812127}{{\tt hep-th/9812127}}.

\bibitem{Douglas:1996sw}
M.~R. Douglas and G.~W. Moore, {\it {D-branes, quivers, and ALE instantons}},
  \href{http://arxiv.org/abs/hep-th/9603167}{{\tt hep-th/9603167}}.

\bibitem{Fiol:2000wx}
B.~Fiol and M.~Marino, {\it {BPS states and algebras from quivers}},  {\em
  JHEP} {\bf 07} (2000) 031, [\href{http://arxiv.org/abs/hep-th/0006189}{{\tt
  hep-th/0006189}}].

\bibitem{Alim:2011ae}
M.~Alim, S.~Cecotti, C.~Cordova, S.~Espahbodi, A.~Rastogi, and C.~Vafa, {\it
  {BPS} quivers and spectra of complete {N=2} quantum field theories},  {\em
  Commun. Math. Phys.} {\bf 323} (2013) 1185--1227,
  [\href{http://arxiv.org/abs/1109.4941}{{\tt arXiv:1109.4941}}].

\bibitem{GV9811}
R.~Gopakumar and C.~Vafa, {\it On the gauge theory / geometry correspondence},
  {\em Adv. Theor. Math. Phys.} {\bf 3} (1999) 1415--1443,
  [\href{http://arxiv.org/abs/hep-th/9811131}{{\tt hep-th/9811131}}].

\bibitem{Aganagic:2000gs}
M.~Aganagic and C.~Vafa, {\it {Mirror symmetry, D-branes and counting
  holomorphic discs}},  \href{http://arxiv.org/abs/hep-th/0012041}{{\tt
  hep-th/0012041}}.

\bibitem{Aganagic:2001nx}
M.~Aganagic, A.~Klemm, and C.~Vafa, {\it {Disk instantons, mirror symmetry and
  the duality web}},  {\em Z. Naturforsch.} {\bf A57} (2002) 1--28,
  [\href{http://arxiv.org/abs/hep-th/0105045}{{\tt hep-th/0105045}}].

\bibitem{Shadchin:2006yz}
S.~Shadchin, {\it {On F-term contribution to effective action}},  {\em JHEP}
  {\bf 08} (2007) 052, [\href{http://arxiv.org/abs/hep-th/0611278}{{\tt
  hep-th/0611278}}].

\bibitem{Witten:1978mh}
E.~Witten and D.~I. Olive, {\it Supersymmetry algebras that include topological
  charges},  {\em Phys. Lett.} {\bf 78B} (1978) 97--101.

\bibitem{Hori:2014tda}
K.~Hori, H.~Kim, and P.~Yi, {\it Witten index and wall crossing},  {\em JHEP}
  {\bf 01} (2015) 124, [\href{http://arxiv.org/abs/1407.2567}{{\tt
  arXiv:1407.2567}}].

\bibitem{Khovanov}
M.~Khovanov, {\it A categorification of the {Jones} polynomial},  {\em Duke
  Math. J.} {\bf 101} (2000) 359--426,
  [\href{http://arxiv.org/abs/math/9908171}{{\tt math/9908171}}].

\bibitem{KhR1}
M.~Khovanov and L.~Rozansky, {\it {Matrix factorizations and link homology}},
  {\em Fund. Math.} {\bf 199} (2008) 1--91,
  [\href{http://arxiv.org/abs/math/0401268}{{\tt math/0401268}}].

\bibitem{KhR2}
M.~Khovanov and L.~Rozansky, {\it {Matrix factorizations and link homology
  II}},  {\em Geom. \& Topol.} {\bf 12} (2008) 1387--1425,
  [\href{http://arxiv.org/abs/math/0505056}{{\tt math/0505056}}].

\bibitem{GNSSS1512}
S.~Gukov, S.~Nawata, I.~Saberi, M.~Stosic, and P.~Sulkowski, {\it Sequencing
  {BPS} spectra},  {\em JHEP} {\bf 03} (2016) 004,
  [\href{http://arxiv.org/abs/1512.07883}{{\tt arXiv:1512.07883}}].

\bibitem{Gukov:2011ry}
S.~Gukov and M.~Stosic, {\it Homological algebra of knots and {BPS} states},
  {\em Proc. Symp. Pure Math.} {\bf 85} (2012) 125--172,
  [\href{http://arxiv.org/abs/1112.0030}{{\tt arXiv:1112.0030}}]. [Geom. Topol.
  Monographs 18 (2012) 309].

\bibitem{GGS1304}
E.~Gorsky, S.~Gukov, and M.~Stosic, {\it Quadruply-graded colored homology of
  knots},  \href{http://arxiv.org/abs/1304.3481}{{\tt arXiv:1304.3481}}.

\bibitem{GLL1604}
S.~Garoufalidis, A.~D. Lauda, and T.~T.~Q. Le, {\it The colored {HOMFLY-PT}
  polynomial is {$q$}-holonomic},  {\em Duke Math. J.} {\bf 167} (2018), no.~3
  397--447, [\href{http://arxiv.org/abs/1604.08502}{{\tt arXiv:1604.08502}}].

\bibitem{FGSS1209}
H.~Fuji, S.~Gukov, P.~Sulkowski, and M.~Stosic, {\it {3d analogs of
  Argyres-Douglas theories and knot homologies}},  {\em JHEP} {\bf 01} (2013)
  175, [\href{http://arxiv.org/abs/1209.1416}{{\tt arXiv:1209.1416}}].

\bibitem{Ng0407}
L.~Ng, {\it Framed knot contact homology},  {\em Duke Math. J.} {\bf 141}
  (2008), no.~2 365--406, [\href{http://arxiv.org/abs/math/0407071}{{\tt
  math/0407071}}].

\bibitem{Ng1010}
L.~Ng, {\it Combinatorial knot contact homology and transverse knots},  {\em
  Adv. Math.} {\bf 227} (2011), no.~6 2189--2219,
  [\href{http://arxiv.org/abs/1010.0451}{{\tt arXiv:1010.0451}}].

\bibitem{GKS1504}
S.~Garoufalidis, P.~Kucharski, and P.~Sulkowski, {\it Knots, {BPS} states, and
  algebraic curves},  {\em Comm. Math. Phys.} {\bf 346} (2016), no.~1 75--113,
  [\href{http://arxiv.org/abs/1504.06327}{{\tt arXiv:1504.06327}}].

\bibitem{Chung:2014qpa}
H.-J. Chung, T.~Dimofte, S.~Gukov, and P.~Sulkowski, {\it 3d-3d correspondence
  revisited},  {\em JHEP} {\bf 04} (2016) 140,
  [\href{http://arxiv.org/abs/1405.3663}{{\tt arXiv:1405.3663}}].

\bibitem{2012arXiv1210.4803N}
L.~{Ng}, {\it {A topological introduction to knot contact homology}},
  \href{http://arxiv.org/abs/1210.4803}{{\tt arXiv:1210.4803}}.

\bibitem{1983NuPhB.222...45D}
A.~{D'adda}, A.~C. {Davis}, P.~{Di Vecchia}, and P.~{Salomonson}, {\it {An
  effective action for the supersymmetric CP $^{n-1}$ model}},  {\em Nuclear
  Physics} {\bf  B222} (July, 1983) 45--70.

\bibitem{Witten:1993yc}
E.~Witten, {\it {Phases of N=2 theories in two-dimensions}},  {\em Nucl. Phys.}
  {\bf B403} (1993) 159--222, [\href{http://arxiv.org/abs/hep-th/9301042}{{\tt
  hep-th/9301042}}]. [AMS/IP Stud. Adv. Math. 1 (1996) 143-211].

\bibitem{Hanany:1997vm}
A.~Hanany and K.~Hori, {\it {Branes and N=2 theories in two-dimensions}},  {\em
  Nucl. Phys.} {\bf B513} (1998) 119--174,
  [\href{http://arxiv.org/abs/hep-th/9707192}{{\tt hep-th/9707192}}].

\bibitem{Hori:2000kt}
K.~Hori and C.~Vafa, {\it {Mirror symmetry}},
  \href{http://arxiv.org/abs/hep-th/0002222}{{\tt hep-th/0002222}}.

\bibitem{Dimofte:2011jd}
T.~Dimofte and S.~Gukov, {\it {Chern-Simons} theory and {S}-duality},  {\em
  JHEP} {\bf 05} (2013) 109, [\href{http://arxiv.org/abs/1106.4550}{{\tt
  arXiv:1106.4550}}].

\bibitem{Lawrence:1997jr}
A.~E. Lawrence and N.~Nekrasov, {\it {Instanton sums and five-dimensional gauge
  theories}},  {\em Nucl. Phys.} {\bf B513} (1998) 239--265,
  [\href{http://arxiv.org/abs/hep-th/9706025}{{\tt hep-th/9706025}}].

\bibitem{Gaiotto:2008cd}
D.~Gaiotto, G.~W. Moore, and A.~Neitzke, {\it Four-dimensional wall-crossing
  via three-dimensional field theory},  {\em Commun. Math. Phys.} {\bf 299}
  (2010) 163--224, [\href{http://arxiv.org/abs/0807.4723}{{\tt
  arXiv:0807.4723}}].

\bibitem{Aharony:1997bx}
O.~Aharony, A.~Hanany, K.~A. Intriligator, N.~Seiberg, and M.~J. Strassler,
  {\it {Aspects of N=2 supersymmetric gauge theories in three-dimensions}},
  {\em Nucl. Phys.} {\bf B499} (1997) 67--99,
  [\href{http://arxiv.org/abs/hep-th/9703110}{{\tt hep-th/9703110}}].

\bibitem{Smo2017}
P.~Smolinski, {\it From topological strings to quantum invariants of knots and
  quivers},  Master's thesis, {University of Warsaw}, 2017.

\bibitem{Aganagic:2012hs}
M.~Aganagic and S.~Shakirov, {\it {Refined Chern-Simons} theory and topological
  string},  \href{http://arxiv.org/abs/1210.2733}{{\tt arXiv:1210.2733}}.

\bibitem{Beem:2012mb}
C.~Beem, T.~Dimofte, and S.~Pasquetti, {\it Holomorphic blocks in three
  dimensions},  {\em JHEP} {\bf 12} (2014) 177,
  [\href{http://arxiv.org/abs/1211.1986}{{\tt arXiv:1211.1986}}].

\bibitem{Hwang:2012jh}
C.~Hwang, H.-C. Kim, and J.~Park, {\it {Factorization of the 3d superconformal
  index}},  {\em JHEP} {\bf 08} (2014) 018,
  [\href{http://arxiv.org/abs/1211.6023}{{\tt arXiv:1211.6023}}].

\bibitem{Bullimore:2016hdc}
M.~Bullimore, T.~Dimofte, D.~Gaiotto, J.~Hilburn, and H.-C. Kim, {\it {Vortices
  and Vermas}},  \href{http://arxiv.org/abs/1609.04406}{{\tt
  arXiv:1609.04406}}.

\bibitem{Gadde:2013wq}
A.~Gadde, S.~Gukov, and P.~Putrov, {\it Walls, lines, and spectral dualities in
  3d gauge theories},  {\em JHEP} {\bf 05} (2014) 047,
  [\href{http://arxiv.org/abs/1302.0015}{{\tt arXiv:1302.0015}}].

\bibitem{Gukov:2007ck}
S.~Gukov, {\it Gauge theory and knot homologies},  {\em Fortsch. Phys.} {\bf
  55} (2007) 473--490, [\href{http://arxiv.org/abs/0706.2369}{{\tt
  arXiv:0706.2369}}].

\bibitem{GSV0412}
S.~Gukov, A.~Schwarz, and C.~Vafa, {\it {Khovanov-Rozansky} homology and
  topological strings},  {\em Lett. Math. Phys.} {\bf 74} (2005) 53--74,
  [\href{http://arxiv.org/abs/hep-th/0412243}{{\tt hep-th/0412243}}].

\bibitem{Wed1602}
P.~Wedrich, {\it Exponential growth of colored {HOMFLY-PT} homology}, {\em Adv.Math.} {\bf 353} (2019) 471--525,
  [\href{http://arxiv.org/abs/1602.02769}{{\tt arXiv:1602.02769}}].

\bibitem{KN1703}
M.~Kameyama and S.~Nawata, {\it Refined large {N} duality for knots},
  \href{http://arxiv.org/abs/1703.05408}{{\tt arXiv:1703.05408}}.

\bibitem{2009arXiv0909.5153G}
M.~{Gross} and R.~{Pandharipande}, {\it {Quivers, curves, and the tropical
  vertex}},  \href{http://arxiv.org/abs/0909.5153}{{\tt arXiv:0909.5153}}.

\bibitem{Galakhov:2013oja}
D.~Galakhov, P.~Longhi, T.~Mainiero, G.~W. Moore, and A.~Neitzke, {\it Wild
  wall crossing and {BPS} giants},  {\em JHEP} {\bf 11} (2013) 046,
  [\href{http://arxiv.org/abs/1305.5454}{{\tt arXiv:1305.5454}}].

\bibitem{Seiberg:1994aj}
N.~Seiberg and E.~Witten, {\it {Monopoles, duality and chiral symmetry breaking
  in N=2 supersymmetric QCD}},  {\em Nucl. Phys.} {\bf B431} (1994) 484--550,
  [\href{http://arxiv.org/abs/hep-th/9408099}{{\tt hep-th/9408099}}].

\bibitem{Witten:1997sc}
E.~Witten, {\it {Solutions of four-dimensional field theories via M theory}},
  {\em Nucl. Phys.} {\bf B500} (1997) 3--42,
  [\href{http://arxiv.org/abs/hep-th/9703166}{{\tt hep-th/9703166}}].

\bibitem{Gaiotto:2009hg}
D.~Gaiotto, G.~W. Moore, and A.~Neitzke, {\it Wall-crossing, {Hitchin} systems,
  and the {WKB} approximation},  \href{http://arxiv.org/abs/0907.3987}{{\tt
  arXiv:0907.3987}}.

\bibitem{Gaiotto:2012rg}
D.~Gaiotto, G.~W. Moore, and A.~Neitzke, {\it {Spectral networks}},  {\em
  Annales Henri Poincare} {\bf 14} (2013) 1643--1731,
  [\href{http://arxiv.org/abs/1204.4824}{{\tt arXiv:1204.4824}}].

\end{thebibliography}

\providecommand{\href}[2]{#2}\begingroup\raggedright\endgroup

\end{document}